%% file: Manuscript_PEPI.tex
\documentclass[article,10pt]{elsarticle}
\usepackage{natbib}
\usepackage{geometry}

\journal{Physics of the Earth and Planetary Interiors}

\usepackage{lineno,hyperref}
\modulolinenumbers[5]

\usepackage{graphicx}
\usepackage{epstopdf, epsfig}
\usepackage{placeins} 
\usepackage{amsmath,amssymb}
\usepackage{xcolor}
\usepackage{ulem}
\usepackage{csvsimple}
\usepackage{booktabs}

\newcommand{\vect}[1]{\ensuremath{\mathbf{#1}}}
\newcommand{\Pm}{\ensuremath{Pm}}
\newcommand{\Rm}{\ensuremath{Rm}}
\newcommand{\Ra}{\ensuremath{{Ra}}}
\newcommand{\Rac}{\ensuremath{{Ra_c}}}
\newcommand{\Ek}{\ensuremath{E}}
\newcommand{\Rol}{\ensuremath{{Ro_\ell}}}
\newcommand{\fdip}{\ensuremath{{f_{dip}}}}

\usepackage{caption}
\graphicspath{{figures/}} 
\usepackage{subfigure} 

\usepackage[T1]{fontenc}
\usepackage[utf8]{inputenc}

\usepackage{lineno}

\begin{document}

\begin{frontmatter}

\title{Magnetic effects on fields morphologies and reversals in geodynamo simulations}

\author[1,2]{M\'elissa D. Menu}

\author[2]{Ludovic Petitdemange}

\author[3]{S\'ebastien Galtier}

\address[1]{Laboratoire de Physique des Plasmas, Universit\'e Paris-Saclay, \'Ecole polytechnique, CNRS, Sorbonne Universit\'e, Observatoire de Paris, F-91128 Palaiseau Cedex, France}
\address[2]{LERMA, CNRS, Observatoire de Paris, PSL Research University, Sorbonne Universit\'e, F-75005 Paris, France}
\address[3]{Laboratoire de Physique des Plasmas, Universit\'e Paris-Saclay, Institut Universitaire de France, \'Ecole polytechnique, CNRS, Sorbonne Universit\'e, Observatoire de Paris, F-91128 Palaiseau Cedex, France}
\date{\today}

\cortext[cor1]{melissa.d.menu@gmail.com}


\begin{abstract}
\input{abstract}
\end{abstract}


\end{frontmatter}


\section{Introduction}

\input{intro}

\section{Model and  input/output parameters}
We follow previous extensive parameter studies \citep{christensen06,olson06,schrinner12,Petitdemange2018} and use the Boussinesq approximation. The fluid freely evolves between an inner and outer spherical shell, respectively at radius $r_i$ and $r_o$, with no-slip boundary conditions. The aspect ratio is defined as $\chi = r_i /  r_o$ and is fixed to 0.35. The evolution is constrained by the classical Boussinesq magneto-hydrodynamic (MHD) equations with rotation (momentum and induction), and an additional equation constraining the temperature evolution. A temperature contrast $\Delta T$ is imposed between the inner and outer shells to generate convection and the magnetic field matches a potential field outside the fluid shell (insulating boundaries). The law considered for the acceleration of gravity is $\vect{g}(r) = g_0 \frac{\vect{r}}{r_o}$.

Length is expressed in boundary shell gap units $D = r_o - r_i$, time in units of $D^2 / \nu$, magnetic field $\vect{B}$ in units of $(\rho \mu \eta \Omega)^{1/2}$, pressure in units of $\rho \nu \Omega$ and temperature in units of $\Delta T$, where $\rho$  is the density, $\Omega$ the rotation rate, $\nu$, $\mu$, $\eta$ and $\kappa$ are the kinematic viscosity, magnetic permeability, magnetic diffusivity and thermal diffusivity respectively.
These units are used to obtain the dimensionless equations :

\begin{equation}
    	\Ek \bigg( \frac{\partial\vect{u}}{\partial t} + \vect{u}\cdot\ \vect{\nabla}\vect{u}  - \nu\nabla^2\vect{u} \bigg)= -\vect{\nabla}P - 2\vect{z}\times\vect{u}+\Ra \frac{\vect{r}}{r_o}T + \frac{1}{\Pm}(\vect{\nabla}\times\vect{B})\times\vect{B}\,,
    	\label{amomentum}\\
\end{equation}

\begin{equation}
 	\frac{\partial\vect{B}}{\partial t} = \nabla\times(\vect{u}\times\vect{B})+\frac{1}{\Pm}\nabla^2\vect{B}\,,  
	\label{ainduction}
\end{equation}

\begin{equation}
	\frac{\partial T}{\partial t} + \vect{u}\cdot\nabla T = \frac{1}{Pr} \nabla^2 T\,,
	\label{atemperature}
\end{equation}
	
\begin{equation}
	\nabla \cdot \vect{u} =0 \,,\ \nabla \cdot \vect{B} =0 ,
	\label{aconservation}
\end{equation}

with $\vect{z}$ the direction of the rotation axis and $\vect{r}= r \vect{e_r}$, where $\vect{e_r}$ is the radial direction and $r\in[r_i,r_o]$ the radius considered.
The nondimensionalization of these equations reveals commonly used dimensionless numbers: 
the Ekman number $\Ek=\nu / \Omega D^2$ which represents the ratio of viscous to Coriolis forces, the Rayleigh number $\Ra = \alpha g_o \Delta T D^3 / \nu \kappa$ which can be seen as the ratio of buoyancy to viscous force, the magnetic Prandtl number $\Pm = \nu /\eta$, and the classical Prandtl number $Pr = \nu /\kappa$ which is fixed to 1 in our study. 

To solve equations \eqref{amomentum} to \eqref{aconservation} in spherical coordinates, we use the 3D pseudo-spectral code PaRoDy with the dimensionless quantities defined above as control parameters (see \citet{PaRoDy} for more details). All the simulations performed have been started with an initial dipolar magnetic field. To reach high Rayleigh number values, some simulations have been initialised from the solution of a previous simulation with similar parameter values. 

The main global output quantities we will use in this article are the Elsasser number, the Rossby number, and the magnetic Reynolds number. The Elsasser number $\Lambda = B_{rms}^2 /2\Omega \rho \mu \eta $ gives information about the importance of the Lorentz force compared to the Coriolis force. The Rossby number $Ro = u_{rms} / \Omega D$ compares inertia to Coriolis  force and the magnetic Reynolds number $\Rm = u_{rms} D / \eta$ measures the ratio of the advection to the diffusive term in the induction equation \eqref{ainduction}. $B_{rms}$ and $u_{rms}$ stand for the root mean square magnetic field and velocity respectively. Moreover we have chosen to use the Rayleigh number normalised by its critical value $\Rac$ for convection, whose values has been taken from \citet{christensen06}.

We also compute the local Rossby number $\Rol$ as introduced by \citet{christensen06} to have a better measurement of the balance between inertia and Coriolis force, taking into account a length scale deduced from the mean spherical harmonic degree of the flow $\bar{\ell_u}$ instead of the shell gap $D$:
\begin{equation}
 	\Rol = \frac{Ro}{L_u}\quad  with\quad L_u = \frac{\pi}{\bar{\ell_u}},\quad where\quad \bar{\ell_u}= \frac{\sum_\ell{\ell \langle \vect{u_\ell} \cdot \vect{u_\ell} \rangle}}{2 E_{kin}}\,,
	\label{Rol}
\end{equation}
with $\ell$ the spherical harmonic degree, $E_{kin} = \frac{1}{2}\int{\langle \vect{u}\cdot\vect{u} \rangle \ dV}$ and the brackets $\langle \cdot \rangle$ denotes an average over time, radial and azimuthal directions. The brackets $\langle\cdot\rangle_i$ will be used to denote an average over the quantity $i$. 

Another characteristic length scale, the kinetic dissipation length scale $L_\nu$, can be defined as: 
\begin{equation}
	L_{\nu}^2= \langle \frac{ \int \vect{u}^2 dV}{\int ( \vect{\nabla} \times \vect{u})^2 dV} \rangle_t .
	\label{Lnu}
\end{equation}
This typical length for dissipation enables us to define another local Rossby number $Ro_{L_\nu} = Ro / L_\nu$.

Similarly, a dissipation length scale for the magnetic field exists $L_{\eta}^2 ~ = ~ \langle \frac{ \int \vect{B}^2  dV}{\int (\vect{\nabla} \times \vect{B})^2 dV} \rangle_t $. \citet{Dormy2016} derived a modified Elsasser number $\Lambda ' $ that is more suitable to measure the relative importance of the Lorentz force compared to the Coriolis force. Indeed, it also takes into account the magnetic dissipation length scale and the magnetic Reynolds number: 
\begin{equation}
	\Lambda '= \dfrac{\Lambda\ D}{\Rm\ L_\eta} .
	\label{Lambdador}
\end{equation}

In order to distinguish dynamos with a dominant dipolar component from the others, \citet{christensen06} have calculated the relative dipole field strength $ f_{dip} $, which is the ratio of the mean magnetic dipole component to the others, at the outer shell:
 \begin{equation}
 	f_{dip} = \langle \dfrac{(B_{l=1}\cdot B_{l=1})^{1/2}}{\sum_{l=1}^{12}(B_l \cdot B_l)^{1/2}} \rangle_{\phi, t} .
\end{equation}
Only the first twelve modes are considered in order to compare simulations with geomagnetic observations, for which core fields at smaller length scales are more challenging to distinguish from crustal fields. The typical value of $f_{dip}$ for the Earth is $0.68$ \citep{christensen06}. A transition from a dipolar state ($f_{dip} \gtrsim 0.5$) to a non-dipolar one occurs around a value of the local Rossby number $\Rol \sim 0.1$. This empirical result, called the local Rossby number criteria, seems quite robust as it is valid for multiple values of $\Ek$, $\Pm$ and $Pr$, each of these parameters being varied over at least two orders of magnitude \citep{christensen06}. This feature is interpreted as a boundary between two regimes : dipolar dynamos and multipolar dynamos.\\

Volume-averaged output parameters enable us to compare models with different input parameters. However, physical interactions and, in particular, force balance can depend on the length scale in turbulent systems. We calculate the different forces as function of the spherical harmonic degree $\ell$. To avoid the pressure gradient influence (or the geostrophic balance), we take the azimuthal average of the $\phi$ component of each force $F_i$ as has been done by \citet{Sheyko2017}. This process also discards the Archimedean force. For sake of clarity, we ignored the numerical shells close to the spherical boundaries where viscous effects dominate, took the absolute value and time averaged, leading to $\langle \left| F_i (\ell) \right|\rangle_t $ with 
\begin{equation}
	F_i (\ell) = \sum_{r\ >\ r_i + 0.15 D}^{r\ <\ r_o - 0.15 D} \langle F_i^\phi (l) \rangle_{\phi}\ .
	\label{FBeq}
\end{equation}

The viscous effects are not calculated as it has been shown by \citet{Sheyko2017} that they are low compared to the three forces computed with this method : Lorentz force, inertia and Coriolis force.
Typical resolutions are 288 points in the radial direction (up to 384 points). The spectral decomposition is truncated at $80 \leq l_{max} \sim m_{max} \leq 256$, in order to observe a drop by a factor 100 or more for the kinetic and the magnetic energy spectra of the spherical harmonic degree $l$ and order $m$ from the maximum to the energy cut-off $l_{max}$ and $m_{max}$. The simulations performed are summarised in \ref{Table}.

\section{Results}\label{Results}
\subsection{Two distinct turbulent dipolar dynamos branches with high $Rm$}
\label{Part1}

  At first, it is important to notice different characteristic behaviours observed on the time evolution of output parameters when turbulent dynamos (high buoyant forcing) with $\Rol$ close to the transitional value 0.12 and high $Rm$ (or $Pm$) are considered. As a result, a need for a more precise definition of dipolar dynamos appears, taking into account the presence, or abscence, of reversals. By using this definition, we will be able to study the stability domain of dipolar dynamos in the next sections.

Considering geodynamo simulations with realistic parameters is still far from being a reality. Except for the magnetic Reynolds number ($Rm\sim 1000$), the other parameters differ from realistic ones by several orders of magnitude. Such a value of $Rm$ was considered by recent studies which employed different strategies. \citet{Dormy2016}, \citet{Petitdemange2018} and \citet{Dormy2018} have shown that the Lorentz force plays a major role when $Rm$ is sufficiently high. However, these studies reach high values of $Rm=Pm\, Re$ by considering sufficiently high values of $Pm$ and low values of the Reynolds number $Re = u_{rms} D / \nu $, meaning that their models cannot be really classified as turbulent models. For instance, \citet{Dormy2016} considered simulations close to the onset of convection $(Ra<4Ra_c)$.
\citet{Yadav2016}, \citet{schaefferJNF17} and \citet{Aubert2017} have also obtained a dominant Lorentz force by considering turbulent models at low $\Pm$ and low Ekman numbers.
In our study, we complete this parameter space and we link the different previous studies by considering turbulent models (high values of $\Ra/\Rac$) with high $\Pm$. To this end, the Rayleigh number has been gradually increased with a fixed $\Pm$ value in order to reach this turbulent regime. \\

\begin{figure}
	\subfigure[Multipolar dynamo at $\Ek=10^{-4}$, $\Ra/\Rac = 43.1$, $\Pm =3$, $\Lambda = 12.6$, $\Rol \sim 0.18$, $\fdip \sim 0.25$.]{
	\includegraphics[width=0.46\linewidth]{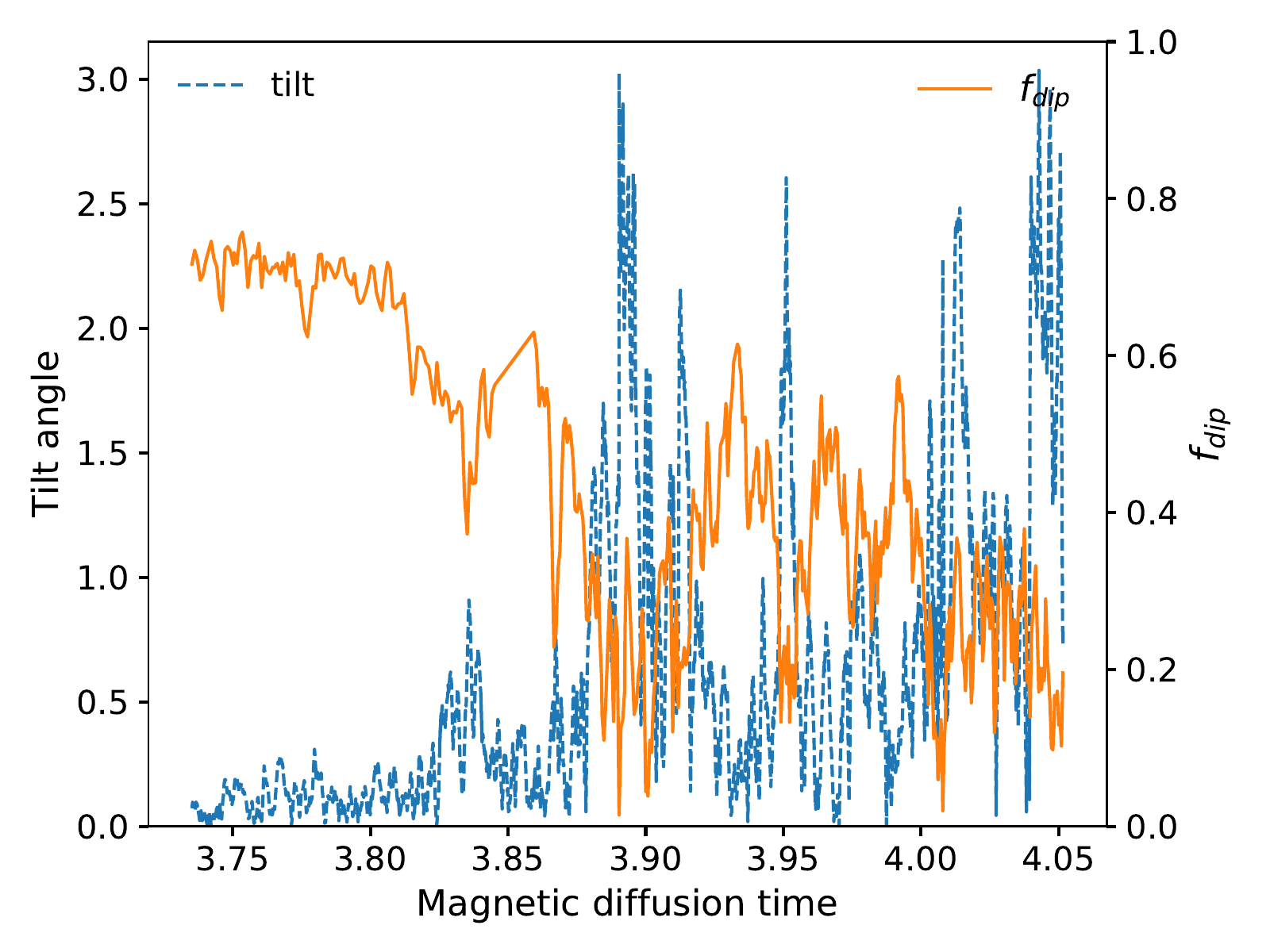} \label{tilta}}
	\hfill
	\subfigure[Stable dipolar dynamo at $\Ek=10^{-4}$, $\Ra/\Rac = 50.2$, $\Pm=5$, $\Lambda = 45.9$, $\Rol \sim 0.18$, $\fdip \sim 0.7$.]{
	\includegraphics[width=0.46\linewidth]{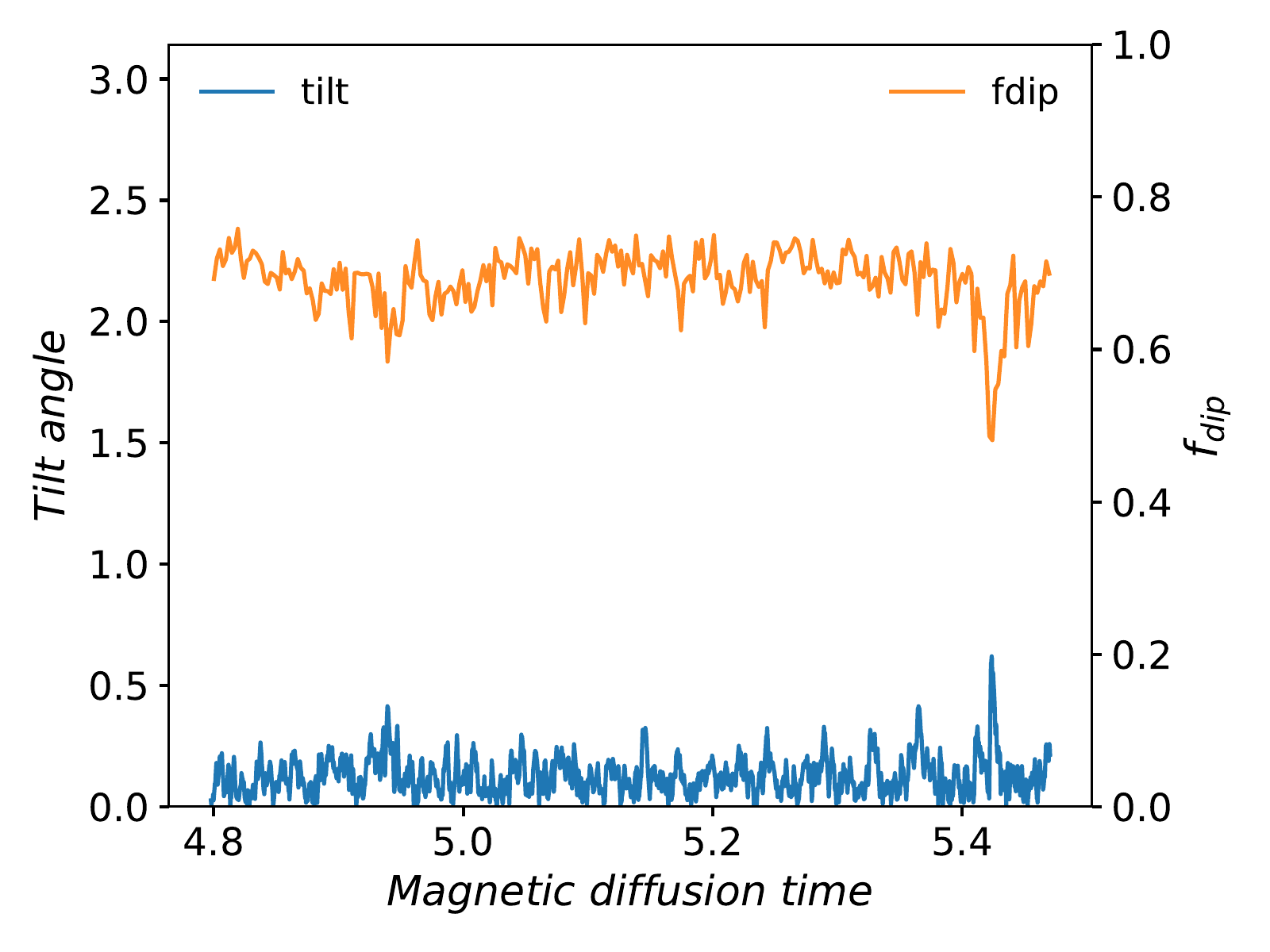} \label{tiltb}}
	\subfigure[Unstable dipolar dynamo at $\Ek = 3 \times 10^{-4}$, $\Ra/\Rac = 24.7$, $\Pm=6$, $\Rol \sim 0.17$, $\fdip \sim 0.56$.]{
	\includegraphics[width=0.46\linewidth]{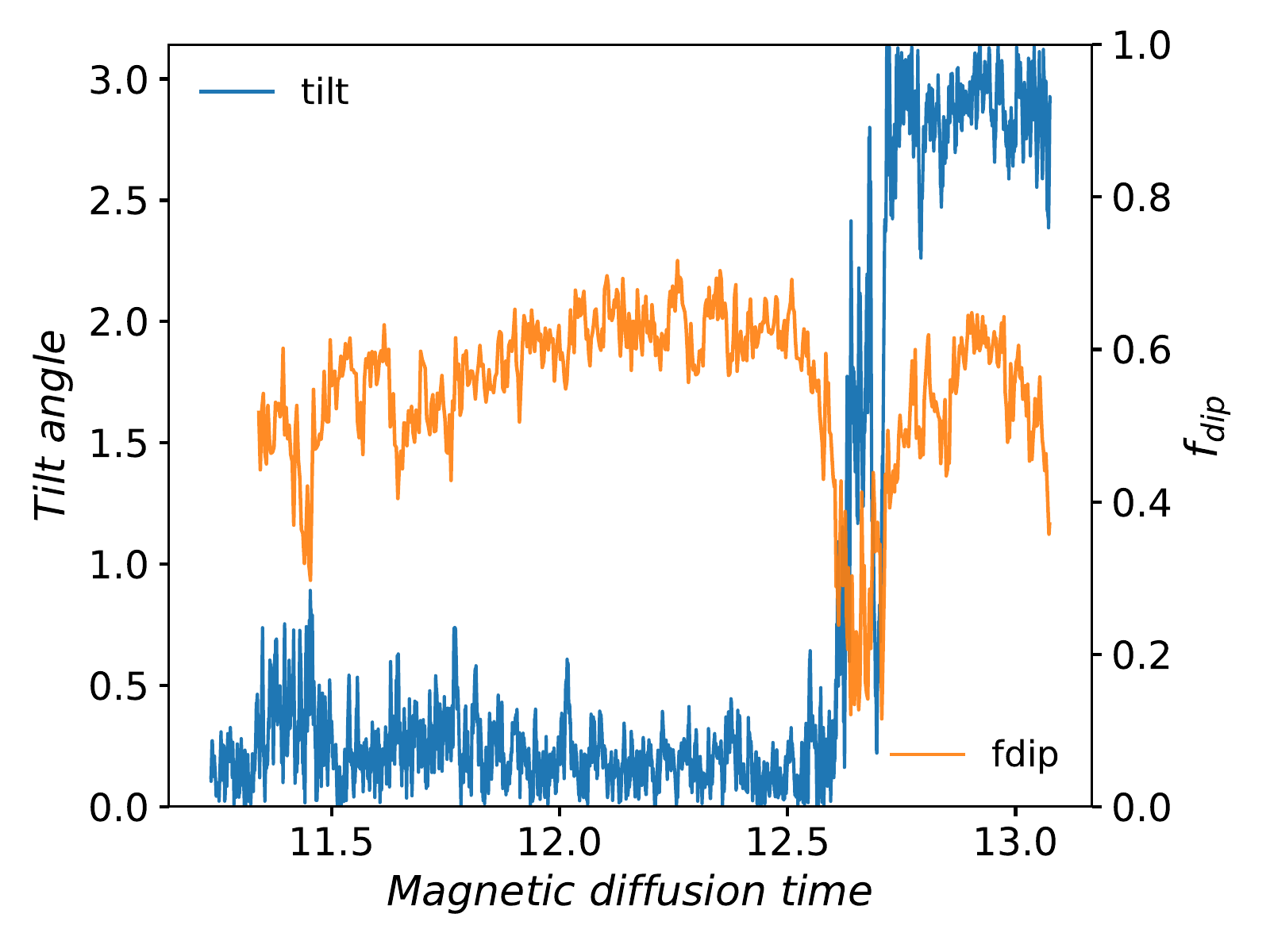} \label{tiltc}}
	\hfill
	\subfigure[Elsasser ($\Lambda$) and Rossby ($Ro$) numbers for the same parameters as figure \ref{tiltc}. The mean values are respectively $\Lambda = 21.2$ and $Ro \sim 0.04$.]{
	\includegraphics[width=0.46\linewidth]{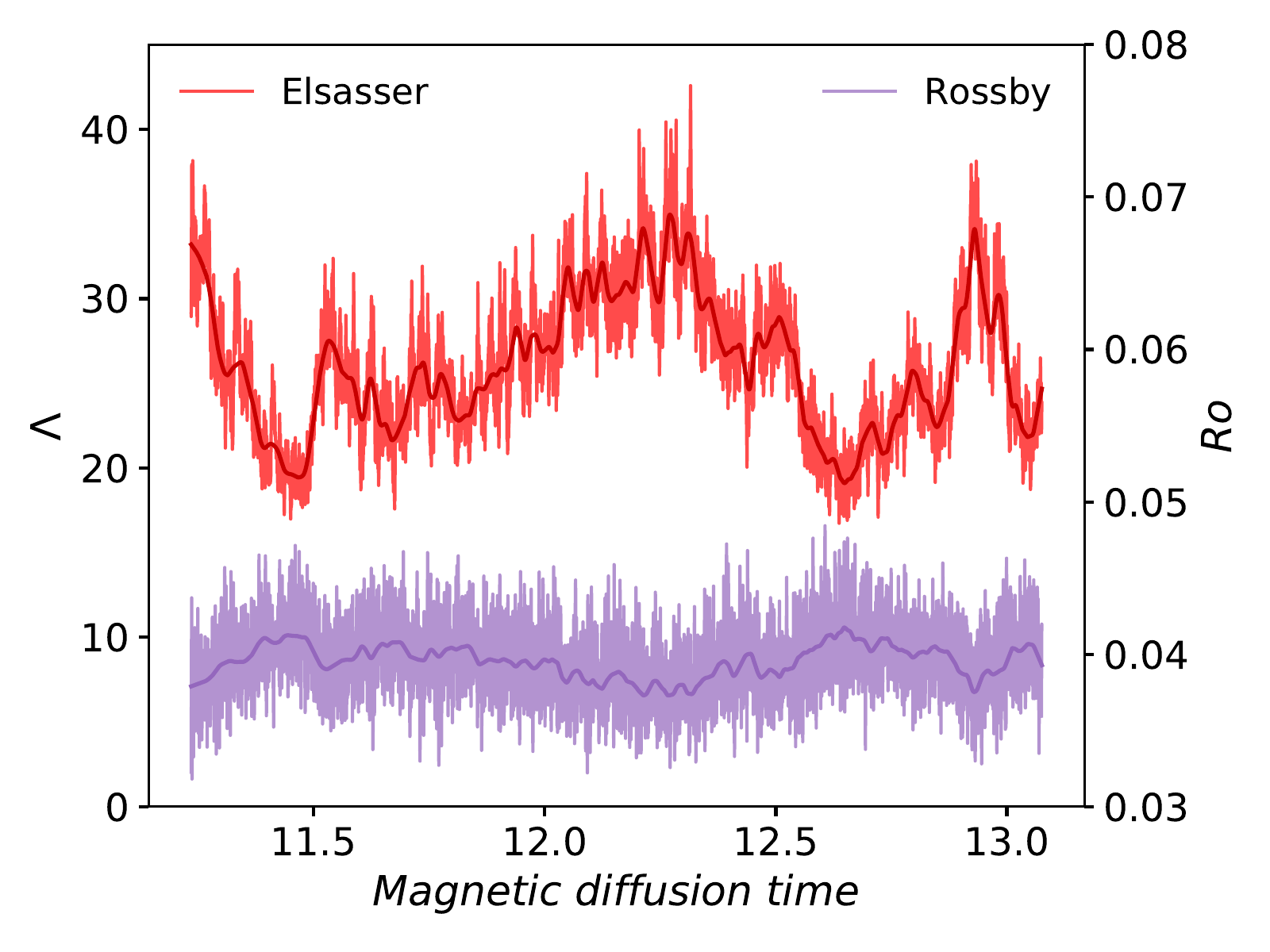} \label{tiltd}}
	\caption{Panels \ref{tilta}, \ref{tiltb} \& \ref{tiltc} : Tilt angle in radians (blue) and $\fdip$ (orange) evolution as a function of time (in magnetic diffusion time units). Panel \ref{tilta} at low $\Pm$ value denotes a multipolar simulation, whereas panel \ref{tiltb} denotes a stable dipolar dynamo with a quite low mean value of $\fdip$. Panel \ref{tiltd} : $\Lambda$ (red) and $Ro$ (mauve) evolution through magnetic diffusion time. This case is very close to a dipolar state, but sometimes explores another configuration.}
	\label{tilt}
\end{figure}

The tilt angle $\theta$, which measures the direction of the magnetic dipole compared to the rotation axis, and the relative dipole field strength $\fdip$ as a function of time are shown for different cases (see  figure \ref{tilt}) with $\Rol$ higher than $0.12$. Figures \ref{tilta} and \ref{tiltb} correspond to simulations with the same Ekman number. The buoyant forcing $Ra/Ra_c$ and the magnetic Prandtl number $\Pm$ are slightly higher for figure \ref{tiltb}. On the one hand, large temporal fluctuations of $\theta$ and $\fdip$ are observed for the model with $Pm=3$ (figure \ref{tilta}). The axial dipole is not a dominant component at the surface of the numerical domain (low $\fdip$ value) and several reversals of this component are obtained ($\theta$ oscillations). On the other hand, figure \ref{tiltb} with $\Pm=5$ does not present such a behaviour: although a higher $Ra/Ra_c$ has been employed, the axial dipole dominates at any time with a constant magnetic polarity (no reversal). The latter model can thus be classified as a dipolar dynamo. However, it is important to note that $\Rol$ exceeds $0.12$. 

Figures \ref{tiltc} and \ref{tiltd} show the evolution of, respectively, $\fdip$, $\theta$ and $\Lambda$, $Ro$ as functions of time for a model at : $E=3 \times 10^{-4}$ and $Pm=6$. After a long period of time (larger than one magnetic diffusion time), the dipole collapses and $\fdip$ becomes lower than $0.2$. At the same time, the polarity of the magnetic dipole measured by $\theta$ fluctuates around $\pi/2$ which indicates that the dominant component of the magnetic dipole is its equatorial dipole component. This situation is only a transient configuration. Then, the axial dipole with an opposite polarity  again dominates the other components. This reversal seems to appear between two long periods in which the axial dipole is dominant. It is important to note that large temporal fluctuations of the magnetic energy as measured by $\Lambda$ are observed for this model. In comparison, relatively constant values of $\Lambda$ are obtained for figures \ref{tilta} and \ref{tiltb} (not shown). The condition $\fdip>0.5$, mentioned by previous studies \citep{christensen06}, does not appear as a sufficient condition in order to identify dipolar models since it indicates that this dynamo model with unpredictable reversals and important temporal variations of the magnetic field strength is a dipolar solution. This particular behaviour has been reported before for dynamos with $\Rol$ close to the transitional value 0.12. Other proxies such as the time evolution of the tilt angle must be taken into account.  \\

\begin{figure}
	\subfigure[$\Ek = 3 \times 10^{-4}$.]{
	\includegraphics[width=0.45\linewidth]{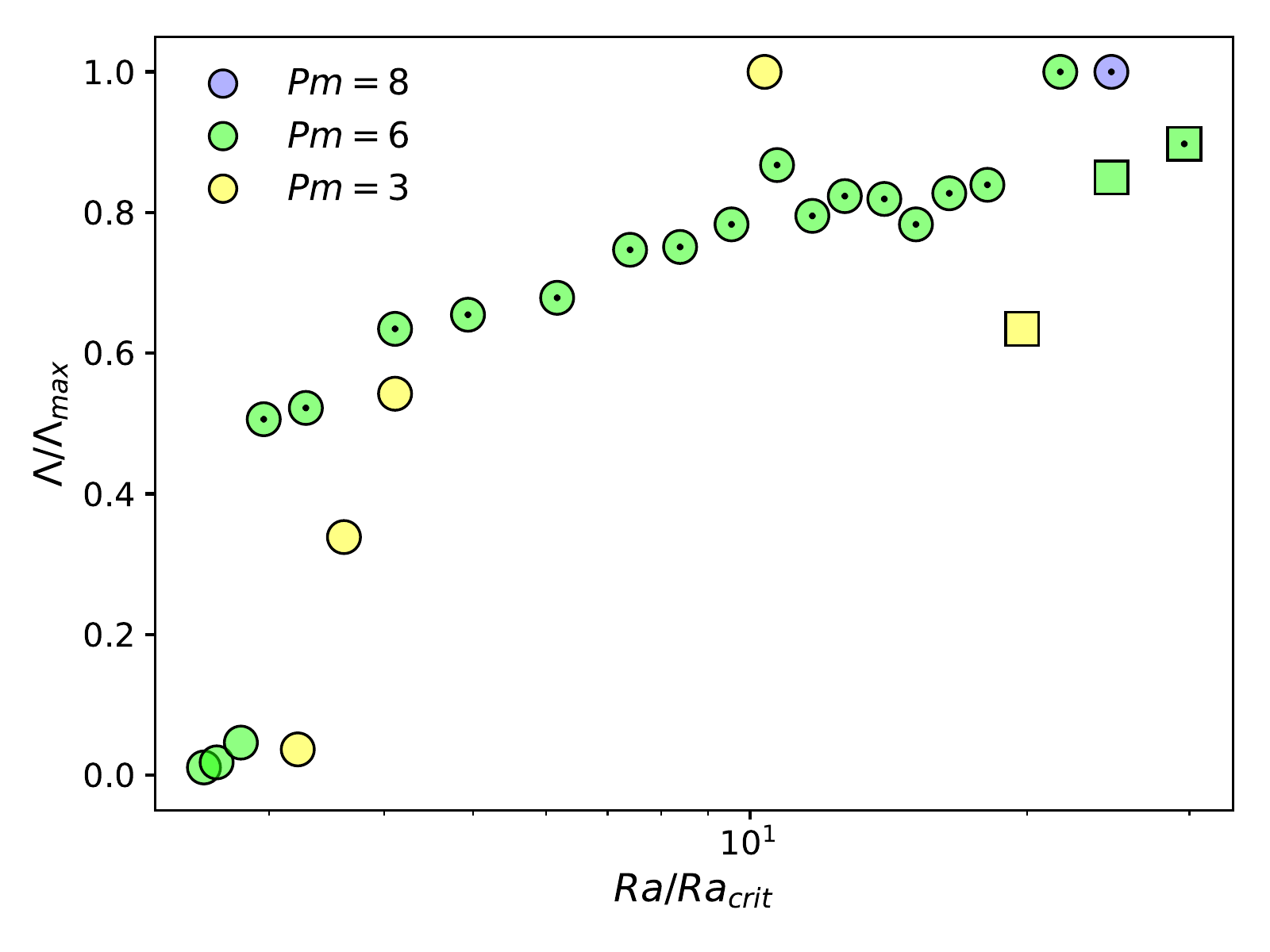}\label{Elsasser1}}
	\hfill
	\subfigure[$\Ek = 1 \times 10^{-4}$.]{
	\includegraphics[width=0.45\linewidth]{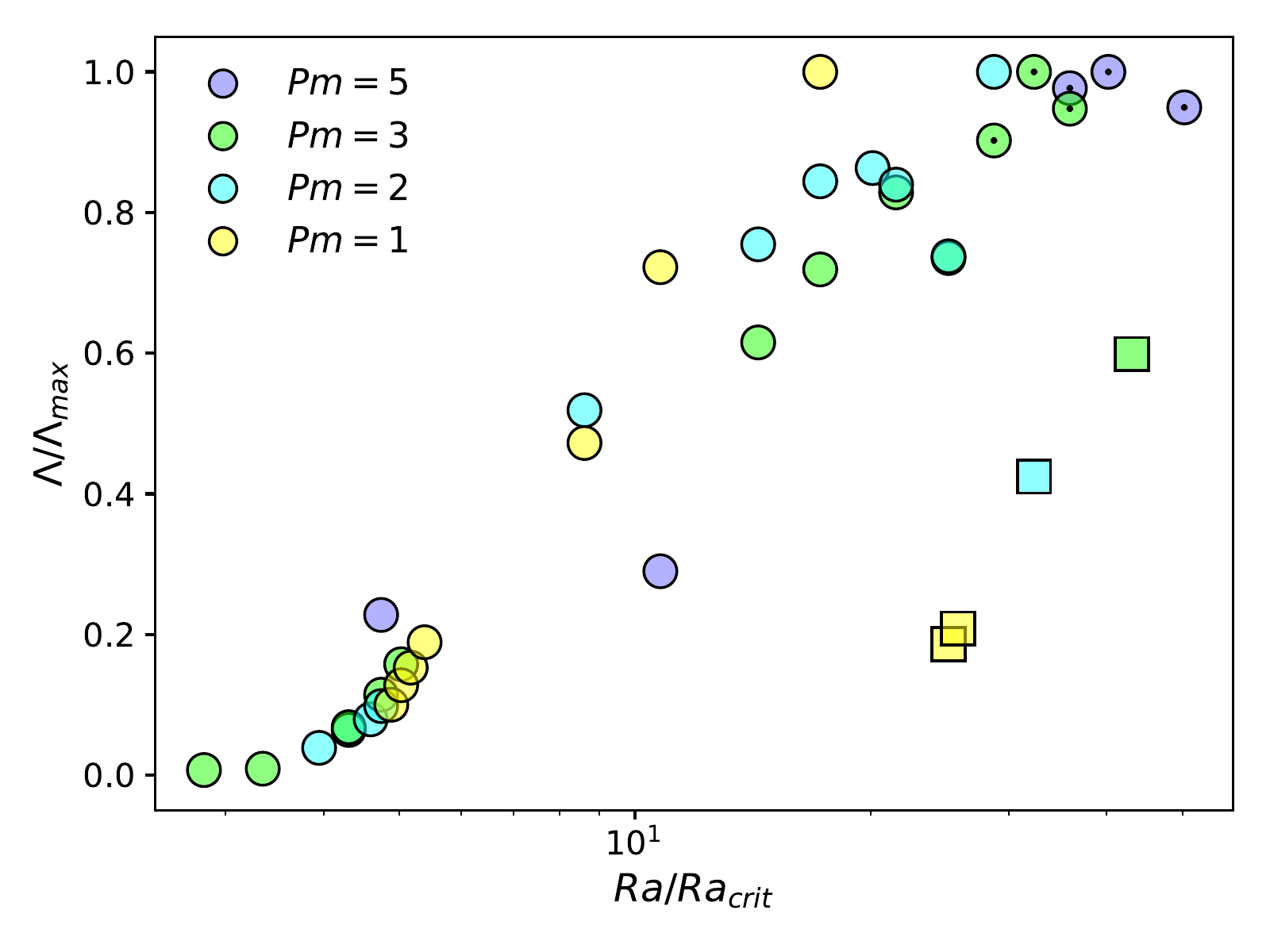} \label{Elsasser2}}
	\subfigure[$\Ek = 3 \times 10^{-5}$.]{
	\includegraphics[width=0.45\linewidth]{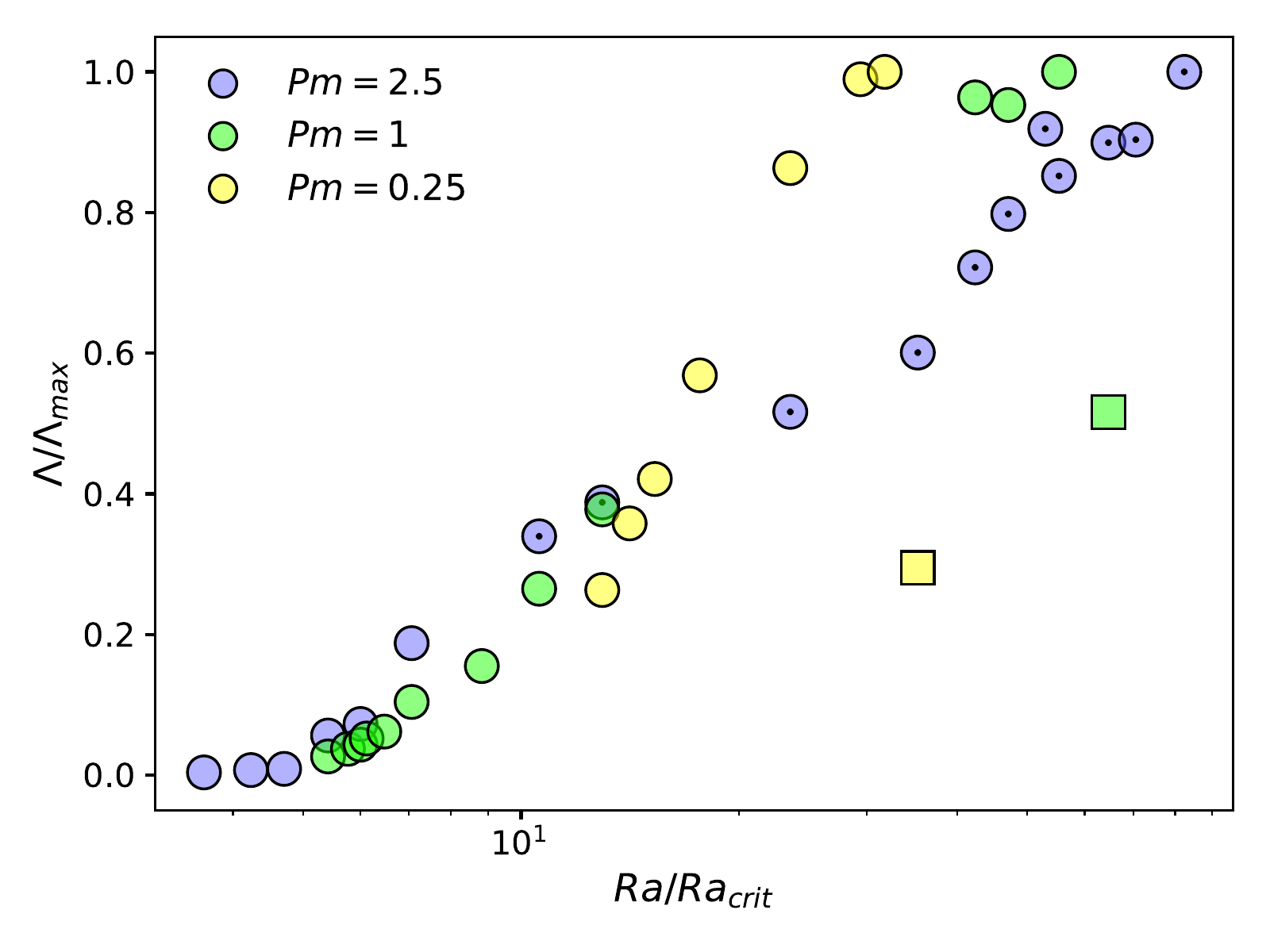} \label{Elsasser3}}
 	\hfill
	\subfigure[$\Ek = 1 \times 10^{-5}$.]{
	\includegraphics[width=0.45\linewidth]{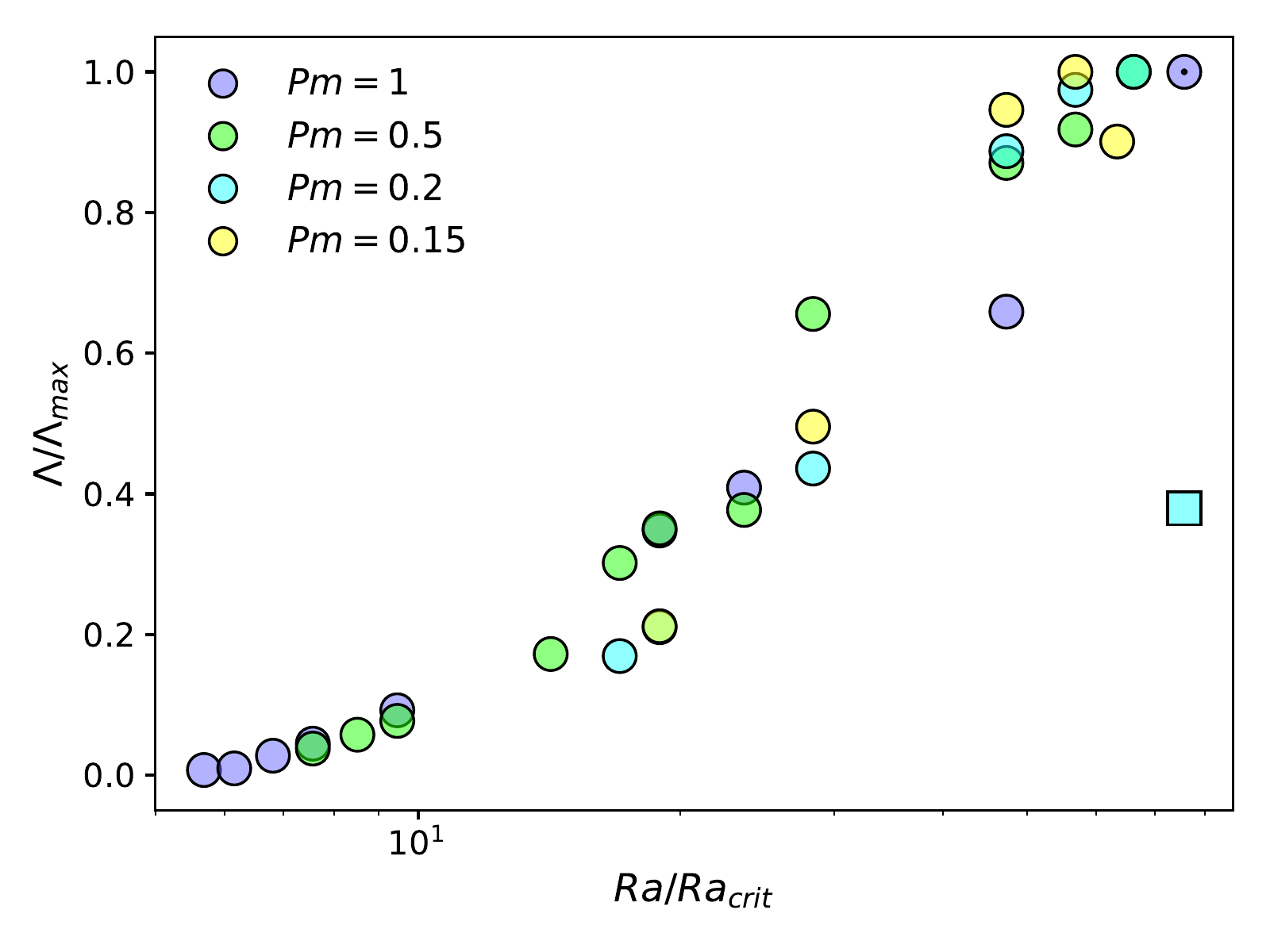} \label{Elsasser4}}
	\caption{$\Lambda$ variations with $\Ra/\Rac$ normalised by the maximum value of each $\Pm$ branch (colours) for four Ekman values. Circles indicate stable dipolar dynamos and squares dynamos at least one reversal. For the same value of $\Pm$, we can observe a sharp decreasing of the Elsasser number for the highest values of $\Ra$, i.e. when the transition to a multipolar state occurs. For a low value of $Pm$ (in yellow - respectively $\Pm=1$ for \ref{Elsasser2} and $\Pm=0.25$ for \ref{Elsasser3}), there is a decline of approximately $75\%$ whereas it is only about $50\%$ for the highest $Pm$ branch (in green). The dot inside some markers distinguishes Lorentz-dominated dynamos (see \S \ref{Part2}).} 
	\label{Elsasser}
\end{figure}

As mentioned, the magnetic energy as measured by the  Elsasser number $\Lambda$ is another useful piece of information. When a transition from a dipolar configuration to a multipolar one occurs, a decrease of $\Lambda$ is observed in most of the cases. In order to highlight the evolution of $\Lambda$ in the parameter space, figure \ref{Elsasser} shows the evolution of $\Lambda$ as a function of $Ra/Ra_c$ for different $\Pm$ and $E$. Since we are interested in the relative variations of $\Lambda$ when a transition occurs, $\Lambda$ is normalised by its maximum value $\Lambda_{max}=max(\Lambda(\Pm))$. Figure \ref{Elsasser} clearly shows that the dipole slump giving rise to multipolar dynamos (square symbols) is accompanied by a decrease of $\Lambda$. This effect (loss of magnetic energy) increases as $\Pm$ decreases. For example, multipolar dynamos at low $\Pm$ values have $\Lambda/\Lambda_{max}$ lower than $0.25$ i.e. the dipole collapse has induced a loss of $75\%$ of the magnetic energy (see yellow markers in figures \ref{Elsasser2} \& \ref{Elsasser3}). Increasing $\Pm$ reduces the difference in $\Lambda$ value between dipolar cases and multipolar cases. Indeed, for the highest $\Pm$ at which a transition is observed (green markers in figure \ref{Elsasser}), it only represents a drop of about $50 \%$ or less. Some symbols marked by a dot will be discussed later in this paper.\\

Figure \ref{Elsasser} shows time averaged quantities. However, such quantities can have large temporal fluctuations as shown in figure~\ref{tilt}, especially when high $\Pm$ are considered. For example, the case represented in figure \ref{tiltc} has a particular behaviour: $\fdip$ temporarily drops below $0.2$ around the period $t\sim 12.7$ while the tilt angle fluctuates around $\pi/2$ at the same time. Later $\fdip$ increases again to reach its original value (see above). This reversal of the magnetic field axis attests that this case is an unstable dipolar solution. Looking at the evolution of $\Lambda$ (figure \ref{tiltd}) confirms that the magnetic field strength has been affected by this behaviour. This is another indication supporting the fact that some care must be taken in order to characterize the morphology of geodynamo models. In particular, the time evolution of $\fdip$, $\theta$ and $\Lambda$ appear as important proxies.

On this basis, it seems crucial to distinguish these cases experiencing intermittent reversals from cases with a purely stable dipole component. Motivated by modelling geomagnetic fields, the terms "stable dipolar" is restricted to dynamos with a dominant dipolar component ($\fdip>0.5$) at any moment of the time integration. The polarity of the dipole measured by the tilt angle $\theta$ has to keep its orientation as well. As mentioned above, the magnetic field strength measured by $\Lambda$ is also a good proxy. A dynamo model is called a stable dipolar dynamo when no important fluctuations of $\Lambda$ are observed during the time integration. This slight nuance in the definition of dipolar dynamos allows to clearly identify unstable dipolar dynamos, having a dominant axial dipole component in average with rare and irregular reversals. The previous general definition introduced by \citet{christensen06} only based on the time averaged of $\fdip$ does not appear as sufficient with our new parameter space and could provide questionable results. For instance, a time average of $\fdip$ could be higher than $0.5$ for a dynamo model with several reversals of its dipole, as the solution presented in figure \ref{tiltc}. Such models are frequently observed close to the transitional value $\Rol=0.12$ or higher, a focus of this study. Here, according to our procedure, such models are classified as unstable dipolar dynamos. Nevertheless, this classification does not completely avoid the problem linked to the time integration dependency : the reversal period of an unstable dipolar dynamo might be larger than the observation window and thus be classified as a stable dipolar dynamo. The models with $\Rol$ higher than $0.1$ have been performed for at least $0.6$ magnetic diffusion time, thus giving an upper limit to the reversal period of unstable dipolar dynamos detected by this study. We can not exclude that all stable dipolar cases reported in this paper do not give rise to unstable dipolar dynamos by considering larger time integration. However, the distinction between dipolar and multipolar dynamos is relatively unaffected by this problem as it is corroborated by the Elsasser number variations as well (see figure \ref{Elsasser}).

\subsection{The failure of the purely hydrodynamic criterion ($Ro_\ell<0,12$) for dipolar dynamos}\label{Part2}

\begin{figure}
	\centering
	\subfigure[$E = 3 \times 10^{-4}$.]{
	\includegraphics[width=0.45\linewidth]{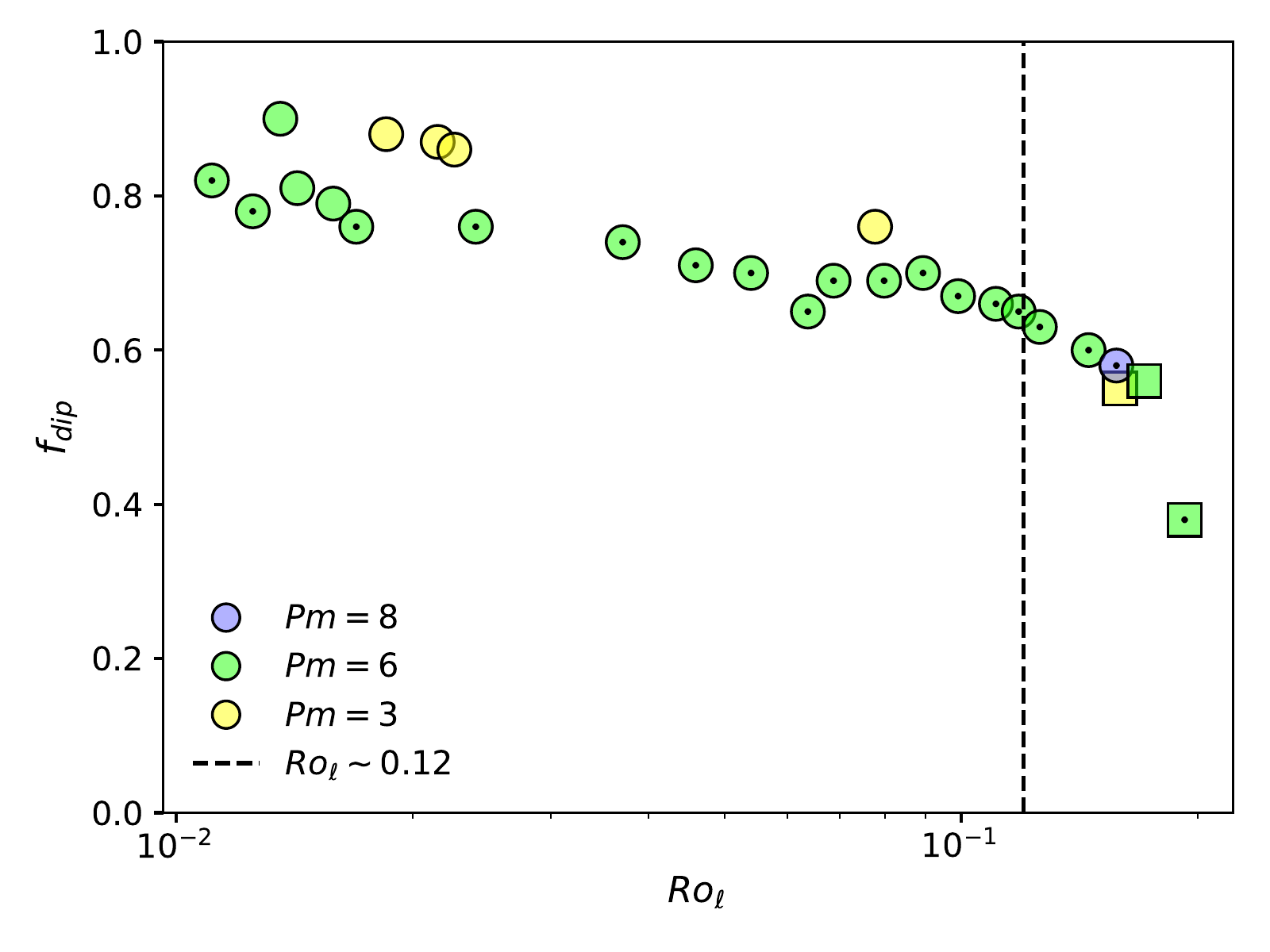}\label{fdip3e4}}
	\hfill
	\subfigure[$E = 10^{-4}$.]{
	\includegraphics[width=0.45\linewidth]{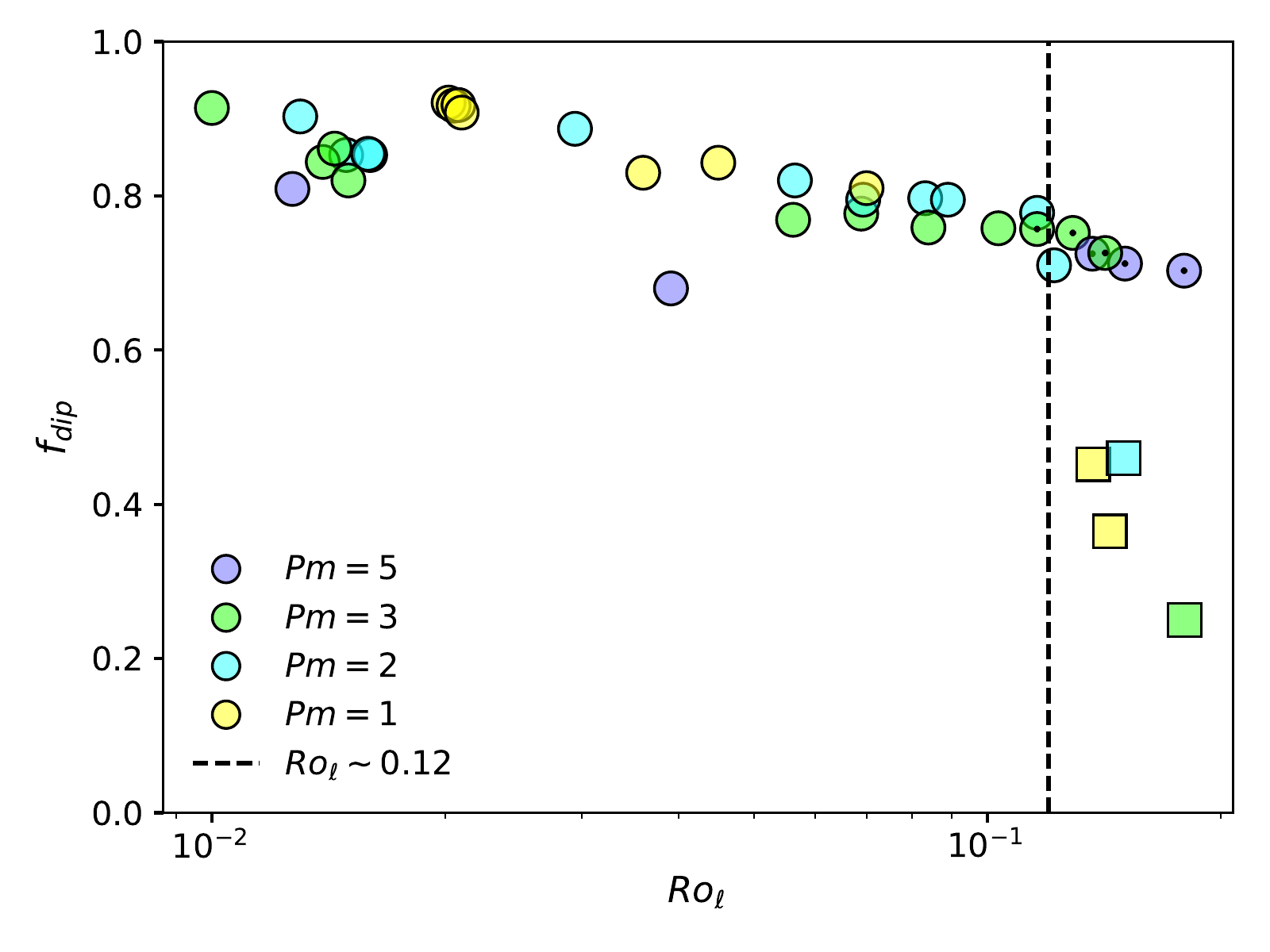}\label{fdip1e4}}
	\subfigure[$E = 3 \times 10^{-5}$.]{
	\includegraphics[width=0.45\linewidth]{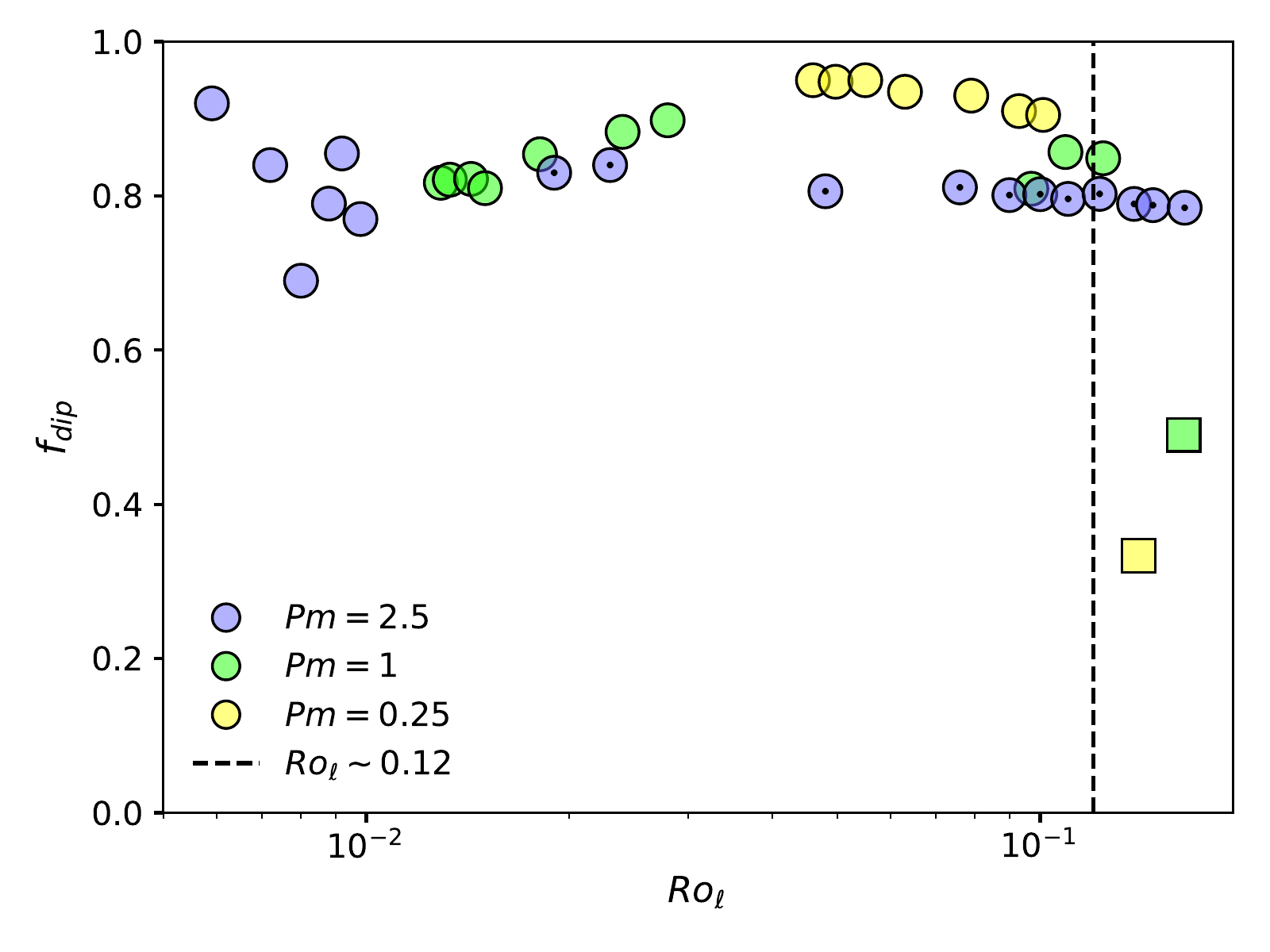}\label{fdip3e5}}
	\hfill
	\subfigure[$E = 10^{-5}$.]{
	\includegraphics[width=0.45\linewidth]{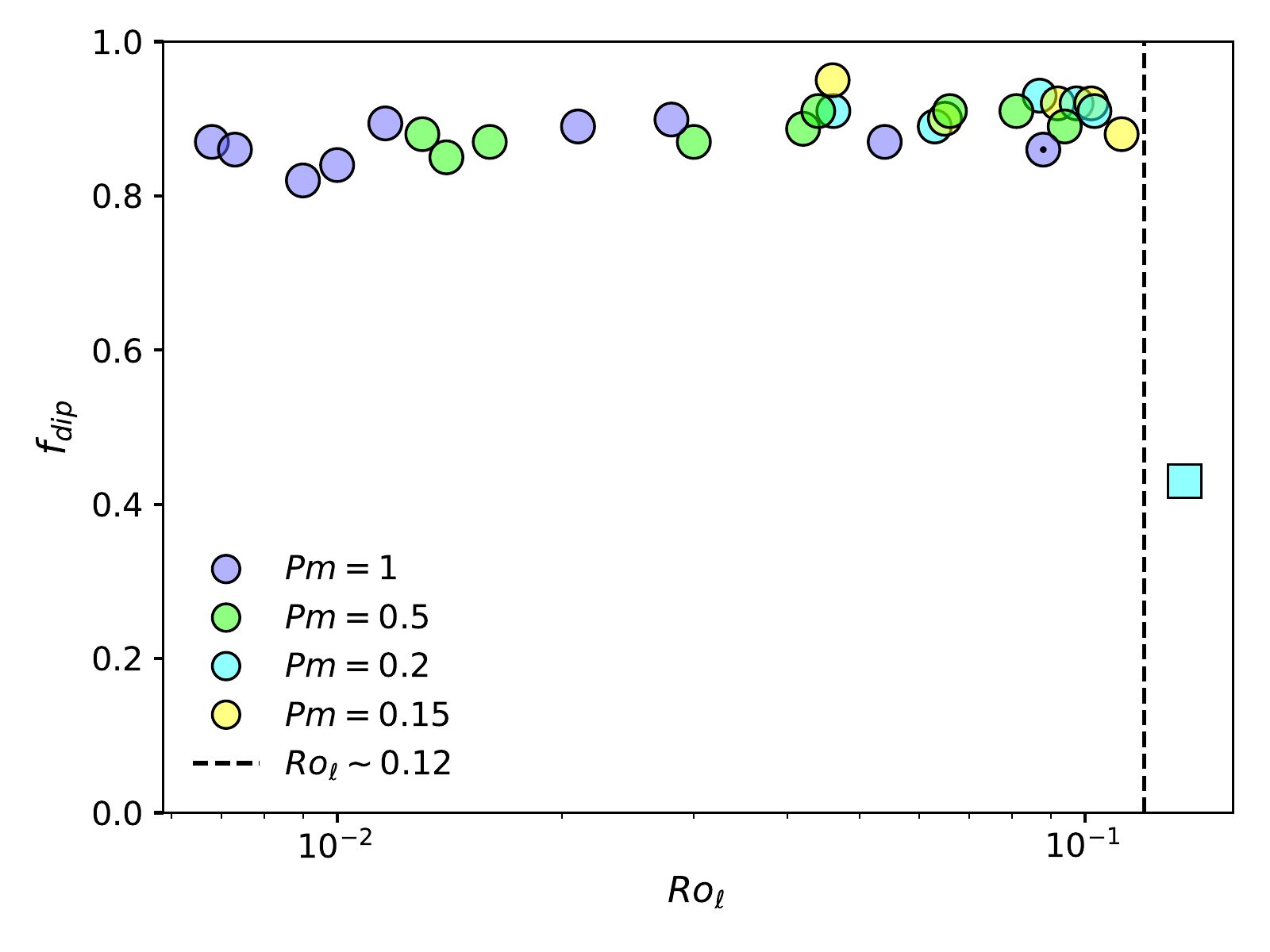}\label{fdip1e5}}
	\caption{Relative dipole field strength $\fdip$ as a function of the local Rossby number $\Rol$ for four values of the Ekman number. As before, the colours on each graph distinguish the magnetic Prandtl number, whereas the marker distinguishes the magnetic field topology : stable dipolar (circles) or unstable dipolar and multipolar (squares) dynamos. Previous studies have only obtained dipolar dynamos when $\Rol$ is lower than $0.12$ (vertical dashed lines). Here, we report stable dipolar dynamos with higher $\Rol$.}
	\label{fdip}
\end{figure}

The dipole field strength $\fdip$ as a function of $\Rol$ is shown in figure \ref{fdip} for our data set. Previous studies have drawn similar figures in order to highlight the critical role of the local Rossby number on the magnetic morphology. We report in figure \ref{fdip} dipolar dynamos with $\Rol$ higher than the critical value $0.12$ marked by a vertical dotted line. If we limit our data set to lower $\Pm$ as considered by previous studies, we also find the same $\Rol<0.12$ criteria for dipolar dynamos. In other words, the $\Pm$ value impacts on the behaviour of the dynamo and in particular on the limit before a transition. Let us discuss our choice for the different symbols corresponding to the magnetic morphology (circles or squares). Dipolar dynamos (circles) have a stable dominant dipolar component ($\fdip>0.5$) with no large temporal fluctuations of $\fdip$, $\theta$ and $\Lambda$, i.e. no reversals. Otherwise, models are classified as unstable dipolar dynamos (squares with $\fdip>0.5$) or multipolar dynamos (squares with $\fdip<0.5$).

Despite this more restrictive definition for stable dipolar dynamos, we report for the first time dipole-dominated solutions with $\Rol$ higher than $0.12$ for dynamo models with fixed temperatures and no-slip boundary conditions. Figure \ref{fdip} shows that stable dipolar dynamos can be maintained above $\Rol>0.12$ only when $\Pm$ is sufficiently high.
Otherwise, regardless of the value of the Ekman number, multipolar dynamos are obtained (see \citet{christensen06}). Here, we question this finding by considering higher $\Pm$ which allows the existence of stable dipolar dynamos in the turbulent regime (i.e. up to $\Rol\sim 0.19$).\\

It is particularly clear with $\Ek =3 \times  10^{-5}$ (see figure \ref{fdip3e5}) that dipolar and multipolar dynamos coexist, when $\Rol$ exceeds $0.1$. In this case, the magnetic morphology depends on the value of $\Pm$ : a collapse of the axial dipole component is obtained when $\Rol$ exceeds $0.12$ with $\Pm=1$ whereas this component is still dominant with $\Pm=2.5$ and higher $\Rol$. \\
Since decreasing the Ekman number is numerically demanding, we limit our systematic parameter study to $\Ek\geq 10^{-5}$. At $\Ek = 10^{-5}$, numerical constraints only enable us to reach sufficiently high values of the Rayleigh number ($\Ra > 70\ \Rac$) to obtain a dipole collapse (a transition from a dipolar to a multipolar morphology) for $\Pm = 0.2$ (see figure \ref{fdip1e5}). Although we do not show dipolar models with $\Rol > 0.12$ at $\Ek = 10^{-5}$, we report for the first time a dipole collapse at $\Ek < 3 \times 10^{-5}$. \\

\subsection{Lorentz-dominated dynamos in the parameter space}\label{Part2}

\begin{figure}
	\center
	\includegraphics[width=1\linewidth]{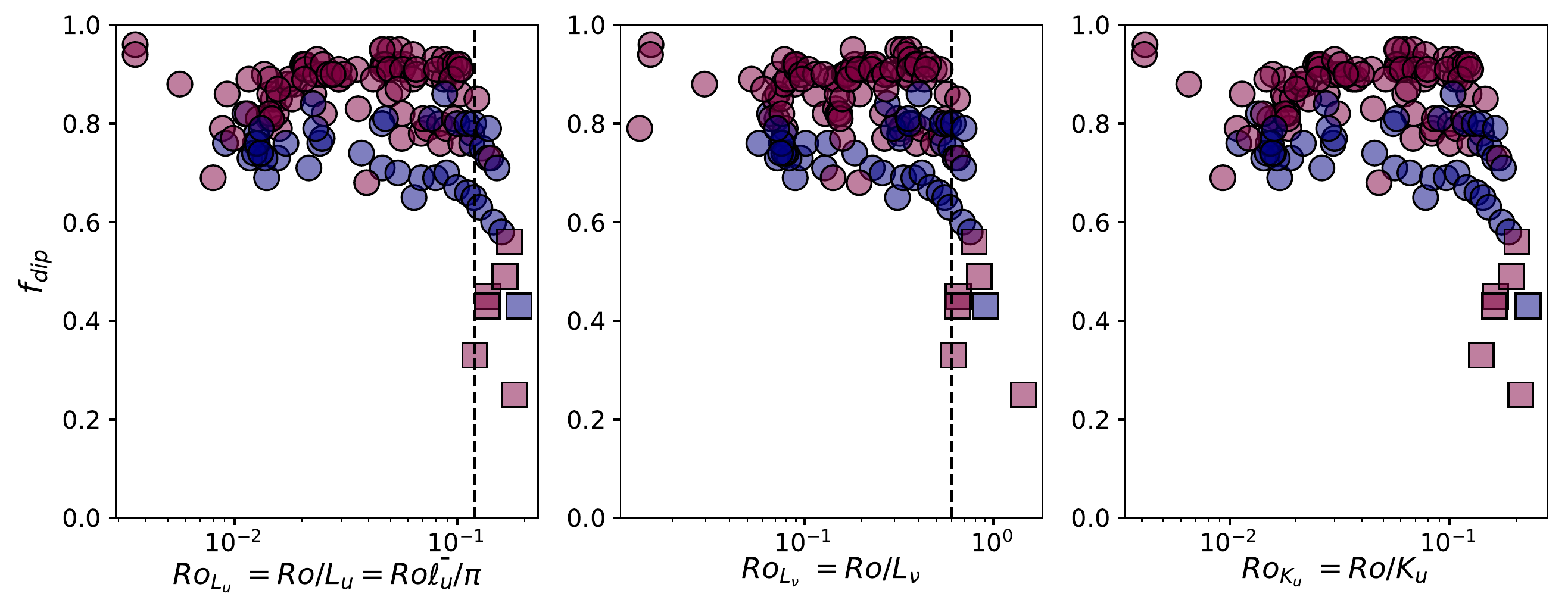}
		\caption{$f_{dip}$ versus $Ro_l$ calculated from the different scales proposed in the literature. Blue denotes Lorentz-dominated dynamos, whereas mauve denotes other cases. None of these scales allow a better distinction between stable dipolar (circles) and unstable dipolar or multipolar (squares) regimes for the Lorentz-dominated branch. Indeed, at a certain $Ro_l$, dipolar and multipolar dynamos coexist.}
	\label{fdipRols}
\end{figure}

In figure \ref{fdipRols}, the left-handed panel represents the classical ($\Rol ,\fdip$)-plane, as in \citet{christensen06} (see equation \eqref{Rol}). For several cases, the critical value proposed by these authors is not appropriate (see \S \ref{Part1} for more details). In the middle sketch, the local Rossby number has been calculated from the kinetic dissipation length scale as defined in equation \eqref{Lnu}. However, this $\Rol$ definition does not allow a better distinction between multipolar and dipolar dynamos. 

Finally, in the right-handed sketch we use a quantity introduced by \citet{Soderlund2012} $k_u$, the characteristic wavenumber of the flow. It has been used to deduce a length scale taking into account both the spherical harmonic degree $\ell$ and order $m$ :
$L_k=\pi / k_u\ with\ k_u = \sqrt{\bar{\ell}_u^2+\bar{m}_u^2}$.  $\bar{m}_u$ is the equivalent of $\bar{\ell}_u$ in terms of order instead of degree. Again, it exhibits no correlation between the $\Rol$ value and the topology of the magnetic field. 
On the contrary, the behaviour of dynamos at higher $Ro$ value strongly depends on the regime studied (depicted by the colours in figure \ref{fdipRols}). It can be interpreted as two regimes with different behaviours when $\Rol$ is sufficiently high (see \ref{Part2}). \\

We also tried to define a $\Rol$ from the $l_{u\ peak}$ which has been presented by \citet{Dormy2018}. However, we have found that the kinetic energy is mainly distributed on one particular length scale in high $\Pm$ simulations only when $Ra$ is sufficiently low. Otherwise, the spectra are much flatter and $l_{u\ peak}$ cannot be calculated. Indeed, $\Lambda ' > 1$ dynamos have been studied only for $\Ra$ values close to the convection threshold by previous studies. In this study, turbulent dipolar dynamos are found at high $\Ra/\Rac$ values, and thus do not present the same features.\\

Regardless of the definition of the volume averaged velocity field and kinetic length scale used to calculate other local Rossby numbers (see figure \ref{fdipRols}), the magnetic morphology of dynamos at $\Rol >0.12$ depends on the magnetic field regime studied : Lorentz-dominated dynamos (in blue) or the others (in mauve). For the latter, the overlap between dipolar and multipolar solutions is wider, which can be explained by the variety of input parameters studied : dipolar and multipolar solutions with similar $\Rol$ values do not belong to the same dynamo branch, as several Ekman and Prandtl numbers sets are merged on this figure.

However, our results are consistent with previous studies at similar values of $\Ra$. Indeed, the empirical value of $\Rol \sim 0.12$ highlighted in \citet{christensen06} corresponds to the limit proposed in \S \ref{Part3} when the Lorentz force plays a minor role.

 The Lorentz force can play a major role in models with high $\Pm$ values. We will use the same modified Elsasser number $\Lambda '$ as \citet{Dormy2016} (see equation \eqref{Lambdador}) in order to distinguish models strongly influenced by the Lorentz force which have $\Lambda ' > 1$. Solutions that meet this criterion are marked with a dot in the figures of \S \ref{Results}. For some cases, $L_\eta$  has not been calculated during the time integration. To complete the missing values, we have used the scaling law $L_{\eta} \propto \Rm^{-1/2}$ introduced by \citet{Soderlund2015}. We checked its validity for our set of simulations (see \ref{AnnA}) and these additional values appear in the Tables (in \ref{Table}) with a particular symbol (*).
\\

\begin{figure}
	\centering
	\subfigure[$E =3 \times 10^{-4}$.]{
	\includegraphics[width=0.45\linewidth]{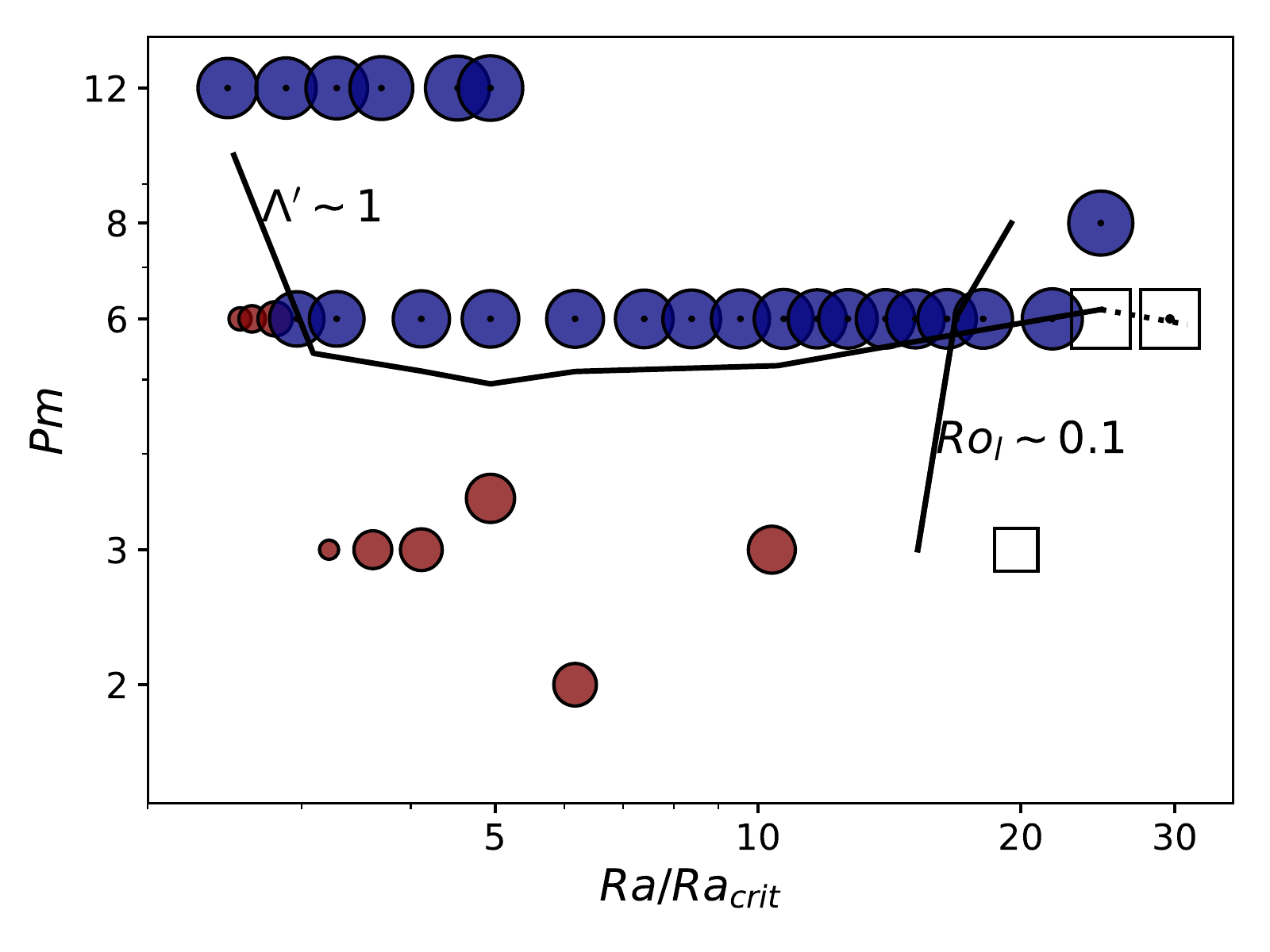}}
	\hfill
	\subfigure[$E = 10^{-4}$.]{
	\includegraphics[width=0.45\linewidth]{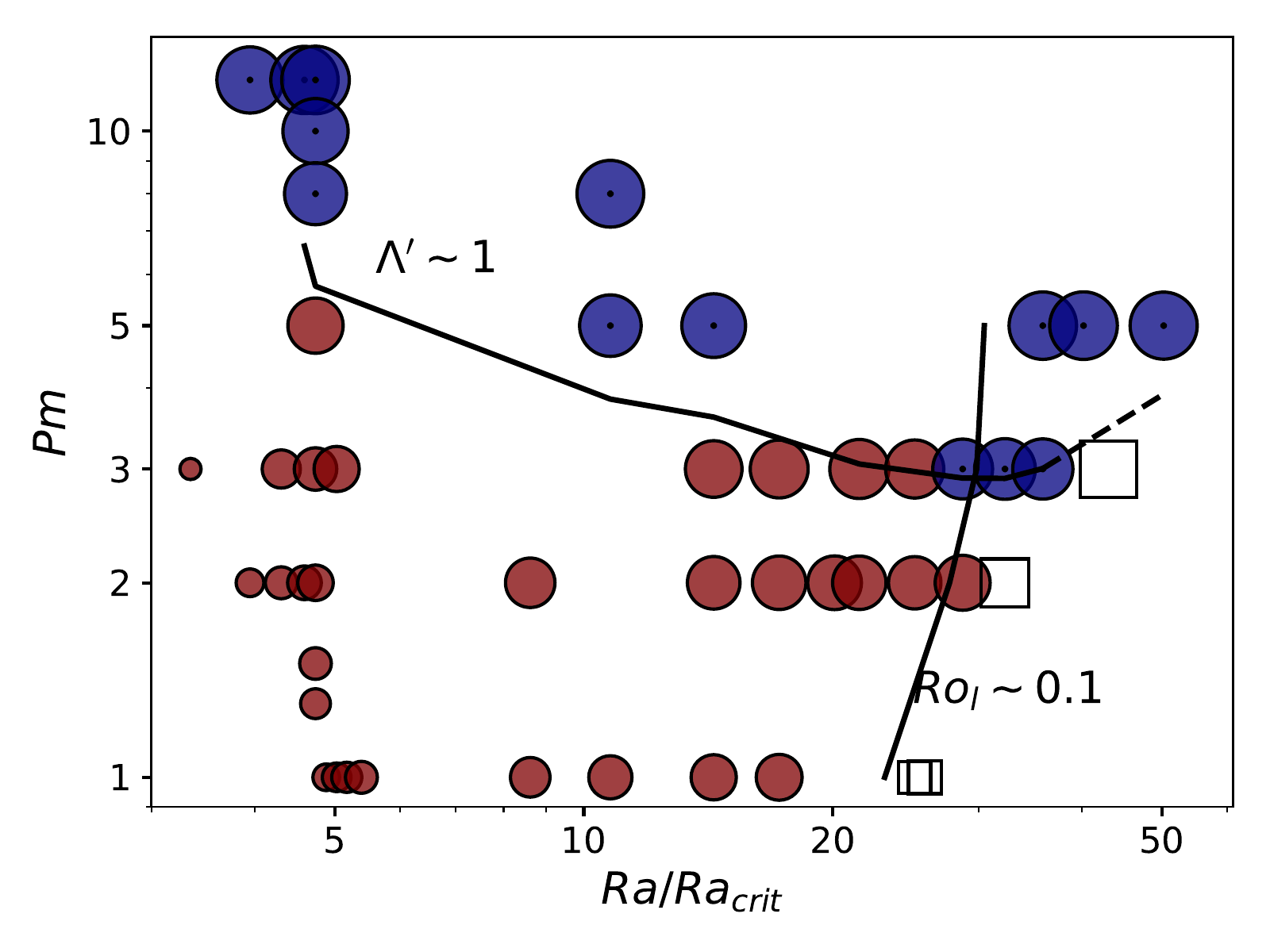}}
	\subfigure[$E =3 \times 10^{-5}$.]{
	\includegraphics[width=0.45\linewidth]{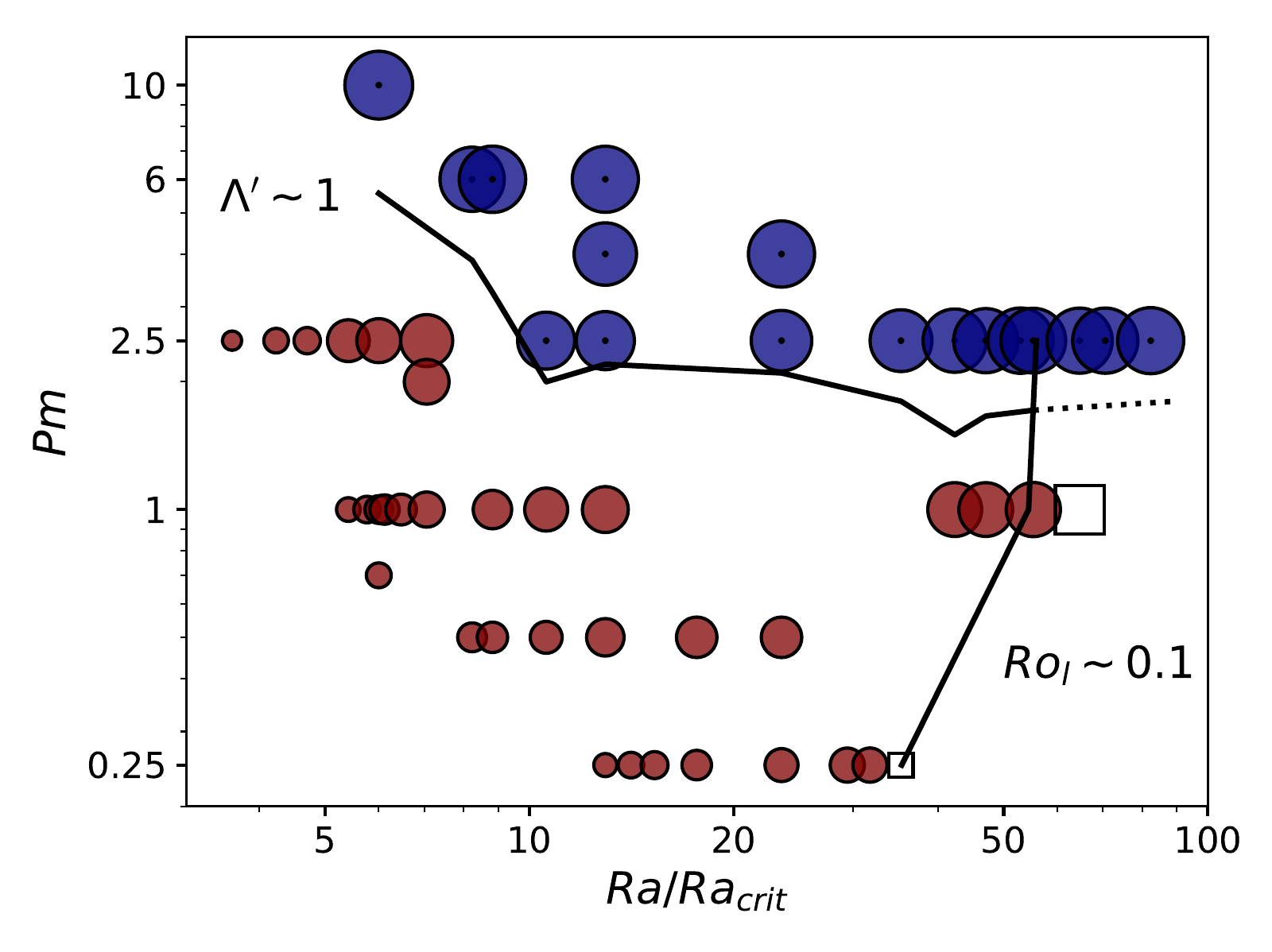}\label{PmRa3e5}}
	\hfill
	\subfigure[$E =10^{-5}$.]{
	\includegraphics[width=0.45\linewidth]{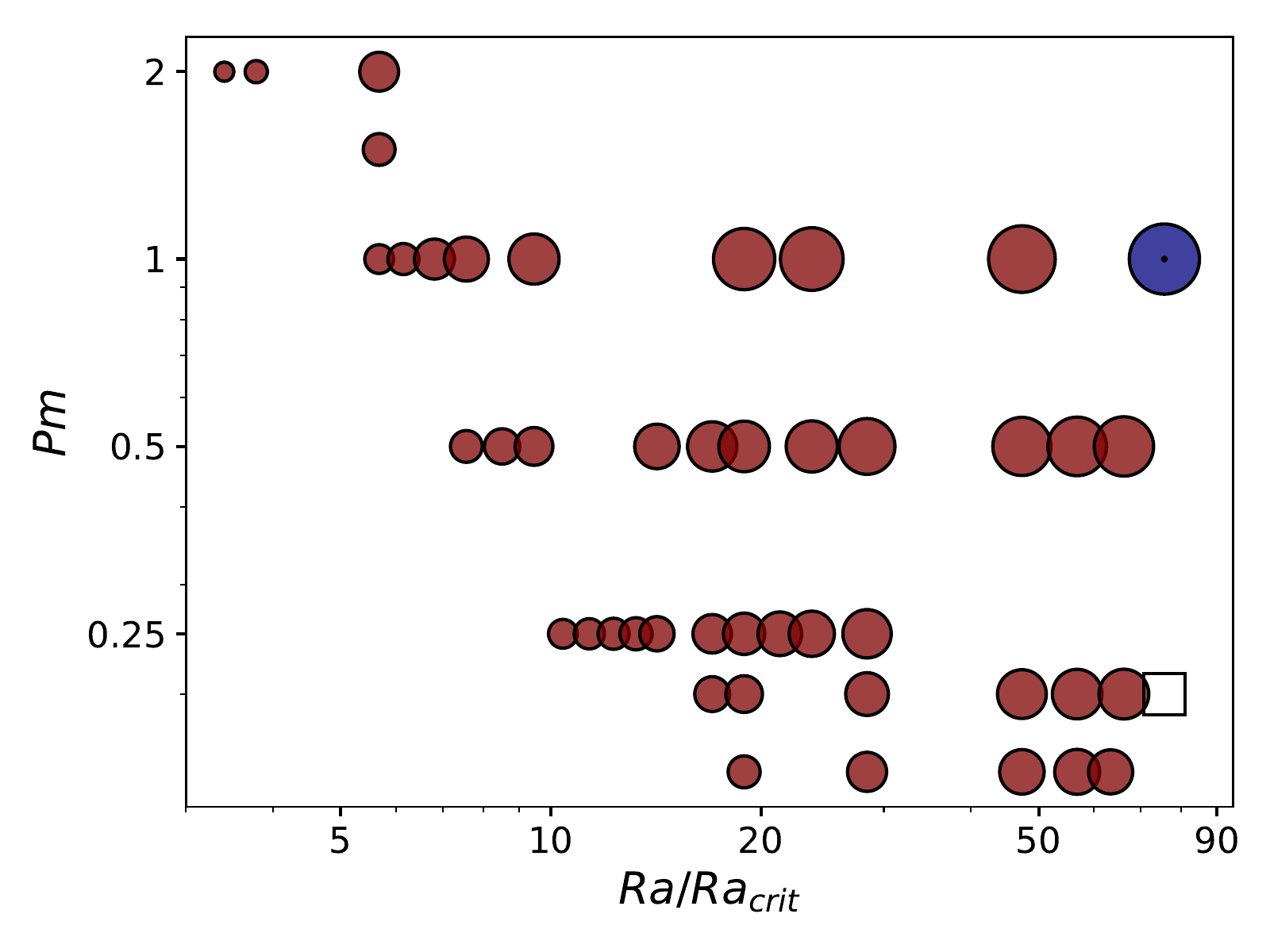}}
	\caption{Regime diagrams at different values of the Ekman number. The markers have the same meaning as in figure \ref{fdipRols}. The colour (mauve or blue) stands for respectively the dipolar dynamos with $\Lambda ' <1$ and $\Lambda ' >1$. The size of the marker is varying with the classical Elsasser number $\Lambda$. When possible, interpolated boundaries  $\Lambda' \sim 1$ and $\Rol \sim 0.12$ have been represented (black lines).}
	\label{PmRa}
\end{figure}

We follow previous studies \citep{christensen06,Petitdemange2018} and show the magnetic field topology in a $(\Pm,\Ra/\Rac)$-plane (see figure \ref{PmRa}). Colours have been added to highlight the existence of  two dipolar branches. In addition, two curves corresponding to $\Lambda'\sim 1$ and $\Rol \sim 0.1$ have been plotted, from values obtained by linear interpolation. These curves are determined by linearly interpolating $\Lambda'$ and $\Rol$ values from our data set. The last part on $\Lambda' \sim 1$ curve has been extrapolated (dashed line). Close to the onset of convection ($Ra<5Ra_c$), the curve $\Lambda'\sim 1$ is a rapidly decreasing function of $Ra$. It means that very high values of $\Pm$ have to be considered in order to model dipolar dynamos with a dominant Lorentz force ($\Lambda'>1$). With higher buoyant forcing ($Ra/Ra_c>10$),  the boundary between these two dipolar regimes is almost constant i.e. it is almost  horizontal in $(Pm,Ra)$-planes. 

The most striking finding visible in figure \ref{PmRa} is that  dipolar dynamos with $\Rol>0.12$ are exclusively dynamos dominated by the Lorentz force (in blue). Hence, the stability domain of dipolar dynamos extends beyond the former limit $\Rol \sim 0.1$, only if $\Lambda ' > 1$. This result seems to be very robust since it is obtained for several Ekman numbers. It is consistent with the behaviour observed for higher $Pm$ in figure \ref{fdip}, where these Lorentz dominated cases are marked by a dot. 

\begin{figure}
	\centering
	\subfigure[$\Ra/\Rac =8.8,\ Pm =1,\ \Rol=0.02,\ f_{dip}=0.85$.]{
	\includegraphics[width=0.45\linewidth]{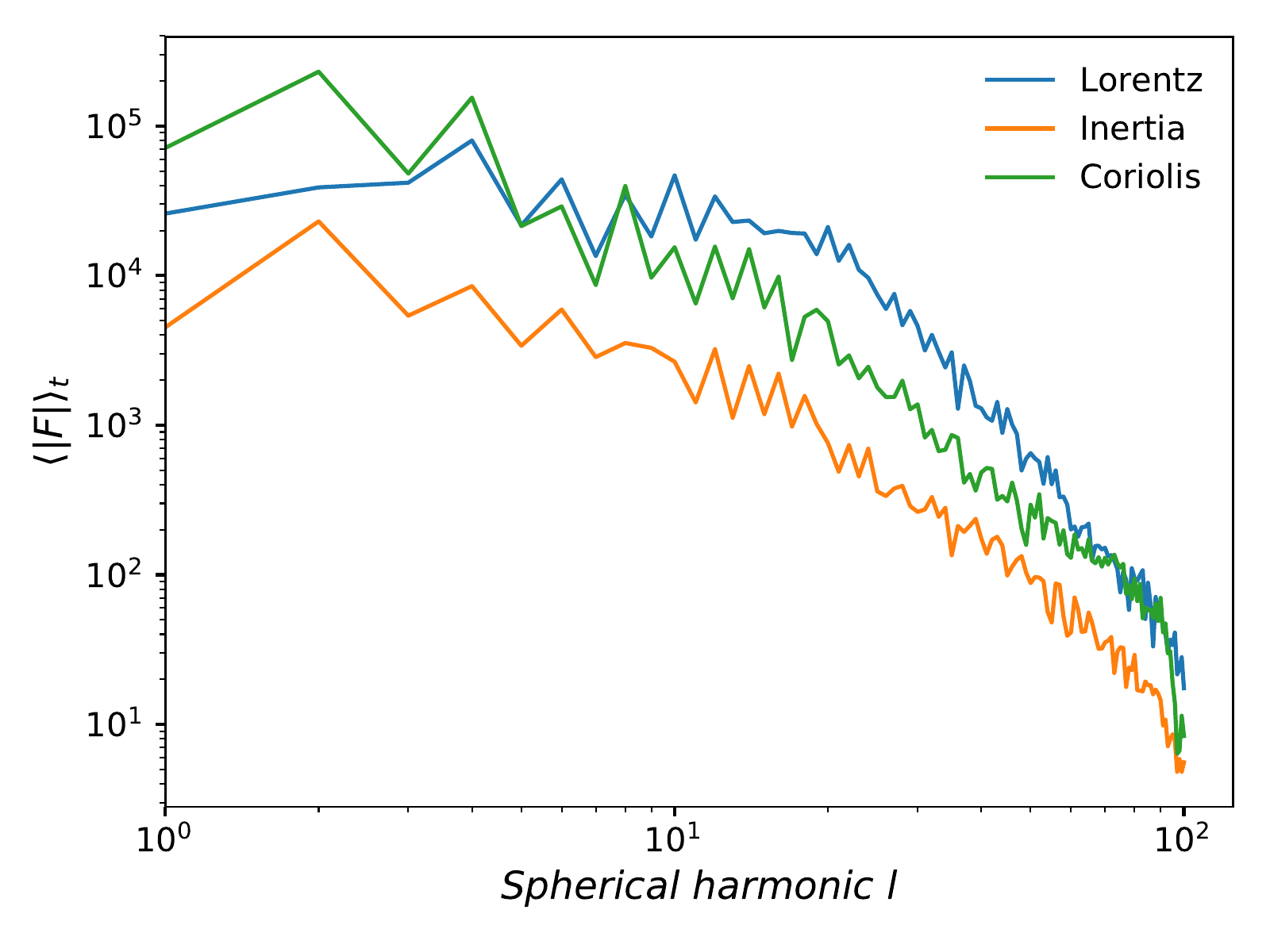}\label{FB71}}
	\hfill
	\subfigure[$\Ra/\Rac =55.3,\ Pm =1,\ \Rol=0.122,\ f_{dip}=0.85$.]{
	\includegraphics[width=0.45\linewidth]{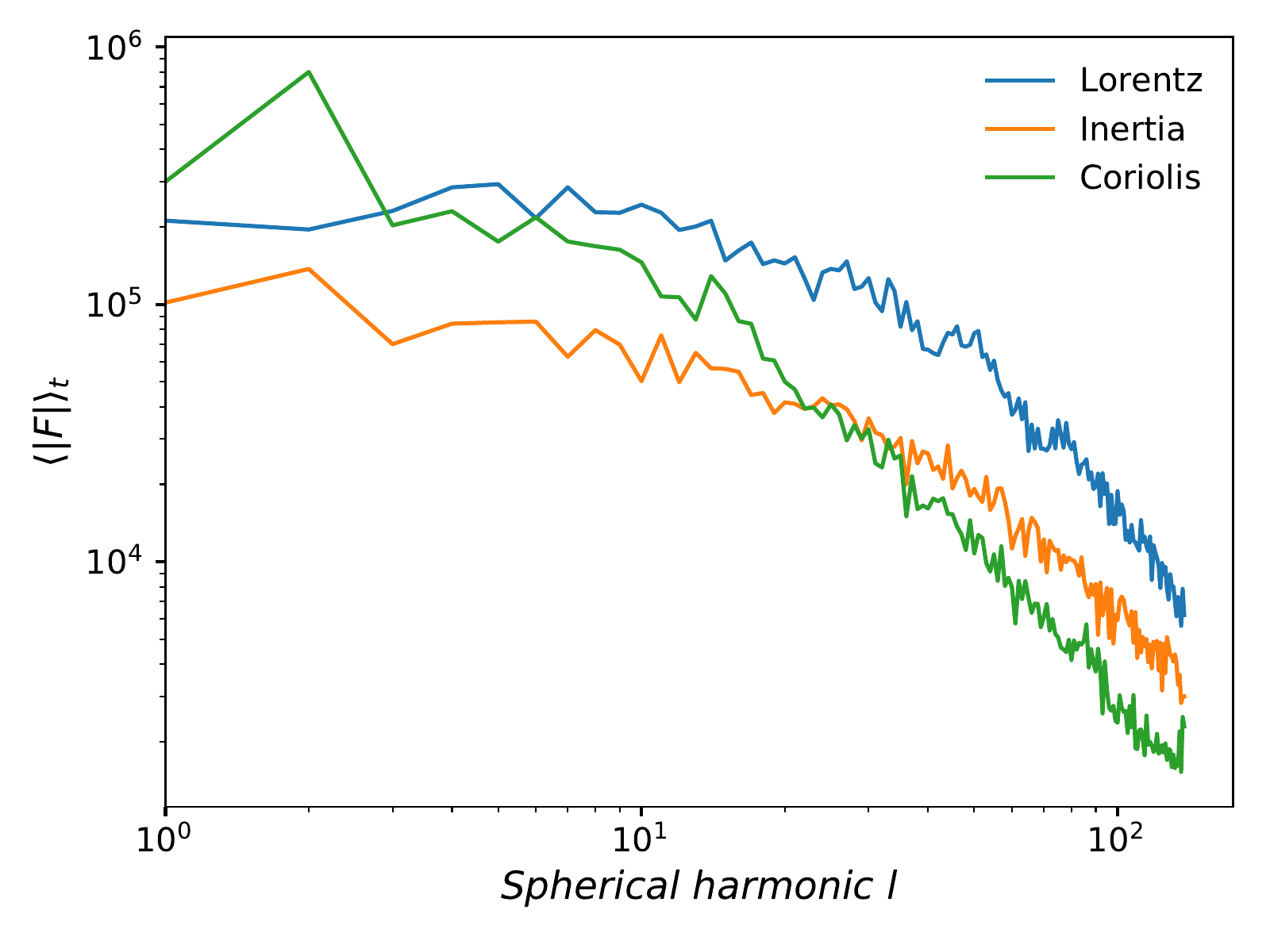}\label{FB471}}
	\subfigure[$\Ra/\Rac =3.5,\ Pm =10,\ \Rol=0.004,\ f_{dip}=0.79$.]{
	\includegraphics[width=0.45\linewidth]{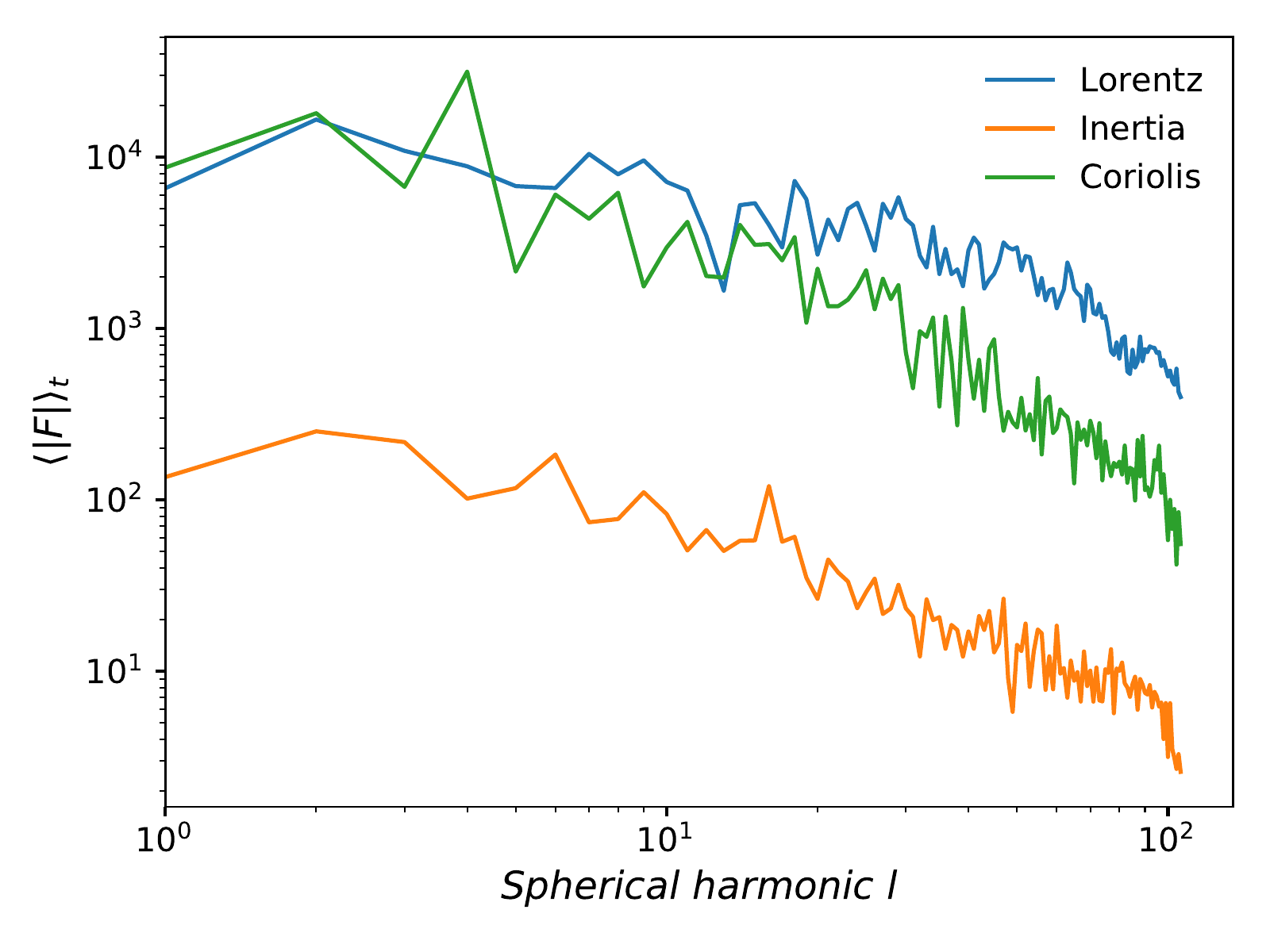}\label{FB310}}
	\hfill
	\subfigure[$\Ra/\Rac=55.3,\ Pm =2.5,\ \Rol=0.118,\ f_{dip}=0.8$.]{
	\includegraphics[width=0.45\linewidth]{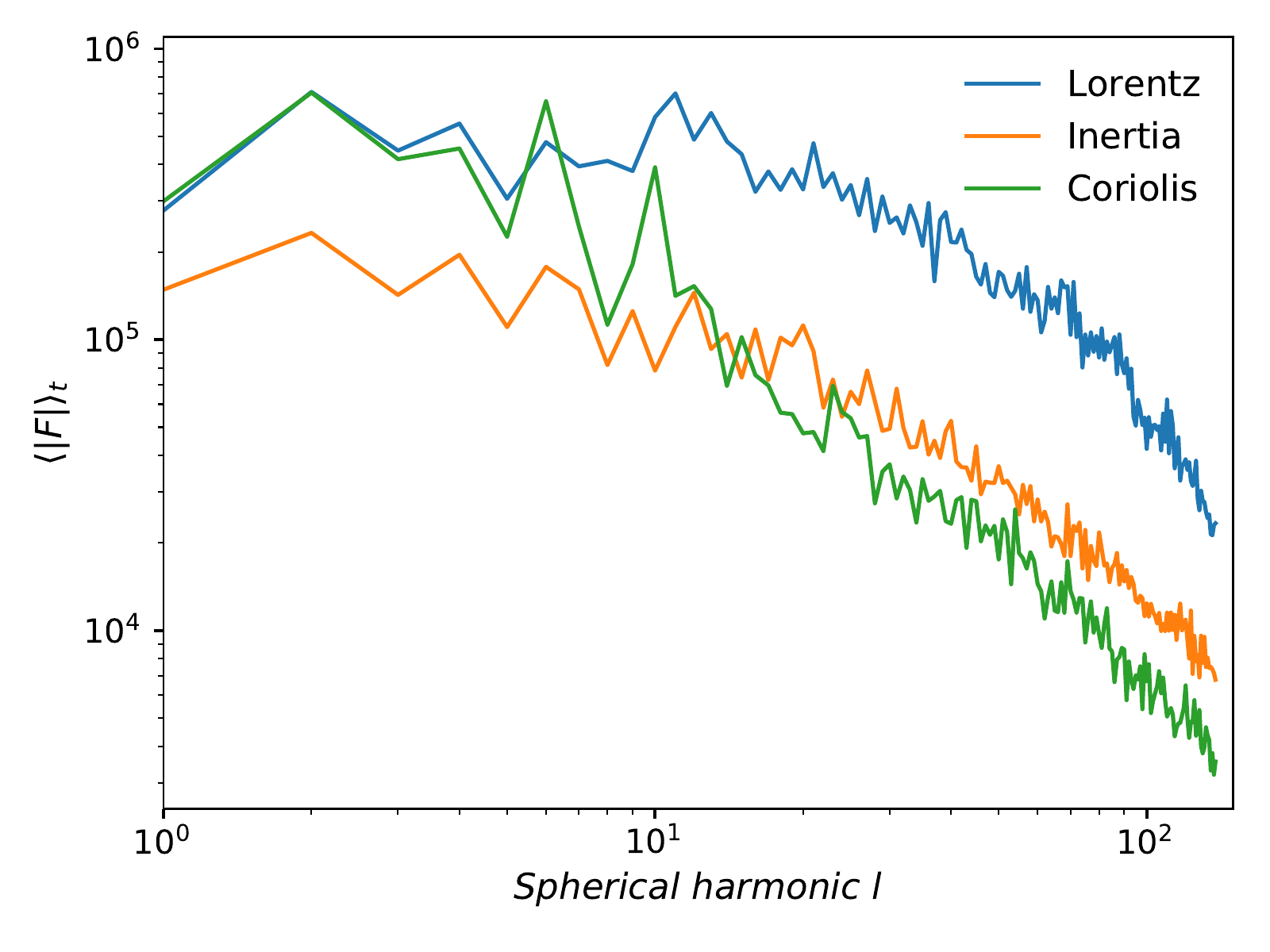}\label{FB4725}}
	\caption{Force balance as a function of the spherical harmonic degree $l$ for several values of $Ra$ and $Pm$ numbers, and a fixed value  $E = 3 \times 10^{-5}$. The three curves distinguish the Lorentz force (blue), inertia (orange) and the Coriolis force (green). On panel \ref{FB4725}, despite a $\Rol > 0.1$, Lorentz and Coriolis forces balance and dominate the large scales.}
	\label{FB}
\end{figure}

In light of the foregoing, Lorentz-dominated dynamos at $\Rol>0.12$ have also been studied in terms of force balance (see figure \ref{FB}). Inertia, Lorentz and Coriolis forces are represented as functions of the spherical harmonic degree $\ell$ for the two different dipolar regimes. 
The forces spectra of well-chosen cases at $\Ek = 3 \times 10^{-5}$ allow us to understand the relative importance of the three forces at each scale (see figure \ref{FB}). The top row (figures \ref{FB71} and \ref{FB471}) represents the force balance for two dipolar solutions not dominated by the Lorentz force on the $\Pm = 1$ branch. The Coriolis force (in green) dominates at the largest scales ($\ell = 1,2$) for both cases despite an important difference in $\Ra/\Rac$, and thus in $\Rol$. Considering the calculation method (equation \eqref{FBeq}), it can be interpreted as a quasi-geostrophic balance, i.e. a near balance between the Coriolis force and the pressure gradient.\\

This needs to be compared to the other dipolar branch, dominated by the Lorentz force (bottom row in figure \ref{FB}). For this branch, Lorentz and Coriolis forces are almost perfectly balanced at large scales, as expected from the values of $\Lambda'$ which exceeds $1$ for these models. This specific balance is similar to the MAC balance (Magnetic-Archimedian-Coriolis) usually obtained for low viscosity models which are numerically very demanding.
The impact of the magnetic Prandtl number can be observed by comparing figures \ref{FB471} and \ref{FB4725}. Indeed these cases have identical input parameter values apart from $Pm$ which has been multiplied by $2.5$ for panel \ref{FB4725}. The main difference observed is the relative importance of the Lorentz force at all scales, whereas inertia and Coriolis force are similarly ditributed. In this way, we can conclude that a slight modification of the magnetic Prandtl number will only affect the Lorentz force magnitude at all scales, the latter increasing with the former.
A comparison of  figure \ref{FB310} and figure \ref{FB4725} enables us to understand the effect of increasing $Ra$. Close to the onset of convection ($Ra/\Rac = 3.5$, figure \ref{FB310}), inertia has a minor role at any length scale. When $Ra/Ra_c$ becomes sufficiently high, the dominance of the Coriolis force on the inertia term depends on the length scale. At large scale, the effects of global rotation and magnetism are dominant. However for lower length scales, inertia dominates the Coriolis force. It means that in models with high $\Pm$ and high $Ra/Ra_c$, the relative importance of inertia (measured globally by the Rossby number) depends on the characteristic length scale. Such behaviour is also expected for the Earth's outer core \citep{NatafS2015}.\\

By comparing the top panels ($\Lambda ' < 1$) with the bottom one ($\Lambda ' > 1$) in figure \ref{FB}, we note qualitative differences at large scales. Although the inertia term seem to play a minor role for the first harmonic degrees $l$ when $\Lambda'$ exceeds 1 - as Coriolis and Lorentz forces are almost perfectly balanced, inertia effects are on an equal footing with Lorentz effects when $\Lambda'<1$. We thus obtain a three-force equilibrium when $\Lambda' <1$, where the Lorentz force and inertia together compensate the non-geostrophic part of the Coriolis force, while the Lorentz force is large enough to equilibrate it when $\Lambda'>1$. Although inertia and viscous effects cannot be completely ignored, models with $\Lambda'>1$  seems to explore a quasi-MAC regime. It is even more interesting to note that even if $\Rol$ exceeds $0.12$ the quasi-MAC balance is still effective at large scales when $\Lambda'>1$ (panel \ref{FB4725}). The strong Lorentz force overshadows the effect of inertia enough to prevent a transition to the multipolar state. This particular spectral distribution of force balance seems to provide a consistent explanation for the particular behaviour of the Lorentz-dominated branch at high $\Rol$ stated in \S \ref{Part1}. \\

The parameter space allowing dipolar dynamos has been extended through Lorentz-dominated dynamos and the threshold with the other regime is quite-well defined ($\Lambda'\sim1$, see figure \ref{PmRa}). The limits of this dipolar regime at high $\Ra/\Rac$ are studied in \S \ref{Part3}.

\subsection{A new criterion for dipolar dynamos taking into account magnetic effects}\label{Part3}

We report for the first time dipolar solutions with $\Rol$ higher than $0.12$ for dynamo models with fixed temperatures and no-slip boundary conditions. We have shown that such dynamos exist only if $\Pm$ is sufficiently high in order to increase sufficiently the effects of the Lorentz force. The role of this magnetic force seems to be critical for preventing inertia effects responsible for dipole collapse when $\Rol$ exceeds $0.12$ with $\Lambda'<1$.

\begin{figure}
	\subfigure[$\Ra/\Rac = 29.6,\ Pm =6,\ \Rol=0.193,\ f_{dip}=0.38,\ \Lambda'=1.11$ .]{
	\includegraphics[width=0.45\linewidth]{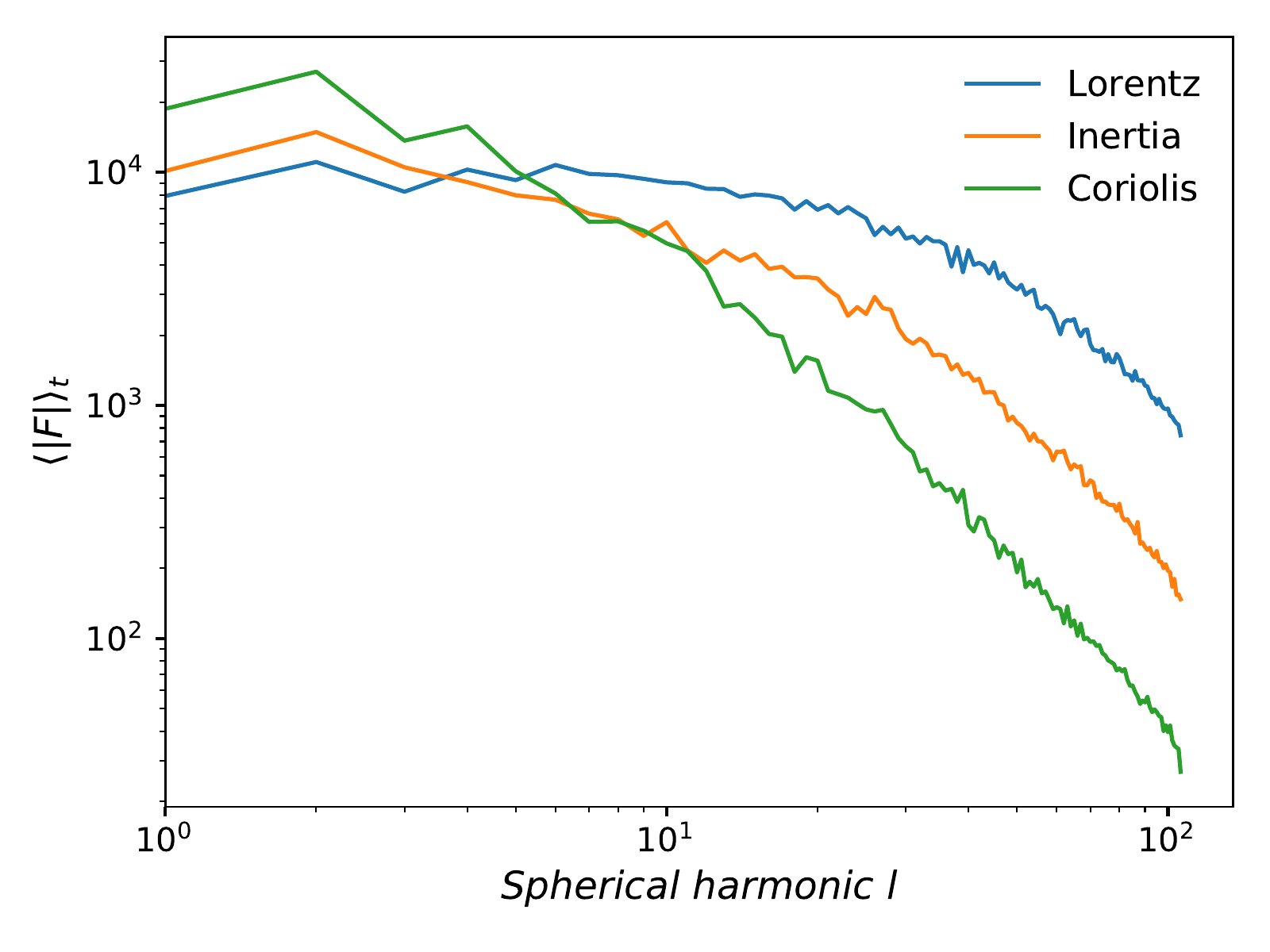}\label{FB186}}
	\hfill
	\subfigure[$\Ra/\Rac = 64.7,\ Pm =1,\ \Rol=0.163,\ f_{dip}=0.49,\ \Lambda'=0.35$ .]{
	\includegraphics[width=0.45\linewidth]{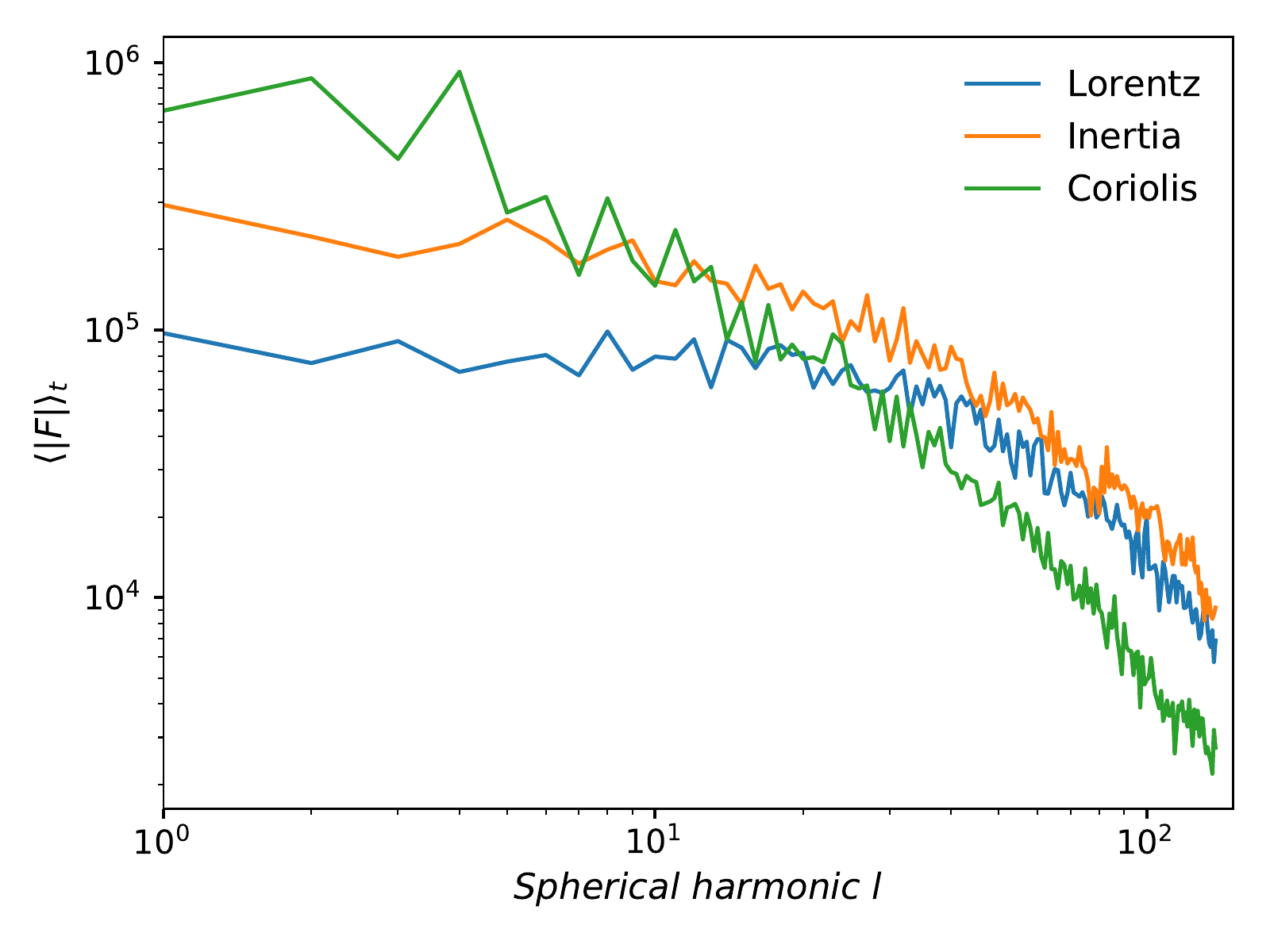}
	\label{FB551}}
	\caption{Force balance for two multipolar cases at $E=3 \times 10^{-4}$ (figure \ref{FB186}) and $E=3 \times 10^{-5}$ (figure \ref{FB551}). The behaviour is different at all scales despite a $\fdip < 0.5$ for both. In case \ref{FB186}, for which $\Lambda'>1$, the Lorentz force is still very important, and dominating small scales.}
\end{figure}

Indeed, the behaviour of the Lorentz-dominated branch at $\Rol > 0.12$ is totally different from the usual behaviour. The force balance of two cases with similar parameters except their $\Pm$ value highlights the different behaviours of the two dipolar branches. At high $\Ra/\Rac$ value, the $\Lambda'<1$ branch becomes multipolar as inertia starts to be more important than the Lorentz force at large scales (see figure \ref{FB551}). However, the $\Lambda'>1$ branch evolves while the Lorentz force maintains the dipolar magnetic field (figure \ref{FB4725}). As shown on figure \ref{FB551}, despite no quasi-MAC balance at large scales, inertia starts to be significant at $\Ra/\Rac > 60$ and $\Rol > 0.16$. These are reasonable grounds to believe that a transition to a multipolar state will occur for higher values of $\Rol$ when $\Lambda'>1$.

Increasing $Ra/\Rac$ for our highest Ekman value leads to $\Rol$ up to two times the critical value, i.e. $0.2$ (see \ref{Table}). For these values of $\Rol$, the transition to the multipolar state occurs. The force balance of a multipolar case at $Rol \sim 0.19$ and $\Lambda'\gtrsim 1$ is represented in figure \ref{FB186}. Comparing it with the classical multipolar force balance (figure \ref{FB551}) - occuring at values of $\Rol$ close to the transitional value $0.12$ and with $\Lambda'<1$ - highlights the importance of Lorentz force at all scales in the former case (figure \ref{FB186}). It is competing with inertia at large scales and dominating medium to small scales, whereas for the non-Lorentz dominated multipolar dynamo (\ref{FB551}) the Lorentz force vanishes at large scales and just reaches inertia at smaller scales. This behaviour is consistent with our interpretation of high-$Pm$ and turbulent dynamos as reaching a force balance in which the Lorentz force cannot be neglected. \\

Our results on force balance depending on the length scale suggest that a minor relative importance of inertia at large scales is a necessary condition for dipolar dynamos with $\Rol>0.12$. In order to better understand this finding, we develop a phenomenological argumentation based on simple force ratios. To get rid of the geostrophic balance i.e. the part of the Coriolis force balanced by the pressure gradient term in the momentum equation, we consider the curl of this equation. The statistical equilibrium between inertia and the Lorentz force can be written as

\begin{equation}
\{\vect{\nabla} \times ((\vect{\nabla} \times \vect{v})\times \vect{v})\} =\{ \vect{\nabla} \times (\frac{1}{\mu \rho}(\vect{\nabla}\times\vect{B})\times\vect{B})\}\,,
\end{equation}

which leads to

\begin{equation}
\frac{v_{rms}^2}{L_{inertia}L_\nu} \sim \frac{B_{rms} ^2}{\mu \rho L_B L_\eta}\,,
\end{equation}
where $L_{inertia}$ and $L_B$ correspond to the length scales of inertia and the Lorentz force respectively. From our analysis on the force balance depending on length scales, we can argue that inertia affects the stability of dipolar dynamos when this term is on an equal footing with the Lorentz force at large scale. As a result, we can focus on the situation $L_{inertia}\sim L_B$. In this case, we obtain:

\begin{equation}
  \frac{L_\eta }{L_\nu} \sim \frac{B_{rms}^2}{\mu \rho v_{rms}^2} .
\end{equation}\label{Scaleq}

Using the definitions of common dimensionless numbers and equations (\ref{Lambdador}) and (\ref{Rol}), this relation can be reduced to :
\begin{equation}
\frac{L_\eta }{L_\nu} \sim \frac{\Lambda}{Ro \ Rm}
\\
\Leftrightarrow \quad  \frac{L_\eta }{L_\nu} \sim \frac{\Lambda '}{Ro_{L_\nu}}\frac{L_\eta }{L_\nu}\ .
\end{equation}

Finally, we find $\Lambda  '\sim Ro_{L_\nu}$.
This expression can also be written as $\Lambda ' \sim 5\ \Rol$ using the scaling (already noticed by \cite{orubaD14GRL}, also checked with our dataset) $L_\nu \sim \bar{\ell_u}/5$ which gives $5\Rol \sim Ro_{L_\nu}$. Consequently, we have found a simple condition corresponding to a balance between inertia and magnetic effects at large scales in which output dimensionless parameters are involved. Let us test this condition using our data set.\\

\begin{figure}
	\centering
	\subfigure[$E = 3 \times 10^{-4}$ .]{
	\includegraphics[width=0.48\linewidth]{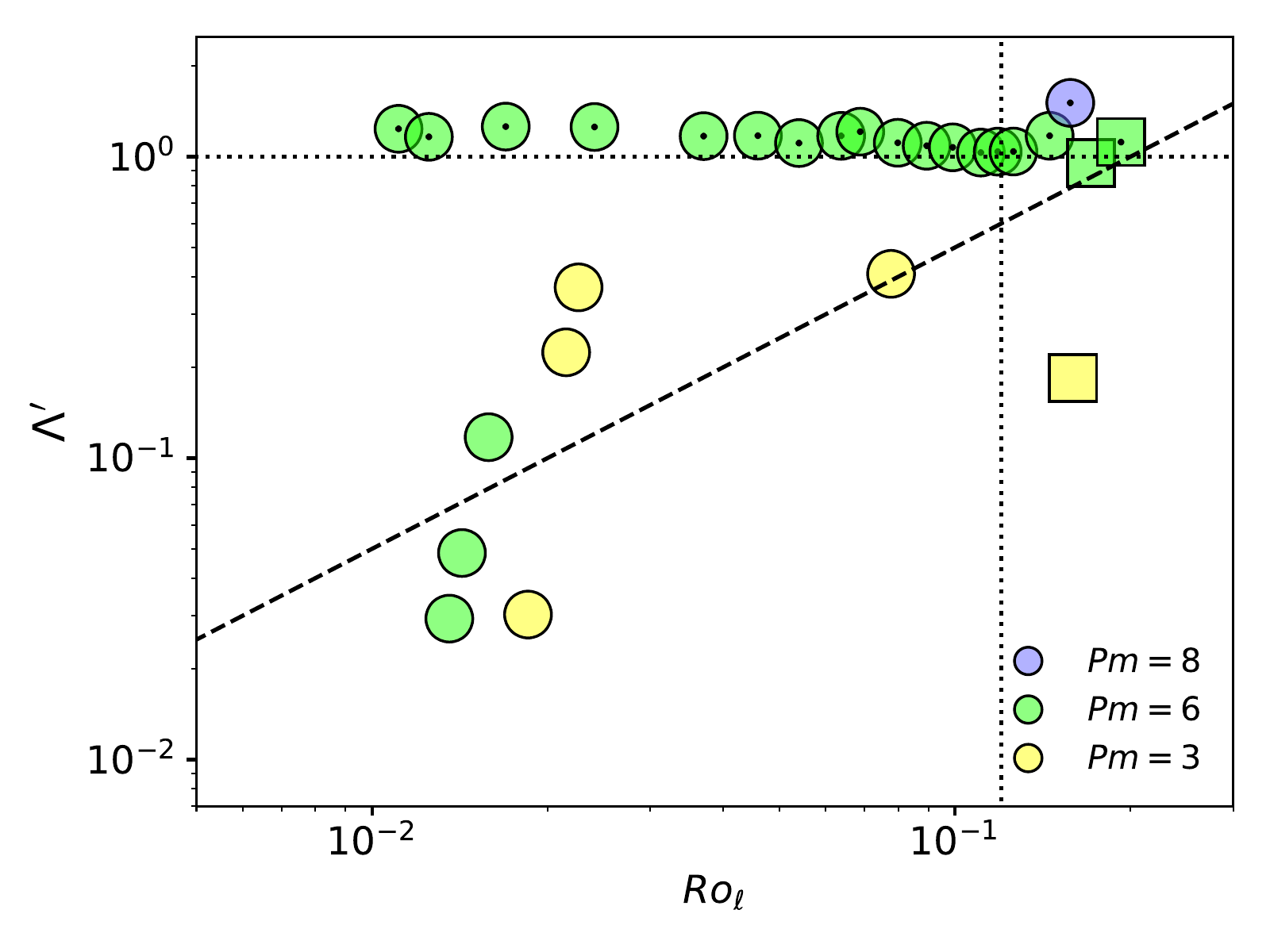}\label{MAC3e4}}
	\hfill
	\subfigure[$E = 10^{-4}$ .]{
	\includegraphics[width=0.48\linewidth]{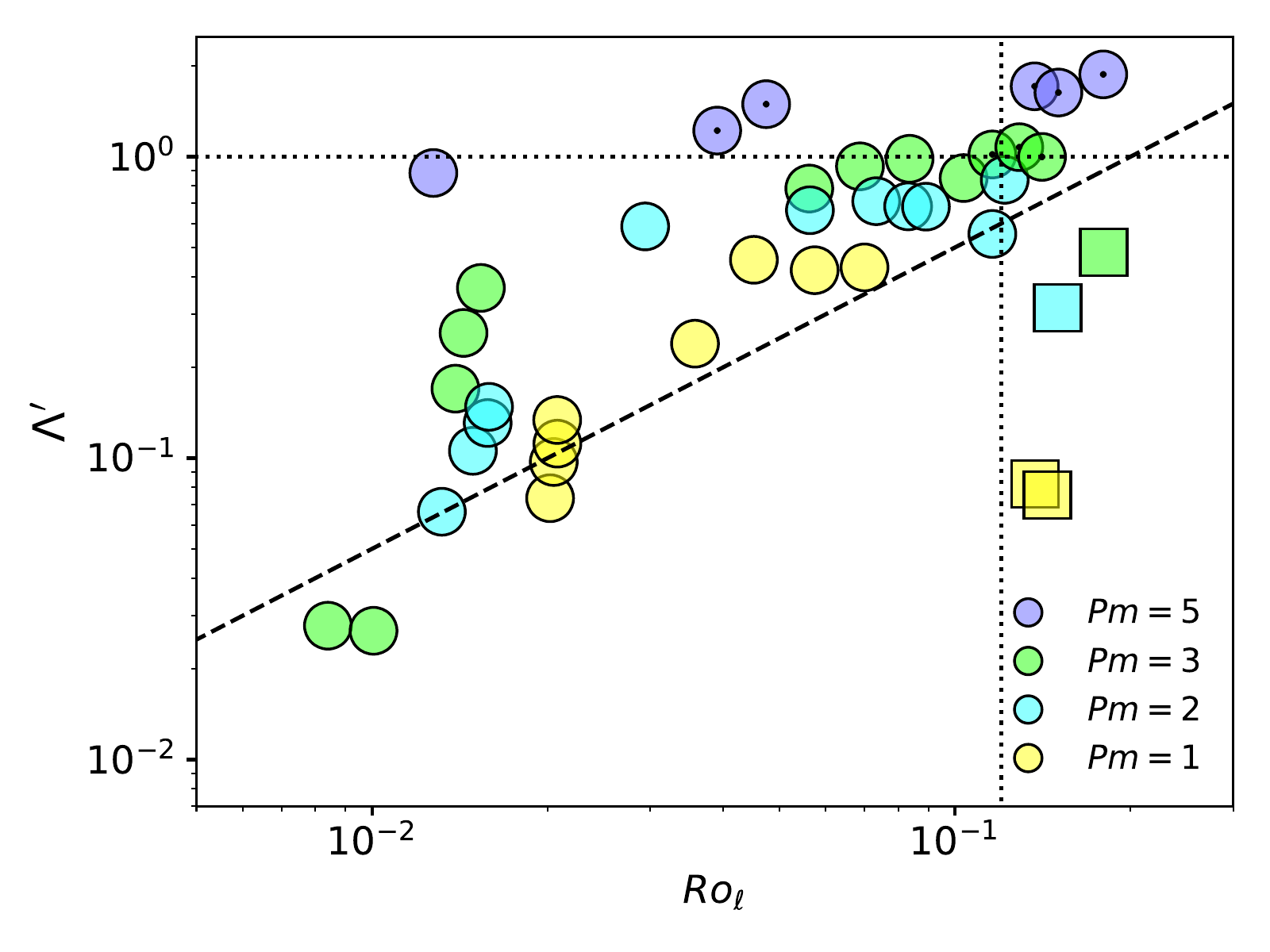}\label{MAC1e4}}
	\subfigure[$E = 3 \times 10^{-5}$ .]{
	\includegraphics[width=0.48\linewidth]{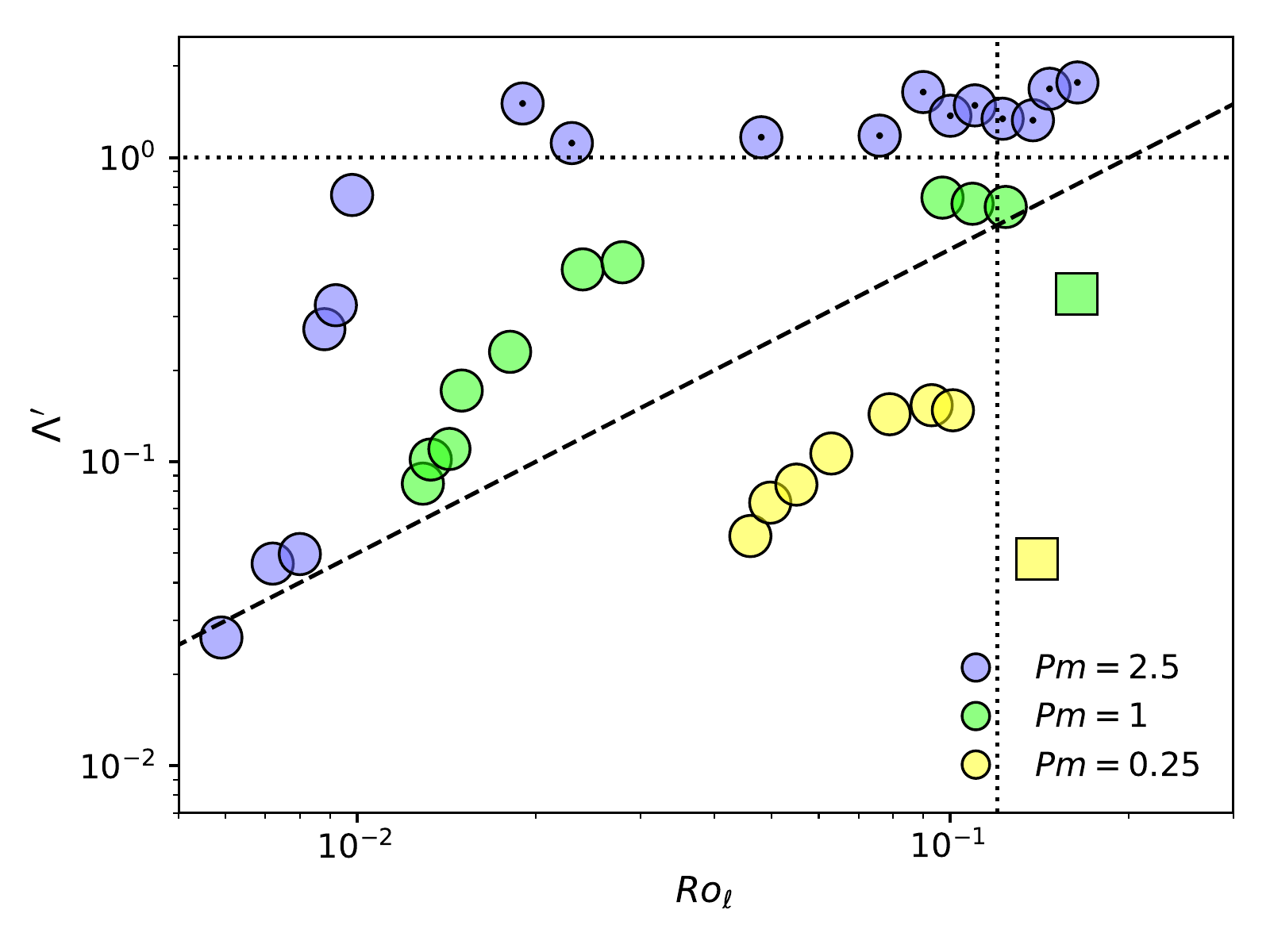}\label{MAC3e5}}
	\hfill
	\subfigure[$E = 10^{-5}$ .]{
	\includegraphics[width=0.48\linewidth]{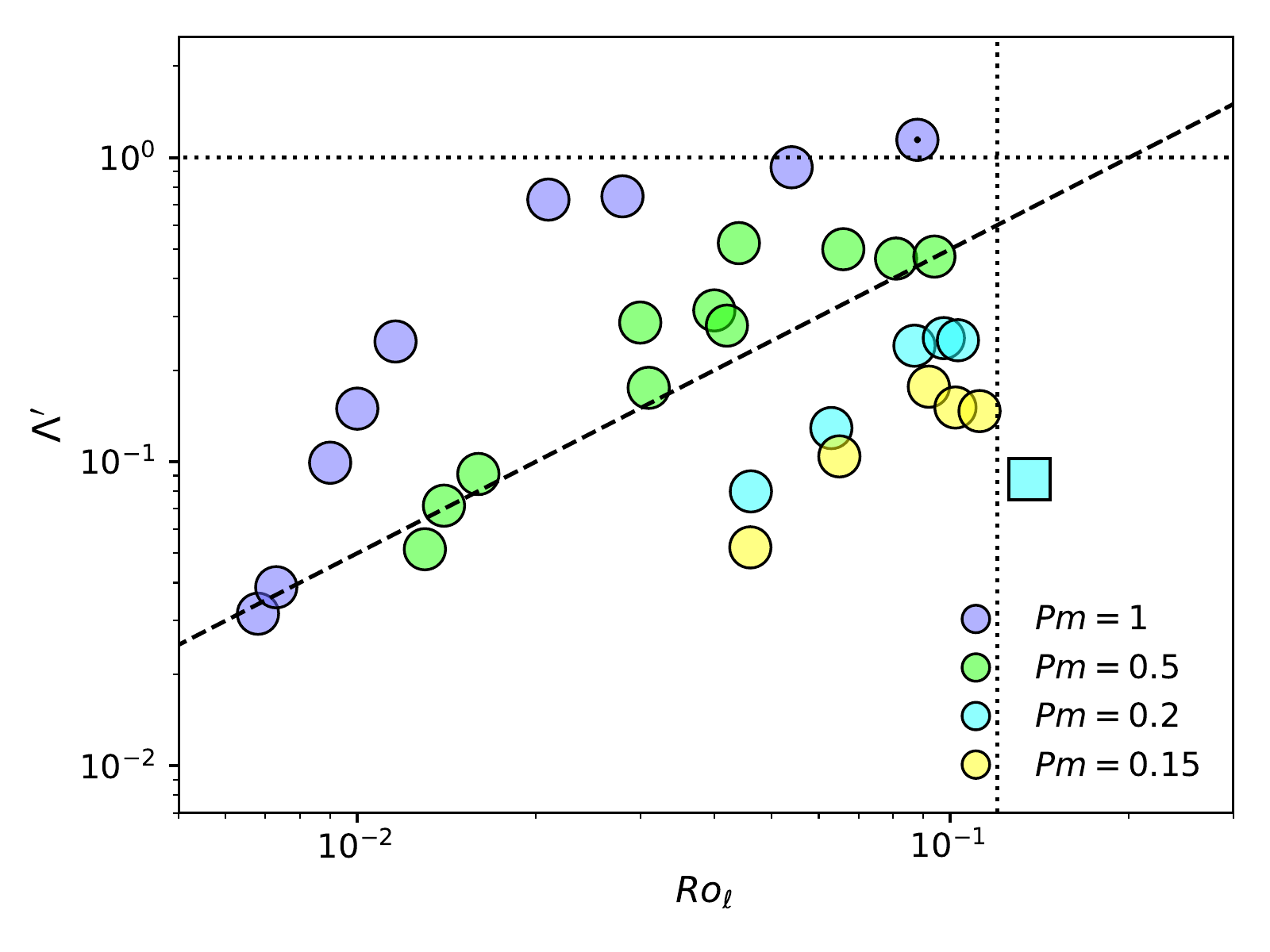}\label{MAC1e5}}
	\caption{$\Lambda ' $ versus $Ro_\ell$ for several Ekman numbers. The colours on each graph allow us to distinguish the magnetic Prandtl numbers while the marker corresponds to the magnetic topology: circles for stable dipolar dynamos and squares for unstable dynamos (multipolar or unstable dipolar). The dotted lines represent $\Rol=0.12$ (vertical line) and $\Lambda '= 1$ (horizontal line). The dashed line shows the proposed limit $\Lambda'  = 5 \Rol $ (see the text).}
	\label{MAC}
\end{figure}

The evolution of a branch at constant $\Pm$ in the parameter space $\Lambda '\ vs\ \Rol$ can be seen for several $\Ek$ numbers in figure \ref{MAC}. The marker legend is the same as in figure \ref{fdip}. Below the horizontal dotted line corresponding to dynamos with $\Lambda'<1$, the dipole collapse is observed when $\Rol$ exceeds $0.12$ (vertical dotted line) as obtained by previous study. The situation differs above $\Lambda'>1$ threshold where dipolar dynamos persist with higher $\Rol$. 

In figure \ref{MAC}, the dashed line represents the condition $\Lambda'=5\Rol$ as obtained from equation \eqref{Scaleq}. For dipolar models above this dashed line, our analysis suggests that inertia effects play a minor role at large scales in comparison with the magnetic and Coriolis effects. Inertia only affects the stability of dipolar fields with $\Lambda'>1$ in the vicinity of the dashed line. Otherwise, dipolar dynamos are maintained even if values of $\Rol$ larger than $0.12$ are considered when $\Lambda'>1$. However, we report here only transitions for $E\geq 10^{-4}$ when $\Lambda'>1$. At this point, additional simulations seem to be needed in order to conclude on the critical role of the condition $\Lambda'>5\Rol$ for Lorentz-dominated dipolar dynamos.

Dynamos approaching this condition (green branches of panels \ref{MAC3e4} and \ref{MAC1e4}) become unbalanced around $0.2$. According to our results, this value is the critical value for dynamos with $\Lambda'=1$ to make the transition to the multipolar state, eventually through an unstable dipolar state.\\

\section{Discussion and conclusion}\label{Discussion}

\input{discussion}

\section*{Acknowledgments}
This work has been done within the LABEX PLAS@PAR project, and received financial state aid managed by the Agence National de la Recherche, 
as part of the Programme ``Investissements d'Avenir'' under the reference ANR-11-IDEX-0004-02.
This work was granted access to the HPC resources of MesoPSL financed
by the Region Ile de France and the project Equip@Meso (reference
ANR-10-EQPX-29-01) of the programme Investissements d’Avenir supervised
by the Agence Nationale pour la Recherche.

\newpage

\appendix

\section{Scaling laws} \label{AnnA}
\begin{figure}[h]
	\centering
	\subfigure[$E = 3 \times 10^{-4}$ .]{
	\includegraphics[width=0.45\linewidth]{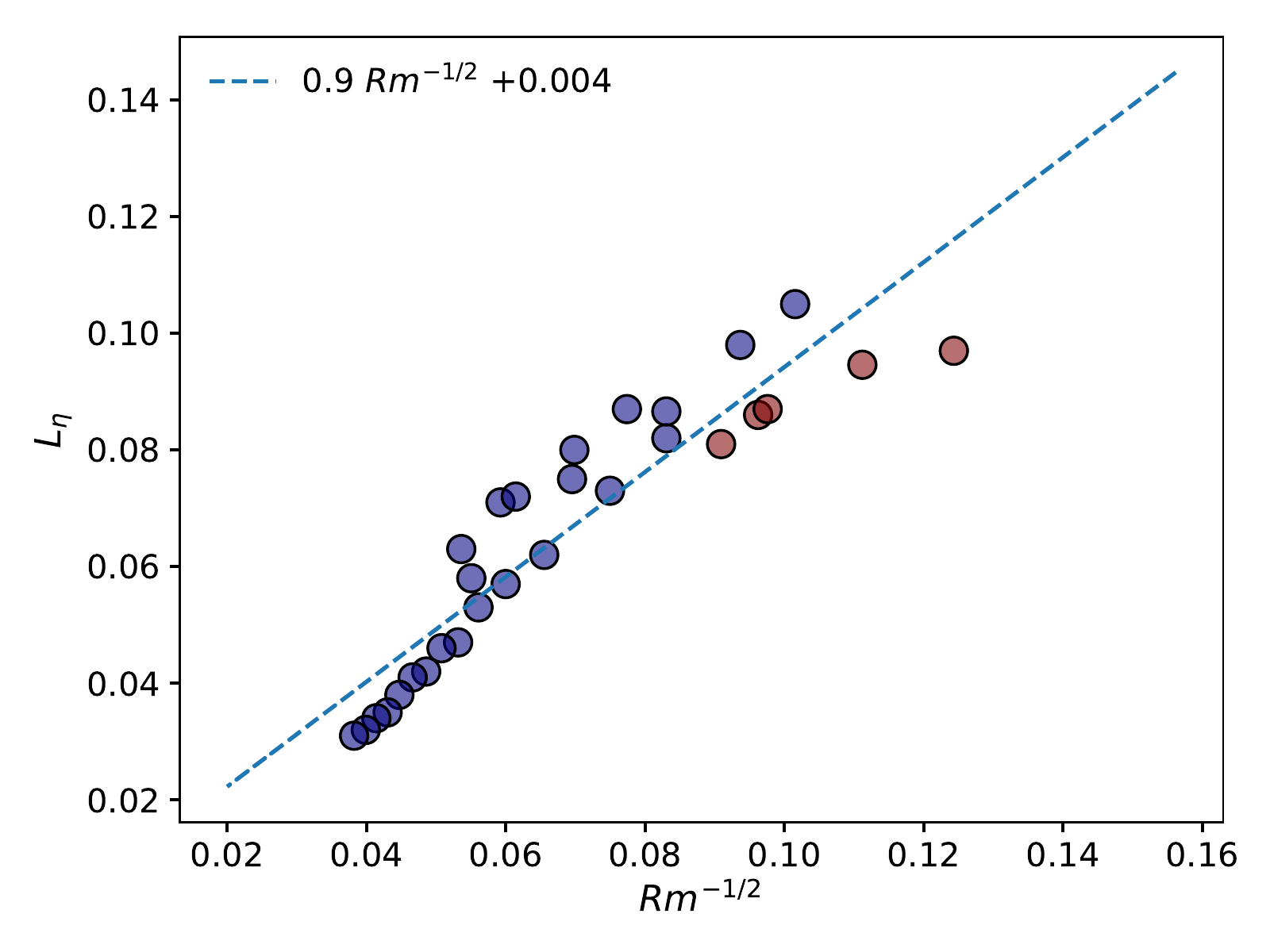}}
	\hfill
	\subfigure[$E = 10^{-4}$ .]{
	\includegraphics[width=0.45\linewidth]{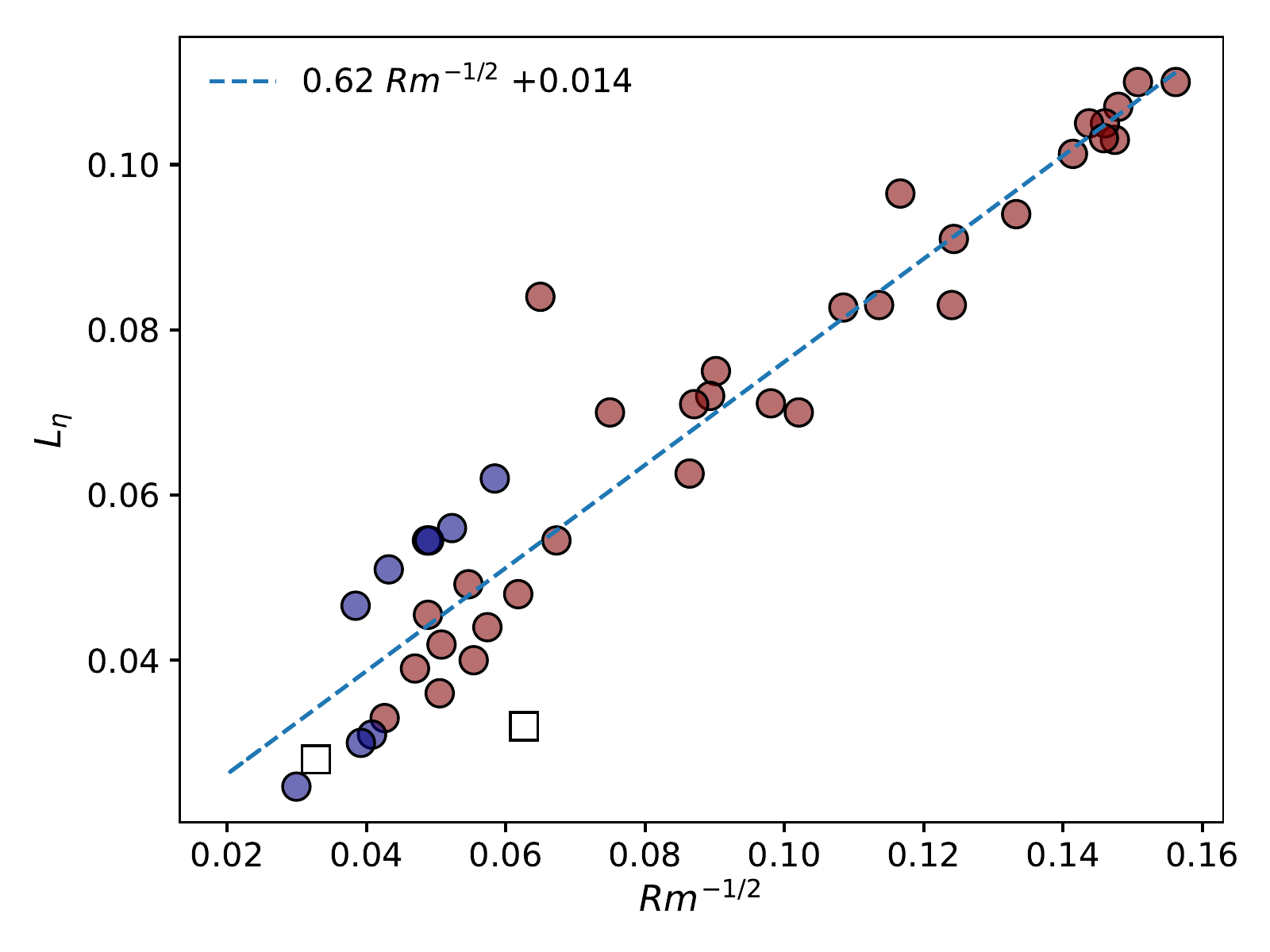}}
	\subfigure[$E = 3 \times 10^{-5}$ .]{
	\includegraphics[width=0.45\linewidth]{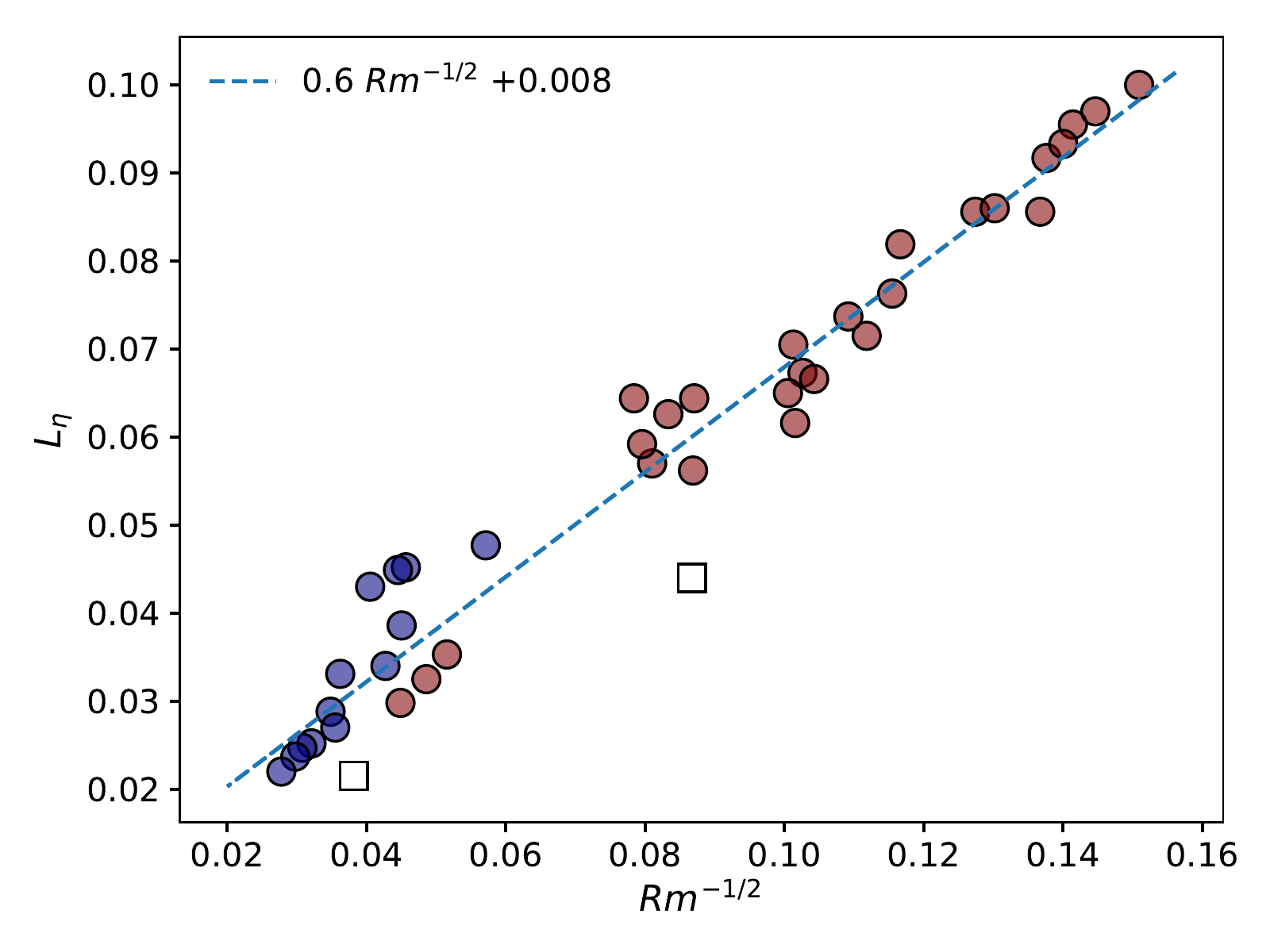}}
	\hfill
	\subfigure[$E = 10^{-5}$ .]{
	\includegraphics[width=0.45\linewidth]{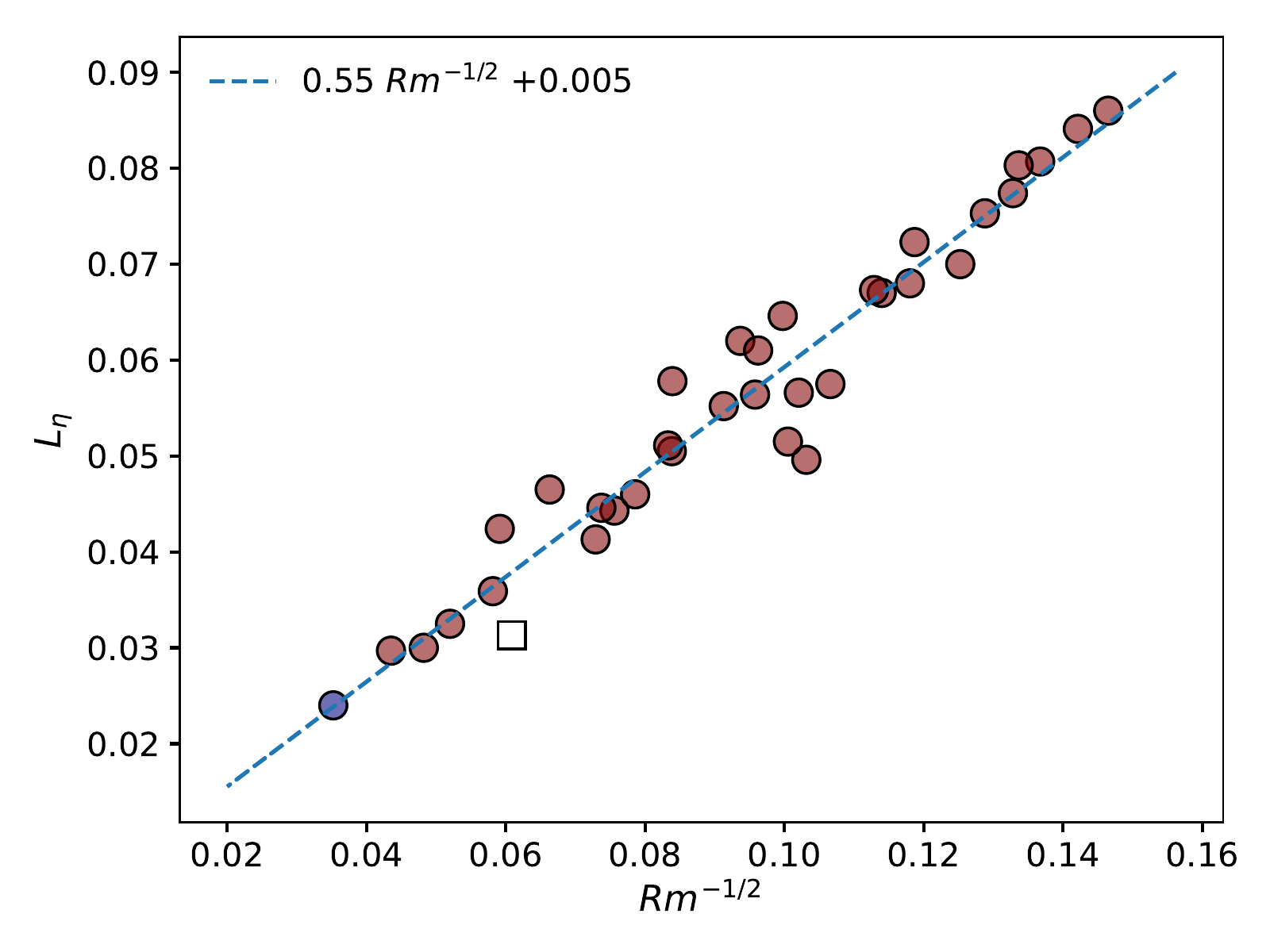}}
	\caption{$L_\eta \propto \Rm^{-1/2}$ for four Ekman values. Mauve  and blue circles stand distinguish the two dipolar behaviours, respectively non-Lorentz dominated and Lorentz-dominated dynamos. Squares are multipolar cases.}
	\label{LbRm}	
\end{figure}

For four values of the Ekman number, we obtain $L_\eta \propto \Rm^{-1/2}$ (figure \ref{LbRm}). The fitting coefficient is indicated on each figure and decreases with Ekman. Both Lorentz dominated dynamos (in blue) and the non-Lorentz dominated dynamos (in mauve) seem to respect this scaling. Multipolar cases (squares) are indicated but not relevant for this length scale. This scaling has been used for $L_\eta$ values not computed (depicted by the symbol * in the Tables, \ref{Table}).

\newpage

\section{Simulation tables}\label{Table}
Table of the simulation presented, with main input and output parameters. (-) stands for a value that has not been computed, stars (*) for values deduced from scaling laws. Nusselt number $Nu$ is defined as the ratio of the total heat flux to the conductive heat flux whereas $f_{ohm}$ is the ratio of ohmic dissipation to total dissipation.\\
\input{Table_pap/Tables}

\clearpage

\bibliographystyle{elsarticle-harv.bst}
\bibliography{refs}

\end{document}

%% file: abstract.tex
The dynamo effect is the most popular candidate to explain the non-primordial magnetic fields of astrophysical objects. Although many systematic studies of parameters have already been made to determine the different dynamical regimes explored by direct numerical geodynamo simulations, it is only recently that the regime corresponding to the outer core of the Earth characterized by a balance of forces between the Coriolis and Lorentz forces is accessible numerically. In most previous studies, the Lorentz force played a relatively minor role. For example, they have shown that a purely hydrodynamic parameter (the local Rossby number $Ro_\ell$) determines the stability domain of dynamos dominated by the axial dipole (dipolar dynamos).

In this study, we show that this result cannot hold when the Lorentz force becomes dominant. We model turbulent geodynamo simulations with a strong Lorentz force by varying the important parameters over several orders of magnitude. This method enables us to question previous results and to argue on the applications of numerical dynamos in order to better understand the geodynamo problem. Strong dipolar fields considerably affect the kinetic energy distribution of convective motions which enables the maintenance of this field configuration. The relative importance of each force depends on the spatial length scale, whereas $Ro_\ell$ is a global output parameter which ignores the spatial dependency. We show that inertia does not induce a dipole collapse as long as the Lorentz and the Coriolis forces remain dominant at large length scales.

%% file: intro.tex
Planetary magnetic fields, maintained over billions of years despite ohmic dissipation, are a precious resource to constrain the laws of physics. The favorite mechanism to explain them, the dynamo effect, explains how a magnetic field can be generated by electromotive forces driven by electrically conductive fluid movements in celestial bodies \citep{moffatt,book_dormy}. Dynamo action is an instability by which a conducting fluid transfers part of its kinetic energy to magnetic energy. Planetary magnetic fields result from such processes thought to be driven by convection in electrically conducting fluid regions. For example, turbulent convective motions in the Earth's outer core, driven by buoyancy, are then subject to two forces: the  Coriolis force resulting from global planetary rotation and the Lorentz force which accounts for the back-reaction of magnetic field on the flow from which it is generated. In order to improve our understanding about the Earth's magnetic field, it is of interest to numerically model turbulent convection in a spherical shell in which the main force balance consists of the Magnetic force, the Buoyancy (Archimede) force and the Coriolis force (the so-called MAC-balance).

Earth-like fields do not only consist of a simple axial dipole field which is maintained over time. Paleomagnetic measurements have allowed us to reconstruct the dynamics of the magnetic field and revealed that the Earth's dipolar field has reversed its polarity several hundred times during the past 160 million years. These polarity reversals are known to be strongly irregular and chaotic, and such events were observed numerically by \citet{gr95} for the first time. 
Since this first fully three-dimensional numerical model \citep[e.g.][]{gr95}, there have been significant advances in understanding the geodynamo. Many features of the Earth's magnetic field have been reproduced numerically \citep{christensen98,christensen99,busse98,taka05,christensen07}.
However the realistic physical properties of the Earth's outer core (fast rotation, low dissipation coefficients\ldots) differ by several orders of magnitude from the values accessible in direct numerical simulations (DNS). For instance, the Ekman number, ratio of the viscous and Coriolis forces, is approximately $E=10^{-15}$ in the Earth's outer core whereas $E\geq 10^{-7}$ can be considered in numerical models (see below for a complete definition of this dimensionless number). 
Progress in both numerical methods and parallel machine architecture has made it possible to explore an extensive parameter space in order to deduce the physical ingredients responsible for the dominance of the axial dipole field \citep{christensen06,king10,schrinner12,Soderlund2012,Petitdemange2018}.  \citet{olson06} have also inferred some of the physical causes associated with field reversals in planetary interiors from numerical studies. Numerical simulations suggest that reversals may arise from the importance of inertia relative to the Coriolis force in Earth's outer core.  \\

Observations and numerical simulations indicate that rapid global rotation and thus the ordering influence of the Coriolis force is of major importance for the generation of coherent magnetic fields \citep{stellmach04,kapyla09,brown10}. \citet{kutzner02} demonstrated the existence of a dipolar and a multipolar dynamo regime, and  \citet{christensen06} showed that the transition between these two regimes is governed by a local Rossby number ($\Rol$), i.e. the influence of inertia relative to the Coriolis force. Similar results were reported by \citet{sreenivasan06}, as well. Dipolar models were found for small local Rossby numbers; separated by a fairly sharp regime boundary from multipolar models, where inertia is more important. The models transition from a dipolar morphology to a multipolar state when the local Rossby number reaches a certain value ($\Rol>0.1$). However, such transitions have been reported by previous systematic parameter studies such as \citet{christensen06} for models that explore a regime in which the Lorentz force does not play a major role (see for instance \citet{KingB13}). \citet{orubaD14GRL} have also confirmed the validity of the criteria $\Rol<0.1$ for dipolar dynamos, but used a data set provided by Christensen, so it cannot be considered as an independant confirmation. Here, we study the stability domain of dipolar dynamos in a unexplored regime in which the Lorentz force is dominant.

In DNS, MAC-balance has been obtained either by considering high values of the magnetic Prandtl number $Pm$ \citep{Dormy2016,Dormy2018,Petitdemange2018}, i.e. increasing the importance of the magnetic force (since $Rm=Pm Re$) ; or by lowering the Ekman number \citep{schaefferJNF17,Sheyko2017,Schwaiger19}, i.e. reducing the effect of viscous forces compared to the Coriolis force. In the former case, turbulence plays a minor role as only low values of the Reynolds number have been used by these studies. By comparison, in the latter situation, high Reynolds numbers are considered and the numerical cost of such models increases significantly. Here, we link these two approaches by increasing the influence of inertia (via the buoyant forcing $Ra$) in models with high magnetic Prandtl number $Pm$. This method enables us to study the influence of the Lorentz force on the local Rossby number criteria. Consequently, we study the stability domain of dipolar dynamo simulations in a regime where the Lorentz force is not negligible.\\

After a brief presentation of the model and control parameters used in this study, the validity of the local Rossby number criteria is tested for these models (see \S \ref{Part1}). Alternative ways to characterise a dipolar model are presented, and used to distinguish two branches of dipolar solutions (\S \ref{Part2}). An explanation of the different behaviours observed for each branch is then proposed (\S \ref{Part3}). Finally these results are compared with observations and previous studies (\S \ref{Discussion}).

%% file: discussion.tex
In this paper we have shown that the distinction between the dipolar and the multipolar regimes cannot be established only through the $\fdip$ value nor through the $\Rol$ criterion. When considering sufficiently high $Pm$ values, time evolution of these quantities seems to be an important parameter as well as the Elsasser number variations. For this study a dipolar dynamo is defined as a solution with a stable tilt angle (no reversals throughout the integration time), a mean dipole field strength $\fdip > 0.5$ and an Elsasser number not subjected to any significant drop. Despite these more restrictive conditions dipolar dynamos with $\Rol>0.12$ are reported (\S \ref{Part1}).\\

By exploring high $Pm$ and $Ra/\Rac$ values, we have been able to define the boundaries of the Lorentz-dominated dynamos area in the parameter space. This regime characterised by a $\Lambda'>1$ can be reached for sufficiently high $Pm$ when $Ra/\Rac\gtrsim 10$. These solutions have proven to be the dipolar dynamos found at high $\Rol$ values. A thorough investigation of the scale dependence of the forces reveals that the large scales balance depends on the regime studied. Indeed Lorentz-dominated dynamos exhibit an almost perfect balance between Lorentz and Coriolis forces at very large scales, whereas it is only partial for the other regime (\S \ref{Part2}).\\

Moving towards higher values of $Ra/\Rac$ allows us to obtain multipolar cases with different force balance. When the Lorentz force does not dominate ($\Lambda'<1$), inertia becomes more important than the magnetic component leading to a transition to the multipolar state around $\Rol \sim 0.12$. However when the $\Lambda'>1$, the strong dipolar component of the magnetic field seems to prevent the growth of inertia at large scales even at high $\Rol$ values. The large scale balance of the Coriolis and Lorentz terms leads to the relation $\Lambda'\sim 5 \Rol$. As long as Lorentz dominates, i.e. $\Lambda'> 5 \Rol$ dipolar dynamos are found even if $\Rol>0.12$. This relation enables a better distinction of the three behaviours studied in the $(\Lambda ' , \Rol)$ representation (\S \ref{Part3}).\\

\begin{figure}
	\center
	\includegraphics[width=0.7\linewidth]{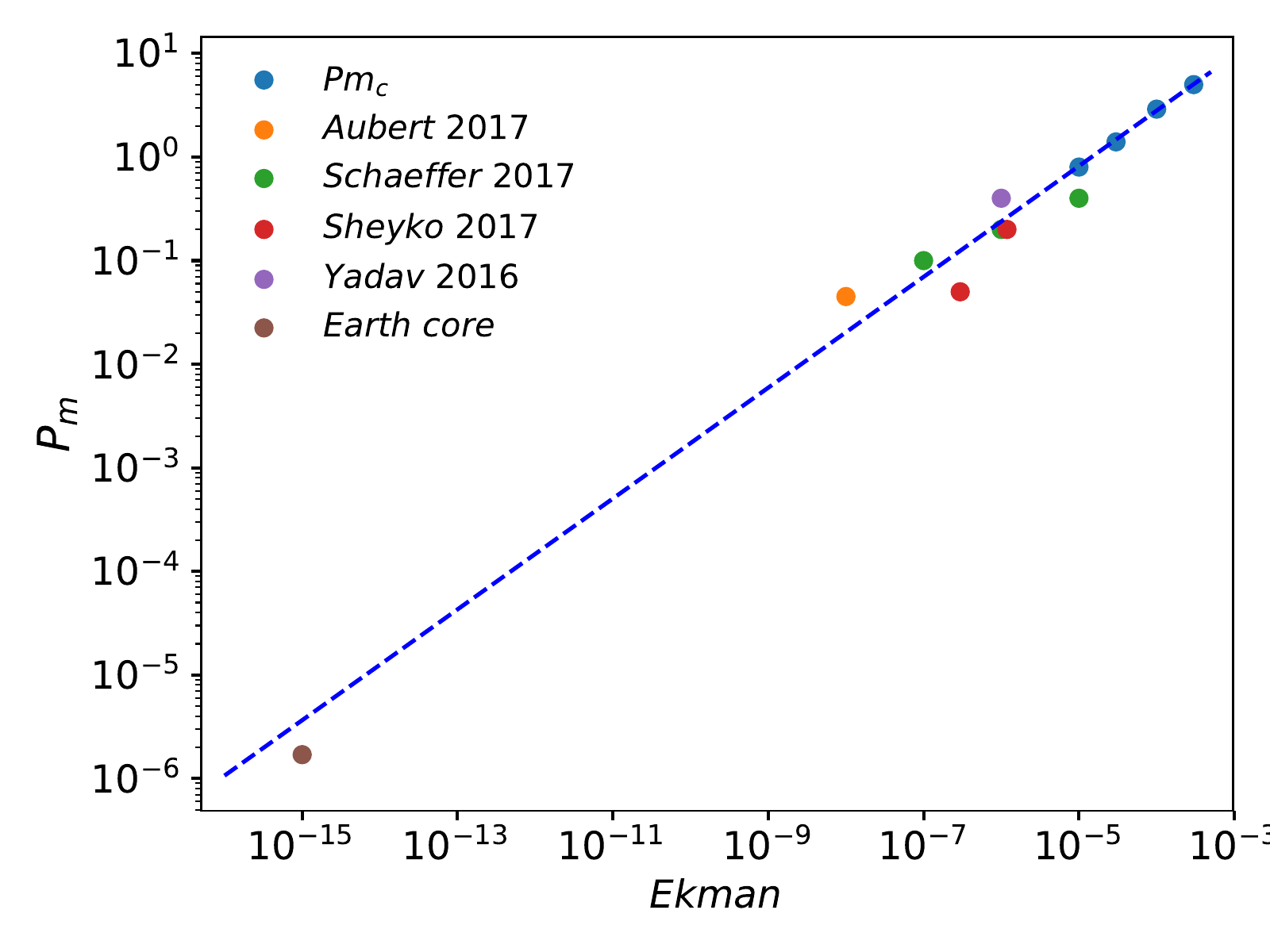}
	\caption{Extrapolation of the critical $\Pm$ number required to have a Lorentz-dominated regime. Other studies and the Earth are also represented. The best fitting law obtained from our four points is $Pm_c = 10^{2.43} \times E^{0.494}$.} \label{Pmc}
\end{figure}

As seen in \S \ref{Part3}, parameters to obtain a transition depend on the importance of the Lorentz force compared to the others. Correlating dipolarity with the classical $\Rol$ number is not relevant for all dynamos as the Lorentz force can prevent the effects of inertia at large scales even if $\Rol$ becomes large. In order to show that the results presented in our study do not depend on the definition of $\Rol$, we show in figure \ref{fdipRols} $\fdip$ as a function of the local Rossby number by using different definitions for this latter parameter. Regardless of this definition, the two regimes - dipolar and multipolar dynamos coexist where a dichotomy was observed by previous studies. We have shown that the local Rossby number is not the only parameter affecting the stability of dipolar dynamos as soon as the magnetic Prandtl number reaches a certain value depending on the Ekman number (see below and figure \ref{Pmc}).\\

As shown in figure \ref{PmRa}, the Lorentz-dominated regime corresponds to higher values of $\Pm$. It is nevertheless important to note that the critical value $Pm_c$ to reach this regime strongly evolves with $\Ek$ and $\Ra$ number. In particular it decreases when reducing the Ekman number. For example $\Ek = 10^{-5}$ and $\Ra/\Rac \sim 80$ allow Lorentz-dominated dynamos at $\Pm$ as low as $1$. When sufficiently high $\Ra/\Rac$ is reached (approximately $10$), $Pm_c$ is almost constant. The $Pm_c$ obtained as a function of $\Ek$ is represented in figure \ref{Pmc} (blue spots). From the four values of $\Ek$ presented in this paper, an estimated slope can be deduced. The extreme values of geodynamo simulations performed by \citet{Aubert2017,schaefferJNF17,Sheyko2017,Yadav2016} (respectively orange, green, red and purple spots) have also been represented, as well as the Earth core (considering $\Ek \sim 10^{-15}$, $\Pm = 2 \times 10^{-6}$, brown spot). \\

According to our data set, the Earth core parameters place it slightly below the critical value ($Pm_c\sim 10^{-5}$ for $E =10^{-15}$). It means that the Lorentz force is a priori not dominant at very large scale, but only from a certain scale as argued by \citet{aurnouK17}. This can be related with force balance obtained for dynamos with $\Pm \lesssim Pm_c$ (see figure \ref{FB471}). However, the excessive extrapolation of the estimated slope to reach these values additioned to the lack of experimental data on these parameters do not exclude the possibility that the Earth might be in the Lorentz-dominated regime, and in particular if we consider local force balance.
A case with parameters above the dashed line on figure \ref{Pmc} has a good chance to be in an almost perfect Lorentz-Coriolis balance at large scales, and thus in the Magnetic-Archimedean-Coriolis (MAC) state. However, as shown on figure \ref{PmRa}, this boundary strongly depends on the value of the Rayleigh number, in particular at low supercriticalities. By extension this regime could be obtained either by decreasing the Ekman number at an almost constant $\Pm$ value (from case S1 to S2 in \citet{schaefferJNF17}) or by increasing the $\Pm$ value at a constant Ekman number as proposed by \citet{Dormy2016}.\\

This analysis emphasises that the Lorentz-dominated regime can not be restricted to high $\Pm$ values, and should be seriously considered when decreasing the Ekman number. Indeed, other studies reach lower $\Ek$ numbers with $\Pm$ above or close to the critical value to obtain a dynamo in which the Lorentz force plays a major role. 
Moreover, we have shown in \S \ref{Part3} that the MAC equilibrium can be studied at higher $\Ek$ and $\Pm$ values, with a behaviour similar to cases at lower values which are numerically demanding. Considering higher Ekman numbers with $\Pm>Pm_c$ appears as an alternative to low-Ekman geodynamo simulations. In fact, this strategy also enables the study of dynamo models with a dominant Lorentz force. \\

\citet{KingB13}, using the database of \cite{christensen06}, highlighted that the extreme planetary values $E$ and $Pm$ could lead to another type of dynamical regime where the Lorentz force can no longer be neglected. The results obtained in \S \ref{Part2} support this possibility, and even bring to light cases showing polarity reversals with a non-negligible role of the magnetic force. Particularly, figure \ref{Pmc} points out that this regime could be relevant even for current leap towards Earth's core values presented in previous works \citep{Yadav16,Aubert2017,schaefferJNF17}.\\

For example the simulation at $E = 10^{-7}$ and $Pm=0.1$ (green dot, S2 simulation of \citet{schaefferJNF17}) is slightly above the $Pm_c$ line (figure \ref{Pmc}). The spatial distribution of the main forces performed in this study \citep{schaefferJNF17} has shown that the force balance depends on the region considered (in or outside of the tangent cylinder). In particular outside the tangent cylinder a magnetostrophic balance emerges, mostly dominated by Lorentz and Coriolis forces with a reduced impact of inertia and buoyancy. This is consistent with our analysis of figure \ref{FB}, i.e. that an important Lorentz force structures the flow and changes the force balance obtained.\\

\begin{figure}
	\subfigure[$Pm=0.5, \Rol=0.1736$.]{
	\includegraphics[width=0.45\linewidth]{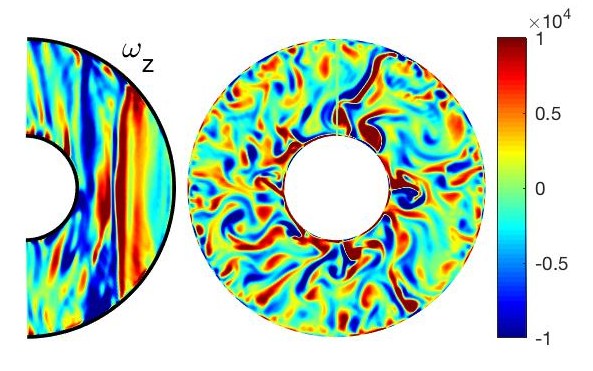}\label{w05}}
	\hfill
	\subfigure[$Pm=3$, $\Rol=0.141$.]{
	\includegraphics[width=0.45\linewidth]{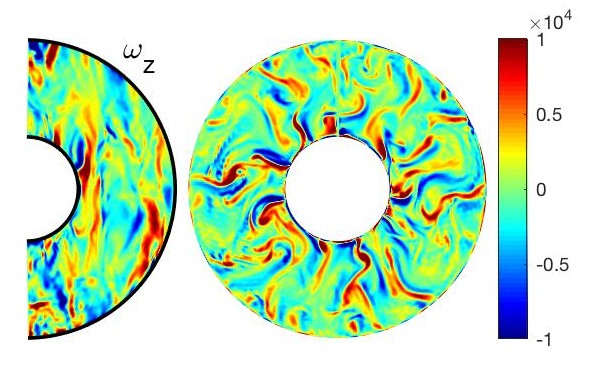}\label{w3}}
	\caption{Snapshots of the vorticity component $\omega_z$ of two simulations at $E= 10^{-4}$ and $Ra/\Rac = 35.9$ which differ only in the $Pm$ value. The simulation at a lower $Pm$ value \subref{w05} is multipolar, whereas the case \subref{w3} is still dipolar. Both of them have $\Rol \geq 0.14$, the other parameters can be found in \ref{Table}. For each, the left panel represents a meridional cut and the right one an equatorial cut.} \label{ws}
\end{figure}

Moreover the vorticity field exhibits different behaviours depending on the branch to which the simulation belongs. For the Lorentz-dominated cases, the field is not organised in Taylor columns as it is for similar parameters at lower $Pm$ values. Figure \ref{ws} compares two identical cases at $Pm=3$ and $Pm=0.5$, therefore exploring the two branches. For the Lorentz-dominated branch (figure \ref{w3}), the meridional cut shows unorganised medium scales structures where the non Lorentz dominated case (figure \ref{w05}) clearly reveals a geostrophic flow. The equatorial cut confirms the difference between the two branches, and in particular corroborates the predictions made from the force equilibrium. Indeed, as the inertia effect is overshadowed by the Lorentz force in higher $Pm$ solutions, the case \ref{w3} shows less large scale structures in favour of medium to small scale structures. These characteristics of Lorentz-dominated dynamos were previously observed in recent studies \citep{Schwaiger19} but only at low Rayleigh number values. \\

\begin{figure}
	\subfigure[$Pm=0.5$.]{
	\includegraphics[width=0.31\linewidth]{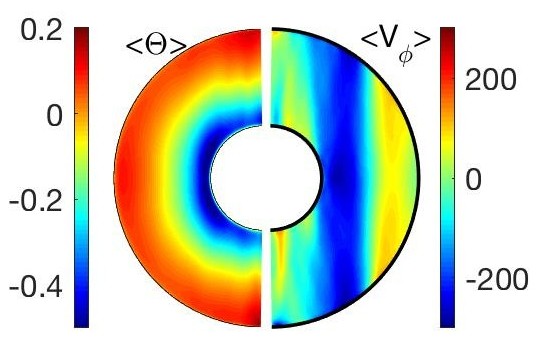}\label{VT05}
	\includegraphics[width=0.15\linewidth]{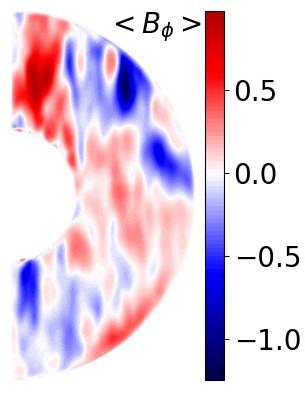}}
	\hfill
	\subfigure[$Pm=3$.]{
	\includegraphics[width=0.31\linewidth]{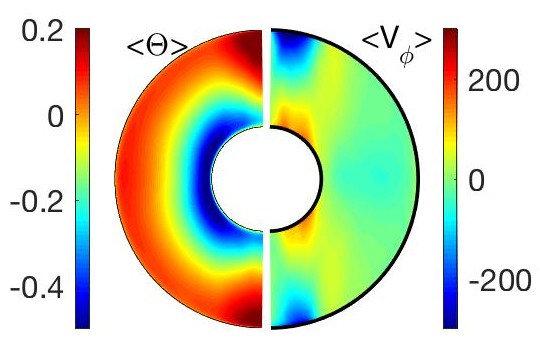}\label{VT3}				\includegraphics[width=0.15\linewidth]{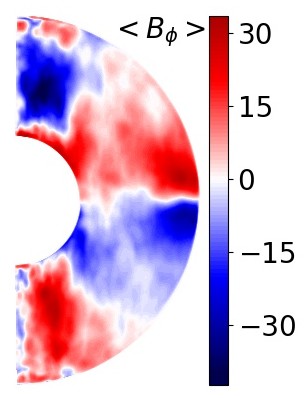}}
	\caption{Temperature $\Theta$ (left), $\phi$ component of the velocity field $V_\phi$ (middle) and $\phi$ component of the magnetic field $B_\phi$ for the same simulations as figure \ref{ws}. The fields have been averaged in time.} \label{VTs}
\end{figure}

Figure \ref{VTs} gives new indications to distinguish the Lorentz-dominated cases from the others at such values of $\Rol$. Indeed, whereas the case $Pm=0.5$ (figure \ref{VT05}) just reconfirms our analysis, the Lorentz-dominated case at $Pm=3$ (figure \ref{VT3}) surprisingly reminds the structures found in previous studies of the MAC-balance (e.g. simulation S1 and S2 of \cite{schaefferJNF17}). Analogous axisymmetric flow is found inside the tangent cylinder and weak dynamics outside. The comparison with the temperature profile illustrates the correlation between the hot spots at the poles and the large scale flow. These similarities with simulations three orders of magnitude lower in Ekman value probably result from a comparable force balance at large scales, and in particular of a non-negligible Lorentz force.\\

The large parameter space explored in our study sheds light on the impact of the Ekman and magnetic Prandtl numbers when considering different magnetic dipolar regimes for highly turbulent flows. We considered a simple convection model with fixed thermal boundary conditions (as \citet{christensen06}) to easily compare our results with previous geodynamo studies. Changing these boundary conditions would probably affect the behaviour observed (as in \citet{sakurabaR09} and \citet{Dharmaraj14}). As strong dipolar magnetic fields have been obtained with more realistic boundary conditions, it would be appropriate to investigate a possible Lorentz-dominated regime. Therefore, the critical values to find behaviours similar to our Lorentz-dominated cases will certainly change with more realistic models. Uniform heat flux at the surface, heterogeneous heat flux \citep{Olson2017} or even compositional convection might be interesting models to explore. It would enable a better understanding of the correlation between a strong dipolar magnetic field and the flow dynamics.

%% file: Table_pap/Tables.tex
\begin{table}[h]
\centering
\begin{filecontents*}{Table_pap/table1_pap3.csv}
E,Ra,Pm,Rm,Elsasser,Ro,Nu,l_uChr,L_nu,L_eta,k_u,fohm,fdip
0.0003,2.4679,12,167,22.65,0.0091,1.495,6.77,--,0.087,8.34,0.66,0.77
0.0003,2.5502,6,104.5,0.27,0.0136,1.29,8.2,0.073,0.088*,9.49,0.063,0.9
0.0003,2.6324,6,108,0.45,0.0143,1.315,8.3,0.0755,0.086,9.6,0.083,0.81
0.0003,2.797,6,121,1.15,0.0159,1.41,8.3,0.076,0.081,9.74,0.17,0.79
0.0003,2.8792,12,205,25.4,0.0114,1.6,7,0.072,0.08,--,0.65,0.75
0.0003,2.9615,6,97,12.6,0.0111,1.56,7.12,0.075,0.105,8.67,0.69,0.82
0.0003,3.2247,3,64.7,0.19,0.0185,1.37,8.95,0.073,0.097,10.69,0.085,0.88
0.0003,3.2906,12,228,30,0.0124,1.675,6.82,--,0.060*,8.27,0.633,0.72
0.0003,3.2906,6,114,13,0.0125,1.64,6.9,0.073,0.098,8.42,0.65,0.78
0.0003,3.6196,3,75.96,1.76,0.0215,1.65,8.9,--,0.103*,11,--,0.87
0.0003,3.7019,12,265,34.65,0.0155,1.85,7.4,0.0695,0.072,9.03,0.623,0.73
0.0003,4.1132,6,145,15.8,0.017,1.87,7.4,0.071,0.0866,9.44,0.64,0.76
0.0003,4.1132,3,80.89,2.82,0.0226,1.73,8.77,0.071,0.0946,10.75,0.46,0.86
0.0003,4.5245,12,330,39.85,0.0215,2.04,8.14,0.065,0.058,9.94,0.59,0.71
0.0003,4.9358,6,178,16.3,0.0241,2.07,8.5,0.067,0.073,10.44,0.6,0.76
0.0003,4.9358,3.5,105,6,0.0252,1.95,8.8,0.07,0.087,10.81,0.54,0.82
0.0003,4.9358,12,390.5,42.38,0.0274,2.23,8.77,--,0.046*,10.7,--,0.7
0.0003,6.1698,6,233,16.9,0.0371,2.37,9.7,0.065,0.062,11.94,0.54,0.74
0.0003,6.1698,2,76,3.1,0.0368,--,10.5,--,0.103*,13.01,--,0.86
0.0003,7.4038,6,278,18.6,0.0459,2.63,10.3,0.0614,0.057,12.64,0.506,0.71
0.0003,8.3909,6,318,18.7,0.054,2.83,10.6,0.062,0.053,12.96,0.47,0.7
0.0003,9.5426,6,354,19.5,0.0638,3.03,11.13,0.058,0.047,13.56,0.45,0.65
0.0003,10.3653,3,200,5.2,0.0777,3.11,12.2,--,0.064*,14.9,--,0.76
0.0003,10.6943,6,388,21.6,0.0688,3.2,11.37,0.057,0.046,13.79,0.44,0.69
0.0003,11.6815,6,424,19.8,0.0797,3.37,11.93,0.0555,0.042,14.42,0.41,0.69
0.0003,12.6686,6,460,20.5,0.0894,3.56,12.21,0.0551,0.041,14.77,0.39,0.7
0.0003,13.9849,6,500,20.4,0.0991,4.75,12.45,0.054,0.038,14.97,0.375,0.67
0.0003,15.1366,6,540,19.5,0.1107,4.95,12.6,0.0535,0.035,15.12,0.36,0.66
0.0003,16.4528,6,583,20.6,0.1183,5.15,12.82,0.0525,0.034,15.34,0.36,0.65
0.0003,18.0981,6,628,20.9,0.1259,5.35,12.76,0.053,0.032,15.26,0.34,0.63
0.0003,19.7433,3,395,3.3,0.1593,5.7,12.83,--,0.045*,15.11,0.165,0.55
0.0003,21.7177,6,684.76,24.9,0.1453,4.77,13.04,0.051,0.031,15.47,0.36,0.6
0.0003,24.6792,8,1020,40,0.1576,6.075,13.03,0.0505,0.026,15.34,0.34,0.58
0.0003,24.6792,6,822.8,21.16,0.1711,6.17,13.11,0.052,0.027,15.49,0.3,0.56
0.0003,29.615,6,763.14,22.35,0.1925,6.53,13.15,0.0507,0.0262,15.44,0.3,0.38
0.0003,34.5509,1.5,314.3,1.02,0.2489,7.17,12.41,--,0.051*,14.48,0.075,0.37
\end{filecontents*}
\csvautobooktabular[table head=\toprule\bfseries $E$ & \bfseries $Ra/\Rac$ & \bfseries $Pm$ & \bfseries $Rm$ & \bfseries $\Lambda$ & \bfseries $Ro_\ell$ & \bfseries $Nu$ & \bfseries $\bar{\ell_u}$ & \bfseries $L_\nu$ & \bfseries $L_\eta$ & \bfseries $k_u$  & \bfseries $f_{ohm}$  & \bfseries $f_{dip}$\\\midrule]{Table_pap/table1_pap3.csv}
\caption{$Ekman = 3 \times 10^{-4}$ .}
\end{table}

\begin{table}
\centering
\begin{filecontents*}{Table_pap/table2_pap3.csv}
E,Ra,Pm,Rm,Elsasser,Ro,Nu,l_uChr,L_nu,L_eta,k_u,fohm,fdip
0.0001,3.3453,3,85.6,0.19,0.0101,1.336,10.9,0.0615,0.083,12.42,0.125,0.91
0.0001,3.9483,12,341,40.7,0.0091,1.95,9.9,--,0.034*,12.045,0.707,0.76
0.0001,3.9483,2,73,0.41,0.0132,1.56,11.5,0.0556,0.0846,13.39,0.27,0.9
0.0001,4.3072,2,77.5,0.68,0.0149,1.65,12,0.056,0.083,14.37,0.344,0.85
0.0001,4.3072,3,117,1.43,0.0139,1.7,11.2,0.0582,0.0719,13.42,0.4,0.84
0.0001,4.5944,12,422,46.1,0.0117,2.21,10.5,0.0487,0.0545,12.83,0.682,0.73
0.0001,4.5944,2,82.8,0.84,0.0158,1.73,12.1,0.0561,0.0776,14.58,0.38,0.85
0.0001,4.738,12,416.5,47.7,0.0124,2.19,11.1,0.046,0.0545,13.47,0.675,0.76
0.0001,4.738,10,366,36.1,0.0122,2.22,10.4,0.0476,0.056,12.58,0.68,0.74
0.0001,4.738,8,293,24.5,0.0126,2.16,10.7,0.049,0.062,13.11,0.685,0.76
0.0001,4.738,5,178,11,0.0127,2.04,11.12,0.054,0.07,13.26,0.7,0.81
0.0001,4.738,3,123,2.4,0.0144,1.81,11,0.057,0.075,13.45,0.47,0.86
0.0001,4.738,2,85,1.04,0.0159,1.77,11.6,0.0573,0.0827,14.06,0.414,0.85
0.0001,4.738,1.5,64.7,0.66,0.0174,1.78,12.7,0.0554,0.091,15.52,0.38,0.88
0.0001,4.738,1.3,56.3,0.53,0.0177,1.77,12.9,0.053,0.094,15.7,0.36,0.89
0.0001,4.8816,1,45.7,0.36,0.0202,1.81,13.8,0.0515,0.107,16.84,0.316,0.92
0.0001,5.0251,3,125.3,3.31,0.0154,1.9,11.5,0.0554,0.072,13.7,0.55,0.82
0.0001,5.0251,1,46,0.46,0.0205,1.85,14,0.052,0.103,17.25,0.36,0.92
0.0001,5.1687,1,46.9,0.55,0.0208,1.9,13.9,0.053,0.105,17.21,0.4,0.92
0.0001,5.3841,1,48.4,0.68,0.0208,1.97,13.6,0.053,0.105,16.79,0.436,0.91
0.0001,8.6145,2,132,5.5,0.0294,2.75,14,0.048,0.071,17.24,0.51,0.89
0.0001,8.6145,1,73.5,1.7,0.0358,2.67,15.41,0.05,0.0965,19.34,--,0.83
0.0001,10.7681,8,670,43.3,0.0382,3.63,14.3,0.0424,0.0385,--,0.56,0.57
0.0001,10.7681,5,419.6,23.3,0.0391,3.53,14.62,0.0432,0.0455,17.84,0.58,0.68
0.0001,10.7681,1,85,2.6,0.0452,3.23,16.7,--,0.067*,20.7,0.52,0.84
0.0001,14.3575,5,524,30.2,0.0474,4.18,14.9,0.0423,0.0386,--,0.566,0.67
0.0001,14.3575,3,335,12.92,0.0562,4.14,16.06,0.0415,0.0492,19.49,0.515,0.77
0.0001,14.3575,2,221,8,0.0564,4.07,16.1,0.0425,0.0545,19.57,0.54,0.82
0.0001,14.3575,1,112.6,3.33,0.0574,3.96,16.11,0.044,0.0704,19.68,0.52,0.89
0.0001,17.229,3,388,15.1,0.0687,4.68,16.6,0.0408,0.0419,20.05,0.52,0.78
0.0001,17.229,2,262,8.95,0.0732,4.64,17.7,0.04,0.048,20.3,0.52,0.79
0.0001,17.229,1,134,3.6,0.0699,4.46,16.9,0.0418,0.0626,20.57,0.48,0.81
0.0001,20.1005,2,304,9.15,0.0831,5.15,17.4,0.04,0.044,21.09,0.48,0.8
0.0001,21.5362,3,453.5,17.4,0.0836,5.46,17.5,0.0384,0.039,20.96,0.49,0.76
0.0001,21.5362,2,326,8.9,0.0891,5.41,17.5,0.039,0.04,21.01,0.465,0.79
0.0001,25.1257,3,551,15.44,0.1034,6.04,18.05,0.0375,0.033,21.52,0.46,0.76
0.0001,25.1257,2,392,7.82,0.1159,6.16,18.2,0.038,0.036,21.87,0.4,0.78
0.0001,25.1257,1,255,0.67,0.1371,6.51,16.57,0.04,0.032,19.51,0.126,0.45
0.0001,25.8435,1,257,0.75,0.144,6.65,17.4,--,0.039*,20.88,0.13,0.37
0.0001,28.715,3,601,18.95,0.1159,6.64,18.2,0.037,0.031,21.62,0.46,0.76
0.0001,28.715,2,416.23,10.6,0.1217,6.57,18.2,--,0.030*,21.63,--,0.71
0.0001,32.3044,3,651,21,0.1289,6.98,18.4,0.037,0.03,21.7,0.45,0.75
0.0001,32.3044,2,529,4.5,0.1499,7.44,18.11,--,0.027*,21.42,--,0.46
0.0001,35.8937,5,1116,47.18,0.1369,7.38,18.7,0.0368,0.0247,21.85,0.44,0.73
0.0001,35.8937,3,725,19.9,0.141,7.5,18.3,0.0375,0.0275,21.74,0.43,0.73
0.0001,35.8937,0.5,176.5,0.25,0.1736,9.04,15.58,0.0412,0.0405,18.05,--,0.45
0.0001,40.2011,5,1222.84,48.3,0.1504,7.77,18.9,0.036,0.0242,22.03,--,0.71
0.0001,43.0725,3,933,12.57,0.1798,8.5,18.22,0.0215,0.028,21.3,0.35,0.25
0.0001,50.2512,5,1513.69,45.86,0.1795,8.755,18.8,--,0.016*,21.85,--,0.7
\end{filecontents*}
\csvautobooktabular[table head=\toprule\bfseries $E$ & \bfseries $Ra/\Rac$ & \bfseries $Pm$ & \bfseries $Rm$ & \bfseries $\Lambda$ & \bfseries $Ro_\ell$ & \bfseries $Nu$ & \bfseries $\bar{\ell_u}$ & \bfseries $L_\nu$ & \bfseries $L_\eta$ & \bfseries $k_u$  & \bfseries $f_{ohm}$  & \bfseries $f_{dip}$\\\midrule]{Table_pap/table2_pap3.csv}
\caption{$Ekman = 1 \times 10^{-4}$ .}
\end{table}

\begin{table}
\centering
\begin{filecontents*}{Table_pap/table3_pap3.csv}
E,Ra,Pm,Rm,Elsasser,Ro,Nu,l_uChr,L_nu,L_eta,k_u,fohm,fdip
3e-05,3.5298,10,335.2,22.5,0.0036,1.81,11.4,0.0398,0.0566,13.7,0.782,0.79
3e-05,3.6475,2.5,102,0.16,0.0059,1.35,15.5,--,0.059*,17.86,0.22,0.92
3e-05,4.2358,2.5,117,0.3,0.0072,1.44,16.11,--,0.055*,18.8,0.294,0.84
3e-05,4.7064,2.5,132.5,0.37,0.008,1.51,15.7,0.0113,0.0562,18.31,0.3,0.69
3e-05,5.4124,2.5,152.4,2.37,0.0088,1.97,15.35,0.0134,0.057,18.9,0.64,0.79
3e-05,5.4124,1,68,0.27,0.0115,1.8,18,0.0143,0.0727,--,0.373,--
3e-05,5.7654,1,72.4,0.38,0.0132,1.92,18.8,0.0153,0.073,--,0.42,--
3e-05,6.0007,10,609,51.8,0.0091,2.9,15.85,0.0314,0.043,19.23,0.754,0.76
3e-05,6.0007,2.5,158,3.06,0.0092,2.07,15.28,0.0128,0.0592,18.88,0.685,0.86
3e-05,6.0007,1,75,0.44,0.013,1.97,18.5,--,0.069*,22.76,0.44,0.82
3e-05,6.0007,0.7,53.5,0.31,0.0144,2.03,19.67,0.0156,0.0856,24.29,0.42,0.89
3e-05,6.1184,1,75.5,0.53,0.0133,2.03,18.1,--,0.069*,22.39,0.477,0.82
3e-05,6.4714,1,80,0.63,0.0143,2.14,18.76,0.016,0.0715,23.28,0.493,0.82
3e-05,7.0596,2.5,162.7,7.9,0.0098,2.57,15.45,0.0126,0.0644,19.25,0.81,0.77
3e-05,7.0596,2,144,3.35,0.0114,2.34,16.27,0.0143,0.0626,20.33,0.704,0.82
3e-05,7.0596,1,83.9,1.06,0.0147,2.36,18.42,0.0162,0.0737,23.09,0.5825,0.81
3e-05,8.2363,6,480,34,0.0131,3.3,17.2,0.0337,0.0452,20.91,0.745,0.79
3e-05,8.2363,0.5,50,0.5,0.023,2.61,24.1,0.0185,0.0955,30.91,0.47,0.9
3e-05,8.8246,6,504.6,44,0.0131,3.46,16.4,0.0335,0.0449,19.7,0.75,0.74
3e-05,8.8246,1,97.5,1.58,0.0181,2.7,19.58,0.0183,0.0705,24.8,0.609,0.85
3e-05,8.8246,0.5,52.8,0.6,0.0248,2.756,24.3,0.0192,0.0917,31.24,0.459,0.9
3e-05,10.5895,2.5,250,14.3,0.0165,3.6,17.3,--,0.038*,--,0.76,0.83
3e-05,10.5895,1,110,2.7,0.0229,3.1,21.76,--,0.057*,27.52,--,0.88
3e-05,10.5895,0.5,61.6,0.72,0.0293,3.07,24.9,0.0214,0.0856,31.97,0.48,0.89
3e-05,12.9427,6,762,42.3,0.0247,4.45,20.4,0.0119,0.0331,24.78,0.765,0.77
3e-05,12.9427,4,493,27.3,0.0231,4.41,19.6,0.0126,0.0386,23.94,0.775,0.79
3e-05,12.9427,2.5,306.5,16.33,0.0225,4.22,19.1,0.0135,0.0477,23.36,0.78,0.84
3e-05,12.9427,1,132,3.85,0.0277,3.92,21.77,0.0142,0.0644,27,0.68,0.9
3e-05,12.9427,0.5,73.5,1.4,0.0353,4.03,25.2,0.0225,0.0819,31.54,0.56,0.91
3e-05,12.9427,0.25,43.9,0.25,0.046,3.85,27.24,0.017,0.1,34.41,0.286,0.95
3e-05,14.1193,0.25,47.8,0.34,0.0497,4.34,27.4,0.0172,0.097,34.38,0.32,0.95
3e-05,15.296,0.25,51,0.4,0.0548,4.8,27.77,0.0172,0.0933,34.56,0.336,0.95
3e-05,17.6491,0.5,95,2,0.0474,5.2,26.1,0.0243,0.0673,31.79,0.57,0.92
3e-05,17.6491,0.25,59,0.54,0.0631,5.61,27.9,0.0206,0.086,34.37,0.366,0.94
3e-05,23.5322,4,824,42.7,0.0458,6.44,23.2,0.0173,0.0288,28.08,0.675,0.8
3e-05,23.5322,2.5,548,21.75,0.0475,6.6,22.6,0.0187,0.034,27.28,0.67,0.81
3e-05,23.5322,0.5,97,2,0.0487,6.5,26.4,0.0254,0.0616,32.06,0.58,0.91
3e-05,23.5322,0.25,75,0.82,0.0791,7.15,27.6,0.0211,0.0763,33.55,0.398,0.93
3e-05,29.4152,0.25,92,0.94,0.0935,8.42,26.7,0.0226,0.0666,32.22,0.394,0.91
3e-05,31.7685,0.25,99,0.95,0.1007,8.82,26.37,0.023,0.065,32.6,0.388,0.91
3e-05,35.2983,2.5,793,25.3,0.0762,8.65,25.2,0.0202,0.027,29.99,0.61,0.81
3e-05,35.2983,0.25,133,0.28,0.1196,10,23.48,0.026,0.044,27.3,0.18,0.33
3e-05,42.358,1,376.4,9.82,0.0966,9.5,25.28,0.0215,0.0353,30.21,0.586,0.81
3e-05,42.358,2.5,950,30.4,0.0903,9.87,25.8,--,0.019*,30.7,0.58,0.8
3e-05,47.0644,2.5,970.57,33.6,0.0997,10.09,26.09,0.0224,0.0252,30.72,0.576,0.8
3e-05,47.0644,1,423.51,9.71,0.1026,10.3,25.37,0.0224,0.0325,30.26,0.575,0.86
3e-05,52.9474,2.5,1054.3,38.7,0.1099,11.06,26.57,0.0225,0.0247,31.18,0.56,0.8
3e-05,55.3007,2.5,1127.47,35.88,0.118,11.55,26.49,0.0229,0.0237,31.1,0.558,0.8
3e-05,55.3007,1,496.6,10.19,0.1221,11.79,25.57,0.0232,0.0298,30.41,0.55,0.85
3e-05,64.7135,2.5,1296.45,37.87,0.1378,12.7,27.06,0.0229,0.022,31.63,0.53,0.79
3e-05,64.7135,1,686.72,5.26,0.1634,14.05,24.44,0.025,0.0215,28.58,0.38,0.49
3e-05,70.5965,2.5,1416.5,38.05,0.1472,13.48,27.2,--,0.016*,31.7,0.53,0.79
3e-05,82.3627,2.5,1576.41,42.11,0.1638,14.59,27.09,--,0.015*,31.54,0.51,0.78
\end{filecontents*}
\csvautobooktabular[table head=\toprule\bfseries $E$ & \bfseries $Ra/\Rac$ & \bfseries $Pm$ & \bfseries $Rm$ & \bfseries $\Lambda$ & \bfseries $Ro_\ell$ & \bfseries $Nu$ & \bfseries $\bar{\ell_u}$ & \bfseries $L_\nu$ & \bfseries $L_\eta$ & \bfseries $k_u$  & \bfseries $f_{ohm}$  & \bfseries $f_{dip}$\\\midrule]{Table_pap/table3_pap3.csv}
\caption{$Ekman = 3 \times 10^{-5}$ .}
\end{table}

\begin{table}
\centering
\begin{filecontents*}{Table_pap/table4_pap3.csv}
E,Ra,Pm,Rm,Elsasser,Ro,Nu,l_uChr,L_nu,L_eta,k_u,fohm,fdip
1e-05,3.4059,2,94,0.051,0.0036,1.26,22.9,0.0318,0.0496,26.5,0.189,0.96
1e-05,3.7843,2,99,0.073,0.0036,1.28,22,0.0333,0.0515,25.26,0.213,0.94
1e-05,5.6764,2,175,0.53,0.0064,1.71,22.8,0.0318,0.0443,--,0.47,0.83
1e-05,5.6764,1.5,131,0.23,0.0064,1.63,23.25,0.0323,0.0493,--,0.365,0.85
1e-05,5.6764,1,88,0.16,0.0068,1.67,24.3,0.0305,0.0575,--,0.38,0.87
1e-05,6.1495,1,96,0.21,0.0073,1.64,24,0.0314,0.0566,--,0.37,0.86
1e-05,6.8117,1,109,0.61,0.009,2.19,25.8,0.0284,0.0564,--,0.575,0.82
1e-05,7.5686,1,120,0.99,0.0102,2.6,26.8,0.0283,0.0552,--,0.61,0.84
1e-05,7.5686,0.5,63.75,0.23,0.0128,2.32,31,0.0267,0.07,--,0.453,0.88
1e-05,8.5147,0.5,71.8,0.35,0.014,2.64,31.5,0.026,0.068,--,0.5,0.85
1e-05,9.4607,1,142,2.04,0.0116,3.26,25.95,0.0267,0.0578,33.1,0.7,0.89
1e-05,9.4607,0.5,77,0.47,0.0156,2.9,32.7,0.0247,0.067,42.1,0.53,0.87
1e-05,10.4068,0.25,46.6,0.16,0.0234,3.12,38.7,0.0243,0.086,49.6,0.35,0.93
1e-05,11.3529,0.25,49.5,0.19,0.0249,3.3,39.1,0.0222,0.0841,50.2,0.36,0.92
1e-05,12.299,0.25,53.5,0.21,0.0265,3.52,39.58,0.0219,0.0807,51,0.37,0.9
1e-05,13.245,0.25,56.7,0.24,0.0294,3.75,40.1,0.0219,0.0774,51.7,0.374,0.91
1e-05,14.1911,0.5,119,1.05,0.0275,4.3,36,--,0.050*,--,0.57,--
1e-05,14.1911,0.25,60.3,0.31,0.0311,4.11,40.69,0.021,0.0753,52.2,0.4,0.9
1e-05,17.0293,0.5,136,1.84,0.0304,5.68,35.4,--,0.047*,44,0.62,0.87
1e-05,17.0293,0.25,77,0.51,0.0387,5.71,40.56,--,0.063*,50.8,0.45,0.89
1e-05,17.0293,0.2,62.6,0.33,0.0401,5.56,40.6,0.0208,0.0765,--,0.39,0.68
1e-05,18.9215,1,227.5,7.7,0.0206,5.82,28.1,0.0237,0.0465,34.5,0.76,0.89
1e-05,18.9215,0.5,153,2.14,0.0335,--,34,--,0.044*,--,0,--
1e-05,18.9215,0.25,84,0.72,0.0424,6.66,39.2,--,0.060*,48.7,0.49,0.89
1e-05,18.9215,0.2,71,0.41,0.0461,6.5,40.2,0.02,0.0723,49.9,0.4075,0.91
1e-05,18.9215,0.15,56,0.235,0.0459,6.31,39,0.0205,0.0803,48.8,0.33,0.95
1e-05,21.2866,0.25,92.8,0.95,0.0452,7.6,38.4,--,0.057*,47.3,0.52,0.9
1e-05,23.6518,1,286,9.05,0.0276,7.02,29.9,0.023,0.0424,36.5,0.77,0.9
1e-05,23.6518,0.5,184,2.3,0.0418,7.4,36.5,0.021,0.0446,44.7,0.61,0.89
1e-05,23.6518,0.25,100.5,1.15,0.0495,8.31,38.9,0.0208,0.0646,47.7,0.54,0.91
1e-05,28.3822,0.5,193,4,0.0442,8.73,35.6,--,0.040*,43.5,0.69,0.91
1e-05,28.3822,0.25,114,1.64,0.0567,9.56,38.7,0.0204,0.062,47.5,0.58,0.91
1e-05,28.3822,0.2,108,0.85,0.0629,9.82,39.5,0.02,0.061,48,0.48,0.89
1e-05,28.3822,0.15,78.5,0.55,0.0654,9.85,39.5,0.0196,0.0673,48,0.424,0.9
1e-05,47.3037,1,528.21,14.59,0.054,12.06,32.02,0.0196,0.0297,38.3,0.75,0.87
1e-05,47.3037,0.5,295.83,5.31,0.0653,12.24,34.75,0.0202,0.0359,42.3,0.67,0.91
1e-05,47.3037,0.2,142.3,1.73,0.087,13.9,38.5,0.0195,0.0505,46.6,0.56,0.93
1e-05,47.3037,0.15,117,1.05,0.0924,14.4,37.2,0.0203,0.0561,44.7,0.49,0.92
1e-05,56.7644,0.5,370,5.6,0.0806,13.9,34.2,0.0206,0.0325,41.4,0.63,0.91
1e-05,56.7644,0.2,162,1.9,0.0975,15.1,37.8,0.0205,0.046,45.5,0.55,0.92
1e-05,56.7644,0.15,144.1,1.11,0.1017,15.81,36.5,0.0193,0.0511,43.7,0.49,0.92
1e-05,63.3869,0.15,153.6,1,0.1117,16.9,35.1,--,0.044*,41.7,0.46,0.88
1e-05,66.2252,0.5,430,6.1,0.0942,16,34.4,0.02,0.03,40.8,0.62,0.89
1e-05,66.2252,0.2,188.2,1.95,0.1034,16.7,36.09,0.0204,0.0413,43.2,0.55,0.91
1e-05,75.6859,0.2,270,0.74,0.1361,20.06,32.88,0.0201,0.0313,38.5,0.31,0.43
1e-05,75.6859,1,805.75,22.14,0.0876,16.47,33.97,0.0201,0.024,40.2,0.694,0.86
\end{filecontents*}
\csvautobooktabular[table head=\toprule\bfseries $E$ & \bfseries $Ra/\Rac$ & \bfseries $Pm$ & \bfseries $Rm$ & \bfseries $\Lambda$ & \bfseries $Ro_\ell$ & \bfseries $Nu$ & \bfseries $\bar{\ell_u}$ & \bfseries $L_\nu$ & \bfseries $L_\eta$ & \bfseries $k_u$  & \bfseries $f_{ohm}$  & \bfseries $f_{dip}$\\\midrule]{Table_pap/table4_pap3.csv}
\caption{$Ekman = 1 \times 10^{-5}$ .}
\end{table}

%% file: Manuscript_PEPI.bbl
\begin{thebibliography}{36}
\expandafter\ifx\csname natexlab\endcsname\relax\def\natexlab#1{#1}\fi
\providecommand{\url}[1]{\texttt{#1}}
\providecommand{\href}[2]{#2}
\providecommand{\path}[1]{#1}
\providecommand{\DOIprefix}{doi:}
\providecommand{\ArXivprefix}{arXiv:}
\providecommand{\URLprefix}{URL: }
\providecommand{\Pubmedprefix}{pmid:}
\providecommand{\doi}[1]{\href{http://dx.doi.org/#1}{\path{#1}}}
\providecommand{\Pubmed}[1]{\href{pmid:#1}{\path{#1}}}
\providecommand{\bibinfo}[2]{#2}
\ifx\xfnm\relax \def\xfnm[#1]{\unskip,\space#1}\fi
\bibitem[{{Aubert} et~al.(2017){Aubert}, {Gastine} and {Fournier}}]{Aubert2017}
\bibinfo{author}{{Aubert}, J.}, \bibinfo{author}{{Gastine}, T.},
  \bibinfo{author}{{Fournier}, A.}, \bibinfo{year}{2017}.
\newblock \bibinfo{title}{{Spherical convective dynamos in the rapidly rotating
  asymptotic regime}}.
\newblock \bibinfo{journal}{Journal of Fluid Mechanics} \bibinfo{volume}{813},
  \bibinfo{pages}{558--593}.
\newblock \DOIprefix\doi{10.1017/jfm.2016.789}.
\bibitem[{{Aurnou} and {King}(2017)}]{aurnouK17}
\bibinfo{author}{{Aurnou}, J.M.}, \bibinfo{author}{{King}, E.M.},
  \bibinfo{year}{2017}.
\newblock \bibinfo{title}{{The cross-over to magnetostrophic convection in
  planetary dynamo systems}}.
\newblock \bibinfo{journal}{Proceedings of the Royal Society of London Series
  A} \bibinfo{volume}{473}, \bibinfo{pages}{20160731}.
\newblock \DOIprefix\doi{10.1098/rspa.2016.0731}.
\bibitem[{{Brown} et~al.(2010){Brown}, {Browning}, {Brun}, {Miesch} and
  {Toomre}}]{brown10}
\bibinfo{author}{{Brown}, B.P.}, \bibinfo{author}{{Browning}, M.K.},
  \bibinfo{author}{{Brun}, A.S.}, \bibinfo{author}{{Miesch}, M.S.},
  \bibinfo{author}{{Toomre}, J.}, \bibinfo{year}{2010}.
\newblock \bibinfo{title}{{Persistent Magnetic Wreaths in a Rapidly Rotating
  Sun}}.
\newblock \bibinfo{journal}{Astrophysical Journal} \bibinfo{volume}{711},
  \bibinfo{pages}{424--438}.
\newblock \DOIprefix\doi{10.1088/0004-637X/711/1/424}.
\bibitem[{{Busse} et~al.(1998){Busse}, {Hartung}, {Jaletzky} and
  {Sommermann}}]{busse98}
\bibinfo{author}{{Busse}, F.}, \bibinfo{author}{{Hartung}, G.},
  \bibinfo{author}{{Jaletzky}, M.}, \bibinfo{author}{{Sommermann}, G.},
  \bibinfo{year}{1998}.
\newblock \bibinfo{title}{{Experiments on thermal convection in rotating
  systems motivated by planetary problems.}}
\newblock \bibinfo{journal}{Dyn. At. O.} \bibinfo{volume}{27},
  \bibinfo{pages}{161}.
\bibitem[{{Christensen} et~al.(1998){Christensen}, {Olson} and
  {Glatzmaier}}]{christensen98}
\bibinfo{author}{{Christensen}, U.}, \bibinfo{author}{{Olson}, P.},
  \bibinfo{author}{{Glatzmaier}, G.A.}, \bibinfo{year}{1998}.
\newblock \bibinfo{title}{{A dynamo model interpretation of geomagnetic field
  structures}}.
\newblock \bibinfo{journal}{Geophysical Research Letters} \bibinfo{volume}{25},
  \bibinfo{pages}{1565--1568}.
\newblock \DOIprefix\doi{10.1029/98GL00911}.
\bibitem[{{Christensen} et~al.(1999){Christensen}, {Olson} and
  {Glatzmaier}}]{christensen99}
\bibinfo{author}{{Christensen}, U.}, \bibinfo{author}{{Olson}, P.},
  \bibinfo{author}{{Glatzmaier}, G.A.}, \bibinfo{year}{1999}.
\newblock \bibinfo{title}{{Numerical modelling of the geodynamo: systematic
  parameter study}}.
\newblock \bibinfo{journal}{Geophys. J. Int.} \bibinfo{volume}{166},
  \bibinfo{pages}{97--114}.
\bibitem[{{Christensen} and {Wicht}(2007)}]{christensen07}
\bibinfo{author}{{Christensen}, U.}, \bibinfo{author}{{Wicht}, J.},
  \bibinfo{year}{2007}.
\newblock \bibinfo{title}{{Numerical dynamo simulations}}, in:
  \bibinfo{editor}{{P. Olson}} (Ed.), \bibinfo{booktitle}{Core Dynamics}, pp.
  \bibinfo{pages}{245--282}.
\bibitem[{{Christensen} and {Aubert}(2006)}]{christensen06}
\bibinfo{author}{{Christensen}, U.R.}, \bibinfo{author}{{Aubert}, J.},
  \bibinfo{year}{2006}.
\newblock \bibinfo{title}{{Scaling properties of convection-driven dynamos in
  rotating spherical shells and application to planetary magnetic fields}}.
\newblock \bibinfo{journal}{Geophy. J. Int.} \bibinfo{volume}{166},
  \bibinfo{pages}{97--114}.
\newblock \DOIprefix\doi{10.1111/j.1365-246X.2006.03009.x}.
\bibitem[{Dharmaraj et~al.(2014)Dharmaraj, Stanley and Qu}]{Dharmaraj14}
\bibinfo{author}{Dharmaraj, G.}, \bibinfo{author}{Stanley, S.},
  \bibinfo{author}{Qu, A.C.}, \bibinfo{year}{2014}.
\newblock \bibinfo{title}{Scaling laws, force balances and dynamo generation
  mechanisms in numerical dynamo models: influence of boundary conditions}.
\newblock \bibinfo{journal}{Geophysical Journal International}
  \bibinfo{volume}{199}, \bibinfo{pages}{514--532}.
\newblock \DOIprefix\doi{10.1093/gji/ggu274}.
\bibitem[{{Dormy}(2016)}]{Dormy2016}
\bibinfo{author}{{Dormy}, E.}, \bibinfo{year}{2016}.
\newblock \bibinfo{title}{{Strong-field spherical dynamos}}.
\newblock \bibinfo{journal}{Journal of Fluid Mechanics} \bibinfo{volume}{789},
  \bibinfo{pages}{500--513}.
\newblock \DOIprefix\doi{10.1017/jfm.2015.747}.
\bibitem[{{Dormy} et~al.(1998){Dormy}, {Cardin} and {Jault}}]{PaRoDy}
\bibinfo{author}{{Dormy}, E.}, \bibinfo{author}{{Cardin}, P.},
  \bibinfo{author}{{Jault}, D.}, \bibinfo{year}{1998}.
\newblock \bibinfo{title}{{MHD flow in a slightly differentially rotating
  spherical shell, with conducting inner core, in a dipolar magnetic field}}.
\newblock \bibinfo{journal}{Earth and Planetary Science Letters}
  \bibinfo{volume}{160}, \bibinfo{pages}{15--30}.
\newblock \DOIprefix\doi{10.1016/S0012-821X(98)00078-8}.
\bibitem[{{Dormy} et~al.(2018){Dormy}, {Oruba} and {Petitdemange}}]{Dormy2018}
\bibinfo{author}{{Dormy}, E.}, \bibinfo{author}{{Oruba}, L.},
  \bibinfo{author}{{Petitdemange}, L.}, \bibinfo{year}{2018}.
\newblock \bibinfo{title}{{Three branches of dynamo action}}.
\newblock \bibinfo{journal}{Fluid Dynamics Research} \bibinfo{volume}{50},
  \bibinfo{pages}{011415}.
\newblock \DOIprefix\doi{10.1088/1873-7005/aa769c}.
\bibitem[{{{Dormy}, E. and {Soward}, A.~M.}(2007)}]{book_dormy}
\bibinfo{editor}{{{Dormy}, E. and {Soward}, A.~M.}} (Ed.),
  \bibinfo{year}{2007}.
\newblock \bibinfo{title}{{Mathematical aspects of natural dynamos}}.
\newblock \bibinfo{publisher}{CRC Press}.
\bibitem[{{Glatzmaier} and {Roberts}(1995)}]{gr95}
\bibinfo{author}{{Glatzmaier}, G.A.}, \bibinfo{author}{{Roberts}, P.H.},
  \bibinfo{year}{1995}.
\newblock \bibinfo{title}{{A three-dimensional self-consistent computer
  simulation of a geomagnetic field reversal}}.
\newblock \bibinfo{journal}{Nature} \bibinfo{volume}{377},
  \bibinfo{pages}{203--209}.
\newblock \DOIprefix\doi{10.1038/377203a}.
\bibitem[{{K{\"a}pyl{\"a}} et~al.(2009){K{\"a}pyl{\"a}}, {Korpi} and
  {Brandenburg}}]{kapyla09}
\bibinfo{author}{{K{\"a}pyl{\"a}}, P.J.}, \bibinfo{author}{{Korpi}, M.J.},
  \bibinfo{author}{{Brandenburg}, A.}, \bibinfo{year}{2009}.
\newblock \bibinfo{title}{{Large-scale dynamos in rigidly rotating turbulent
  convection}}.
\newblock \bibinfo{journal}{Astrophysical Journal} \bibinfo{volume}{697},
  \bibinfo{pages}{1153}.
\bibitem[{{King} and {Buffett}(2013)}]{KingB13}
\bibinfo{author}{{King}, E.M.}, \bibinfo{author}{{Buffett}, B.A.},
  \bibinfo{year}{2013}.
\newblock \bibinfo{title}{{Flow speeds and length scales in geodynamo models:
  The role of viscosity}}.
\newblock \bibinfo{journal}{Earth and Planetary Science Letters}
  \bibinfo{volume}{371}, \bibinfo{pages}{156--162}.
\newblock \DOIprefix\doi{10.1016/j.epsl.2013.04.001}.
\bibitem[{{King} et~al.(2010){King}, {Soderlund}, {Christensen}, {Wicht} and
  {Aurnou}}]{king10}
\bibinfo{author}{{King}, E.M.}, \bibinfo{author}{{Soderlund}, K.M.},
  \bibinfo{author}{{Christensen}, U.R.}, \bibinfo{author}{{Wicht}, J.},
  \bibinfo{author}{{Aurnou}, J.M.}, \bibinfo{year}{2010}.
\newblock \bibinfo{title}{{Convective heat transfer in planetary dynamo
  models}}.
\newblock \bibinfo{journal}{Geochemistry, Geophysics, Geosystems}
  \bibinfo{volume}{11}, \bibinfo{pages}{Q06016}.
\newblock \DOIprefix\doi{10.1029/2010GC003053}.
\bibitem[{{Kutzner} and {Christensen}(2002)}]{kutzner02}
\bibinfo{author}{{Kutzner}, C.}, \bibinfo{author}{{Christensen}, U.R.},
  \bibinfo{year}{2002}.
\newblock \bibinfo{title}{{From stable dipolar towards reversing numerical
  dynamos}}.
\newblock \bibinfo{journal}{Physics of the Earth and Planetary Interiors}
  \bibinfo{volume}{131}, \bibinfo{pages}{29--45}.
\newblock \DOIprefix\doi{10.1016/S0031-9201(02)00016-X}.
\bibitem[{{Moffatt}(1978)}]{moffatt}
\bibinfo{author}{{Moffatt}, H.K.}, \bibinfo{year}{1978}.
\newblock \bibinfo{title}{{Magnetic field generation in electrically conducting
  fluids}}.
\newblock \bibinfo{publisher}{Cambridge: Cambridge University Press}.
\bibitem[{Nataf and Schaeffer(2015)}]{NatafS2015}
\bibinfo{author}{Nataf, H.C.}, \bibinfo{author}{Schaeffer, N.},
  \bibinfo{year}{2015}.
\newblock \bibinfo{title}{8.06 - turbulence in the core}, in:
  \bibinfo{editor}{Schubert, G.} (Ed.), \bibinfo{booktitle}{Treatise on
  Geophysics (Second Edition)}. \bibinfo{edition}{second edition} ed..
  \bibinfo{publisher}{Elsevier}, \bibinfo{address}{Oxford}, pp.
  \bibinfo{pages}{161 -- 181}.
\newblock \DOIprefix\doi{https://doi.org/10.1016/B978-0-444-53802-4.00142-1}.
\bibitem[{{Olson} and {Christensen}(2006)}]{olson06}
\bibinfo{author}{{Olson}, P.}, \bibinfo{author}{{Christensen}, U.R.},
  \bibinfo{year}{2006}.
\newblock \bibinfo{title}{{Dipole moment scaling for convection-driven
  planetary dynamos}}.
\newblock \bibinfo{journal}{Earth and Planetary Science Letters}
  \bibinfo{volume}{250}, \bibinfo{pages}{561--571}.
\newblock \DOIprefix\doi{10.1016/j.epsl.2006.08.008}.
\bibitem[{{Olson} et~al.(2017){Olson}, {Landeau} and {Reynolds}}]{Olson2017}
\bibinfo{author}{{Olson}, P.}, \bibinfo{author}{{Landeau}, M.},
  \bibinfo{author}{{Reynolds}, E.}, \bibinfo{year}{2017}.
\newblock \bibinfo{title}{{Dynamo tests for stratification below the
  core-mantle boundary}}.
\newblock \bibinfo{journal}{Physics of the Earth and Planetary Interiors}
  \bibinfo{volume}{271}, \bibinfo{pages}{1--18}.
\newblock \DOIprefix\doi{10.1016/j.pepi.2017.07.003}.
\bibitem[{{Oruba} and {Dormy}(2014)}]{orubaD14GRL}
\bibinfo{author}{{Oruba}, L.}, \bibinfo{author}{{Dormy}, E.},
  \bibinfo{year}{2014}.
\newblock \bibinfo{title}{{Transition between viscous dipolar and inertial
  multipolar dynamos}}.
\newblock \bibinfo{journal}{\grl} \bibinfo{volume}{41},
  \bibinfo{pages}{7115--7120}.
\newblock \DOIprefix\doi{10.1002/2014GL062069}.
\bibitem[{{Petitdemange}(2018)}]{Petitdemange2018}
\bibinfo{author}{{Petitdemange}, L.}, \bibinfo{year}{2018}.
\newblock \bibinfo{title}{{Systematic parameter study of dynamo bifurcations in
  geodynamo simulations}}.
\newblock \bibinfo{journal}{Physics of the Earth and Planetary Interiors}
  \bibinfo{volume}{277}, \bibinfo{pages}{113--132}.
\newblock \DOIprefix\doi{10.1016/j.pepi.2018.02.001}.
\bibitem[{{Sakuraba} and {Roberts}(2009)}]{sakurabaR09}
\bibinfo{author}{{Sakuraba}, R.}, \bibinfo{author}{{Roberts}, P.},
  \bibinfo{year}{2009}.
\newblock \bibinfo{title}{{Generation of a strong magnetic field using uniform
  heat flux at the surface of the core}}.
\newblock \bibinfo{journal}{Nature Geoscience} \bibinfo{volume}{211},
  \bibinfo{pages}{802--805}.
\bibitem[{{Schaeffer} et~al.(2017){Schaeffer}, {Jault}, {Nataf} and
  {Fournier}}]{schaefferJNF17}
\bibinfo{author}{{Schaeffer}, N.}, \bibinfo{author}{{Jault}, D.},
  \bibinfo{author}{{Nataf}, H.C.}, \bibinfo{author}{{Fournier}, A.},
  \bibinfo{year}{2017}.
\newblock \bibinfo{title}{{Turbulent geodynamo simulations: a leap towards
  Earth's core}}.
\newblock \bibinfo{journal}{Geophysical Journal International}
  \bibinfo{volume}{211}, \bibinfo{pages}{1--29}.
\newblock \DOIprefix\doi{10.1093/gji/ggx265}.
\bibitem[{{Schrinner} et~al.(2012){Schrinner}, {Petitdemange} and
  {Dormy}}]{schrinner12}
\bibinfo{author}{{Schrinner}, M.}, \bibinfo{author}{{Petitdemange}, L.},
  \bibinfo{author}{{Dormy}, E.}, \bibinfo{year}{2012}.
\newblock \bibinfo{title}{{Dipole Collapse and Dynamo Waves in Global Direct
  Numerical Simulations}}.
\newblock \bibinfo{journal}{\apj} \bibinfo{volume}{752}, \bibinfo{pages}{121}.
\newblock \DOIprefix\doi{10.1088/0004-637X/752/2/121}.
\bibitem[{Schwaiger et~al.(2019)Schwaiger, Gastine and Aubert}]{Schwaiger19}
\bibinfo{author}{Schwaiger, T.}, \bibinfo{author}{Gastine, T.},
  \bibinfo{author}{Aubert, J.}, \bibinfo{year}{2019}.
\newblock \bibinfo{title}{Force balance in numerical geodynamo simulations: a
  systematic study}.
\newblock \bibinfo{journal}{Geophysical Journal International}
  \DOIprefix\doi{10.1093/gji/ggz192}.
\bibitem[{{Sheyko} et~al.(2017){Sheyko}, {Finlay}, {Favre} and
  {Jackson}}]{Sheyko2017}
\bibinfo{author}{{Sheyko}, A.}, \bibinfo{author}{{Finlay}, C.},
  \bibinfo{author}{{Favre}, J.}, \bibinfo{author}{{Jackson}, A.},
  \bibinfo{year}{2017}.
\newblock \bibinfo{title}{Strong field, scale separated, ultra low viscosity
  dynamos}.
\newblock \bibinfo{journal}{ArXiv e-prints} .
\bibitem[{{Soderlund} et~al.(2012){Soderlund}, {King} and
  {Aurnou}}]{Soderlund2012}
\bibinfo{author}{{Soderlund}, K.M.}, \bibinfo{author}{{King}, E.M.},
  \bibinfo{author}{{Aurnou}, J.M.}, \bibinfo{year}{2012}.
\newblock \bibinfo{title}{{The influence of magnetic fields in planetary dynamo
  models}}.
\newblock \bibinfo{journal}{Earth and Planetary Science Letters}
  \bibinfo{volume}{333}, \bibinfo{pages}{9--20}.
\newblock \DOIprefix\doi{10.1016/j.epsl.2012.03.038}.
\bibitem[{{Soderlund} et~al.(2015){Soderlund}, {Sheyko}, {King} and
  {Aurnou}}]{Soderlund2015}
\bibinfo{author}{{Soderlund}, K.M.}, \bibinfo{author}{{Sheyko}, A.},
  \bibinfo{author}{{King}, E.M.}, \bibinfo{author}{{Aurnou}, J.M.},
  \bibinfo{year}{2015}.
\newblock \bibinfo{title}{{The competition between Lorentz and Coriolis forces
  in planetary dynamos}}.
\newblock \bibinfo{journal}{Progress in Earth and Planetary Science}
  \bibinfo{volume}{2}, \bibinfo{pages}{24}.
\newblock \DOIprefix\doi{10.1186/s40645-015-0054-5}.
\bibitem[{{Sreenivasan} and {Jones}(2006)}]{sreenivasan06}
\bibinfo{author}{{Sreenivasan}, B.}, \bibinfo{author}{{Jones}, C.A.},
  \bibinfo{year}{2006}.
\newblock \bibinfo{title}{{The role of inertia in the evolution of spherical
  dynamos}}.
\newblock \bibinfo{journal}{Geophysical Journal International}
  \bibinfo{volume}{164}, \bibinfo{pages}{467--476}.
\newblock \DOIprefix\doi{10.1111/j.1365-246X.2005.02845.x}.
\bibitem[{{Stellmach} and {Hansen}(2004)}]{stellmach04}
\bibinfo{author}{{Stellmach}, S.}, \bibinfo{author}{{Hansen}, U.},
  \bibinfo{year}{2004}.
\newblock \bibinfo{title}{{Cartesian convection driven dynamos at low Ekman
  number}}.
\newblock \bibinfo{journal}{Physics Review E} \bibinfo{volume}{70},
  \bibinfo{pages}{056312}.
\newblock \DOIprefix\doi{10.1103/PhysRevE.70.056312}.
\bibitem[{{Takahashi} et~al.(2005){Takahashi}, {Matsushima} and
  {Honkura}}]{taka05}
\bibinfo{author}{{Takahashi}, E.}, \bibinfo{author}{{Matsushima}, A.},
  \bibinfo{author}{{Honkura}, Y.}, \bibinfo{year}{2005}.
\newblock \bibinfo{title}{{Simulation of a quasi-Taylor state magnetic field
  including polarity reversals on the Earth Simulator''}}.
\newblock \bibinfo{journal}{Science} \bibinfo{volume}{309},
  \bibinfo{pages}{459--461}.
\bibitem[{{Yadav} et~al.(2016a){Yadav}, {Gastine}, {Christensen}, {Duarte} and
  {Reiners}}]{Yadav16}
\bibinfo{author}{{Yadav}, R.K.}, \bibinfo{author}{{Gastine}, T.},
  \bibinfo{author}{{Christensen}, U.R.}, \bibinfo{author}{{Duarte}, L.D.V.},
  \bibinfo{author}{{Reiners}, A.}, \bibinfo{year}{2016}a.
\newblock \bibinfo{title}{{Effect of shear and magnetic field on the
  heat-transfer efficiency of convection in rotating spherical shells}}.
\newblock \bibinfo{journal}{Geophysical Journal International}
  \bibinfo{volume}{204}, \bibinfo{pages}{1120--1133}.
\newblock \DOIprefix\doi{10.1093/gji/ggv506}.
\bibitem[{{Yadav} et~al.(2016b){Yadav}, {Gastine}, {Christensen}, {Wolk} and
  {Poppenhaeger}}]{Yadav2016}
\bibinfo{author}{{Yadav}, R.K.}, \bibinfo{author}{{Gastine}, T.},
  \bibinfo{author}{{Christensen}, U.R.}, \bibinfo{author}{{Wolk}, S.J.},
  \bibinfo{author}{{Poppenhaeger}, K.}, \bibinfo{year}{2016}b.
\newblock \bibinfo{title}{{Approaching a realistic force balance in geodynamo
  simulations}}.
\newblock \bibinfo{journal}{Proceedings of the National Academy of Science}
  \bibinfo{volume}{113}, \bibinfo{pages}{12065--12070}.
\newblock \DOIprefix\doi{10.1073/pnas.1608998113}.

\end{thebibliography}
